\documentclass[11pt,a4paper]{article}
\pdfoutput=1
\usepackage{jheppub}
\usepackage{bbm}
\usepackage{amsfonts}
\usepackage[vcentermath]{youngtab}
\usepackage{booktabs}
 \usepackage{pdflscape}
 \usepackage{rotating}

\newcommand{\bZ}{\ensuremath{\mathbb{Z}}}

\newcommand{\fraksl}{\ensuremath{\mathfrak{sl}}}

\newcommand{\Tr}{\mbox{Tr}}

\def\beaa{\begin{eqnarray*}}
\def\eeaa{\end{eqnarray*}}
\def\bee{\begin{equation*}}
\def\eee{\end{equation*}}
\def\bea{\begin{eqnarray}}
\def\eea{\end{eqnarray}}
\def\be{\begin{equation}}
\def\ee{\end{equation}}
\def\ba{\begin{align}}
\def\ea{\end{align}}

\newcommand{\bem}{\begin{pmatrix}}
\newcommand{\eem}{\end{pmatrix}}

\def\={\;  = \;}
\def\+{\, + \,}

\def\bar{\overline}

\def\rt2{\sqrt{2}}

\newcommand{\mdot}{\raise1.5pt \hbox{.}}

\preprint{NIKHEF-2013-006}

{
\title{Colored HOMFLY polynomials from Chern-Simons theory}

\author[1]{Satoshi Nawata}
\author[2]{P. Ramadevi}
\author[2]{Zodinmawia}
\affiliation[1]{NIKHEF theory group,\\ Science Park 105,
1098 XG Amsterdam, The Netherlands \vspace{.2cm}}

\affiliation[2]{Department of Physics, Indian Institute of Technology Bombay,\\
 Mumbai, India, 400076\vspace{.2cm}}

\emailAdd{s.nawata@nikhef.nl}
\emailAdd{ramadevi@phy.iitb.ac.in}
\emailAdd{zodin@phy.iitb.ac.in}

\abstract{We elaborate the Chern-Simons field theoretic method to obtain colored HOMFLY invariants of knots and links. Using multiplicity-free quantum $6j$-symbols for $U_q(\fraksl_N)$, we present explicit evaluations of the HOMFLY invariants colored by symmetric representations for a variety of knots, two-component links and three-component links.
}


\begin{document}

\maketitle

\section{Introduction}

For the last few decades, we have seen tremendous developments on knot theory, a subject where diverse areas  in mathematics and physics interact in beautiful ways. The interplay between mathematics and physics involving knot theory was triggered by the seminal paper of Witten \cite{Witten:1988hf} which shows that Chern-Simons theory provides a natural framework to study link invariants. In particular, the expectation value of Wilson loop along a link $\cal L$ in $S^3$ gives a topological invariant of the link depending on the representation of the
gauge group. For a representation $R$ of $SU(2)$ gauge group, an invariant corresponds to a colored Jones polynomial $J_R({\cal L};q)$. Besides, one can relate an $SU(N)$ invariant with representation $R$ to a colored HOMFLY invariant $P_R({\cal L};a,q)$.
While the systematic procedure to compute $SU(N)$ invariants in $S^3$ is investigated in \cite{Kaul:1991vt, Kaul:1992rs, RamaDevi:1992dh}, it is very difficult to carry out explicit computations in general.
Even in mathematics,  although the definition \cite{morton1993invariants,lin2010hecke} of colored HOMFLY polynomials was provided,  explicit calculations for non-trivial knots and links are far from under control.

Nevertheless, there have been spectacular developments on computations of colored HOMFLY polynomials in recent years. 
For torus knots and links, the HOMFLY invariants colored by arbitrary representations can be, in principle, computed by using the generalizations \cite{lin2010hecke,Stevan:2010jh,Brini:2011wi} of the Rosso-Jones formulae \cite{Rosso:1993vn}. In addition, Kawagoe has lately formulated a mathematically rigorous procedure based on the linear skein theory to calculate HOMFLY invariants colored by symmetric representations for some non-torus knots and links \cite{Kawagoe:2012bt}. Furthermore, the explicit closed formulae of the colored HOMFLY polynomials $P_{[n]}({\cal K};a,q)$ with symmetric representations ($R=\raisebox{-.1cm}{\includegraphics[width=1.4cm]{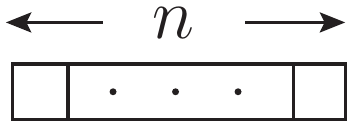}}$) were provided for the $(2,2p+1)$-torus knots \cite{Fuji:2012pm} and the twist knots \cite{Itoyama:2012fq,Nawata:2012pg,Kawagoe:2012bt}. 

In this paper, we shall demonstrate the computations of the HOMFLY polynomials colored by symmetric representations in the framework of Chern-Simons theory.
Exploiting the connection between Chern-Simons theory and the two-dimensional Wess-Zumino-Novikov-Witten (WZNW) model, the prescription to evaluate expectation values of Wilson loops was formulated entirely in terms of the fusion and braid operations on conformal blocks of the WZNW model \cite{Kaul:1991vt, Kaul:1992rs, RamaDevi:1992dh}. Therefore, the procedure inevitably involves the  $SU(N)$ quantum Racah coefficients (the quantum $6j$-symbols for $U_q(\fraksl_N)$), which makes explicit computations hard. The first step along this direction has been made in \cite{Zodinmawia:2011ud}: using the properties the $SU(N)$ quantum Racah coefficients should obey, the explicit expressions involving first few symmetric representations are determined. This result as well as the closed formulae of the twist knots motivated us to explore a closed form expression for the $SU(N)$ quantum Racah coefficients. We succeeded in writing the expression for multiplicity-free representations \cite{Nawata:2013ppa} which enables us to compute the colored HOMFLY polynomials carrying symmetric representations. To consider more complicated knots and links than the ones treated in \cite{Zodinmawia:2011ud}, we make use of the TQFT method developed in \cite{Kaul:1991vt, Kaul:1992rs}.

With this method, the expressions of the twist knots, the Whitehead links, the twist links and the Borromean rings \cite{Kawagoe:2012bt,Nawata:2012pg,Gukov:2013} have been reproduced up to 4 boxes. Even apart from these classes of knots and links,
the validity of our procedure is checked from the complete agreement with the results obtained in \cite{Itoyama:2012qt, Itoyama:2012re}. Furthermore, the explicit evaluations of multi-colored link invariants shed a new light on the general properties of colored HOMFLY invariants of links and provide meaningful implications on homological invariants of links.

The plan of the paper is as follows. In \S \ref{sec:CS}, we briefly review $U(N)$ Chern-Simons theory. In particular, we present the 
list of building blocks and the corresponding states which are necessary for calculations 
of colored HOMFLY polynomials. In \S \ref{sec:knots}, we compute the colored HOMFLY polynomials of seven-crossing knots and ten-crossing thick knots. 
In \S \ref{sec:links}, multi-colored HOMFLY invariants for two-component and
three-component links are expressed. We summarize and present several open problems in \S \ref{sec:conclusion}. For convenience, we explicitly show $SU(N)$ quantum Racah coefficients for some representations in Appendix \ref{sec:fusion}. Finally, we should mention that a Mathematica file with colored HOMFLY invariants whose expressions are too lengthy for the main text is linked on the arXiv page as an ancillary file.

\section{Invariants of knots and links in Chern-Simons theory}\label{sec:CS}

We shall briefly discuss $U(N)$ Chern-Simons theory necessary for computing invariants of framed knots and links. We refer the reader to \cite{Kaul:1991vt, Kaul:1992rs,RamaDevi:1992dh} for more details. The action for $U(N)\simeq U(1)\times SU(N)$ Chern-Simons theory is given by 
\begin{equation*}
 S=\frac{k_1}{4 \pi} \int_{S^3} B \wedge dB+
\frac{k}{4 \pi} \int_{S^3} \Tr\left(A \wedge dA+ \frac{2}{3}A \wedge A \wedge A \right)~,
\end{equation*}
where $B$ is the $U(1)$ gauge connection and $A$ is the 
$SU(N)$ matrix valued gauge connection with Chern-Simons coupling (also referred as Chern-Simons level)  $k_1$ and $k$ respectively. 
The Wilson loop observable for an arbitrary framed  link $\mathcal{L}$ made up of $s$-components $\{\mathcal{K}_{\beta}\}$,  
with framing number $f_{\beta}$, 
is the trace of the holonomies of the components ${\cal K}_{\beta}$:
\begin{equation*}
 W_{(R_1,n_1),(R_2,n_2),\ldots
(R_s,n_s)
}[{\cal L}] = \prod_{\beta=1}^{s}\Tr_{R_{\beta}}
U^A[{\cal K}_{\beta}]
 \Tr_{n_{\beta}}U^B[{\cal K}_{\beta}]~,
\end{equation*}
where the holonomy of the gauge field $A$ around a component knot ${\cal K}_{\beta}$,
carrying a representation $R_{\beta}$, of 
an  $s$-component link is denoted by 
$U^A[{\cal K}_{\beta}]=P[\exp \oint_{{\cal K}_{\beta}}A]$
and $n_{\beta}$ is the $U(1)$ charge carried by the component knot ${\cal K}_{\beta}$. Note that the framing number $f_{\beta}$ for the component knot ${\cal K}_\beta$ is the difference between the total number of left-handed crossings
and  that of right-handed crossings. 
The expectation values of these Wilson loop operators are the framed link invariants:
\begin{equation}\label{feynman}
V_{R_1,\ldots R_s}^{\{SU(N)\}}[{
\cal L}]
V_{n_1,\ldots ,n_s}^{\{U(1)\}}[{\cal L}]
=
\langle W_{(R_1,n_1),\ldots,(R_s,n_s)}[{\cal L}]\rangle=
 \frac{\int[{\cal D}B][{\cal D}A] e^{iS}W_{(R_1,n_1),\ldots ,(R_s,n_s)}
[{\cal L}]}{\int [{\cal D}B][{\cal D}A] e^{iS}}~.
\end{equation}
The $SU(N)$ invariants will be 
rational functions in the variable $q=\exp\left(\frac{2 \pi i}{k+N}\right)$ with the following choice for $U(1)$ charge and coupling $k_1$ \cite{Marino:2001re, Borhade:2003cu} , 
\begin{equation*}
 n_{\beta}=\frac{\ell^{(\beta)}}{\sqrt N} ~;~k_1=k+N~,
\end{equation*}
where $\ell^{(\beta)}$ is the total number of boxes in the 
Young Tableau representation $R_{\beta}$.
The $U(1)$ invariant involves only linking numbers $\{{\rm Lk}_{\alpha \beta}\}$  between the component knots and the framing numbers $\{f_{\beta}\}$
of each component knot. That is,
{\small
\begin{equation}
 V_{\frac{\ell^{(1)}}{\sqrt N},\ldots ,\frac{\ell^{(s)}}{\sqrt N}}^{\{U(1)\}}[
{\cal L}]
=(-1)^{\sum_{\beta} \ell^{(\beta)} f_{\beta}}
\exp\left(\frac{i \pi}{k+N}\sum_{\beta=1}^s
\frac{(\ell^{(\beta)})^2 f_{\beta}}{N}\right) \exp\left( \frac{i \pi}{k+N} \sum_{\alpha \neq \beta}
\frac{\ell^{(\alpha)}\ell^{(\beta)} {\rm Lk}_
{\alpha \beta}}{N} \right)~. \label{u1}
\end{equation}
}
Although the expectation values of Wilson loops \eqref{feynman} involve infinite-dimensional functional integrals,  one can obtain $SU(N)$  invariants non-perturbatively by  utilizing  the relation between  $SU(N)$ Chern-Simons  theory and the $SU(N)_k$ 
WZNW model \cite{Witten:1988hf}. The path integral of Chern-Simons  theory on a three-manifold with boundary defines an element in the quantum Hilbert space on the boundary, which is isomorphic to the space of conformal blocks of the WZNW model.  Using this fact, the evaluations of the expectation values of Wilson loops can be reduced to the braiding and fusion operations on conformal blocks once a link diagram is appropriately drawn in $S^3$ \cite{Kaul:1991vt, Kaul:1992rs, RamaDevi:1992dh}.

\begin{figure}[htbp]
\centering
\includegraphics[scale=.5]{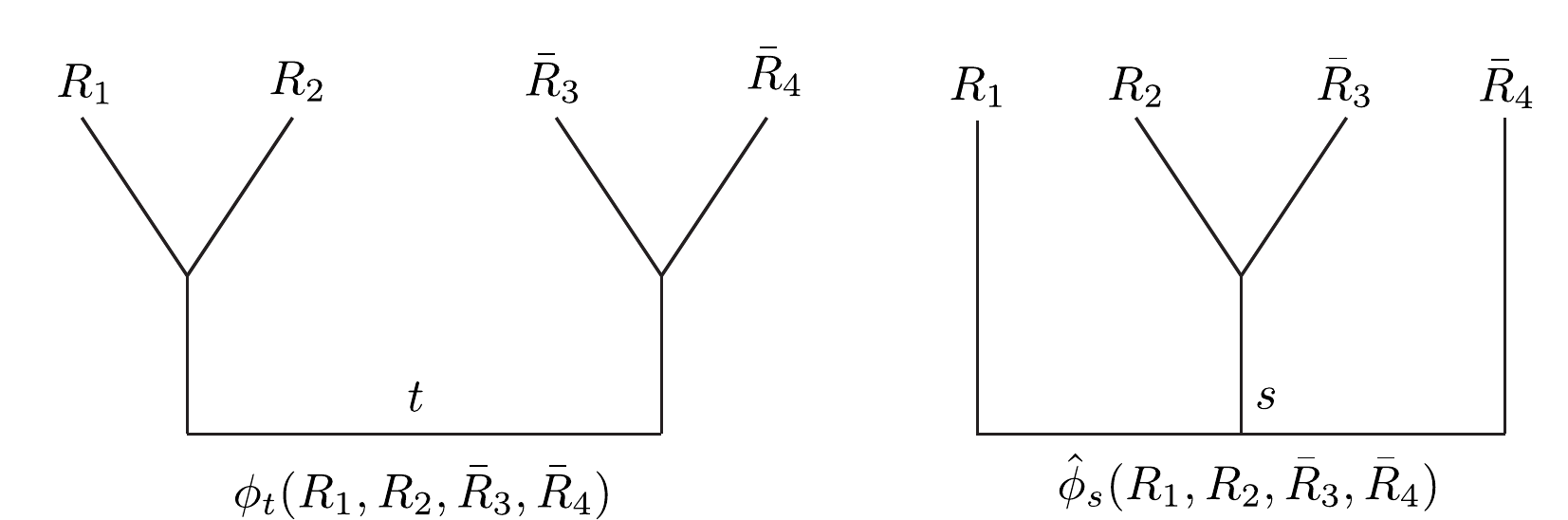}
\caption{Two bases for four-point conformal blocks}
\label{figs:conf}
\end{figure}

The Chern-Simons functional integral over a  three-ball with a four-punctured $S^2$ boundary  is  given by a state in the Hilbert space spanned by 
four-point conformal blocks. There are two different bases for four-point conformal blocks as shown in Figure \ref{figs:conf}
where the internal representations satisfy the fusion rules $t \in (R_1 \otimes R_2) \cap ( R_3 \otimes R_4)$ 
and
$s \in (R_2 \otimes \bar R_3) \cap (\bar R_1 \otimes R_4)$.
The conformal block $\vert \phi_t(R_1,R_2,\bar R_3, \bar R_4)  \rangle$ is suitable
for the  braiding operators $b_1^{(\pm)}$ and $b_3^{(\pm)}$ where $b_i$ denotes 
right-handed half-twist or braiding between the $i^{th}$ and the $(i+1)^{th}$ strand. Here the superscripts $(+)$ and $(-)$ denote the braidings on two strands in parallel orientations and in anti-parallel orientations respectively. Similarly, the braiding in the middle two strands  involving the operator
$b_2^{(\pm)}$ requires the conformal block
$\vert \hat {\phi}_s(R_1,R_2,\bar R_3,\bar R_4) \rangle$. In other words, these states become the eigenstates of the braiding operators
\begin{eqnarray*}
~b_1^
{(\pm)}\vert \phi_t(R_1,R_2,\bar R_3, \bar R_4)  \rangle
&=&
\lambda_t^{(\pm)}(R_1, R_2)
\vert \phi_t(R_2,R_1,\bar R_3,\bar R_4) \rangle~,\\
 b_2^{(\pm)} \vert \hat {\phi}_s(R_1,R_2,\bar R_3,\bar R_4) \rangle
&=&
\lambda_s^{(\pm)}(R_2,\bar R_3)
\vert \hat {\phi}_s(R_1,\bar R_3,R_2,\bar R_4) \rangle
~,\\
b_3^{(\pm)}
\vert \phi_t(R_1,R_2,\bar R_3,\bar R_4) \rangle&=&
\lambda_t^{(\pm)}(\bar R_3,\bar R_4)\vert \phi_t(R_1,R_2,\bar R_4,\bar R_3) \rangle~,
\end{eqnarray*}
where the braiding eigenvalues 
$\lambda_t^{(\pm)}(R_1,R_2)$ in the vertical framing are 
\begin{equation*}
 \lambda_t^{(\pm)}(R_1,R_2)=\epsilon^{(\pm)}_{t;R_1,R_2} \left
(q^{\frac{C_{R_1}+C_{R_2}-C_{R_t}}{2}}\right
)^{\pm 1}~,\label {brev}
\end{equation*}
where  $\epsilon^{(\pm)}_{t;R_1,R_2}=\pm 1$ (See (3.9) in \cite{Zodinmawia:2011ud}).
Here the quadratic Casimir 
for the representation $R$ is denoted by 
\begin{equation*}
C_R= \kappa_R - \frac{\ell^2}{2N}~,~\kappa_R=\frac{1}{2}[N\ell+\ell+\sum_i (\ell_i^2-2i\ell_i)]~, \label{casi}
\end{equation*}
where $\ell_i$ is the number of boxes in the $i^{\rm th}$ row of the Young Tableau corresponding to the representation
$R$ and $\ell$ is the total number of boxes.
The two bases in Figure \ref{figs:conf} are related by a fusion matrix $a_{ts}$ as follows:
\begin{equation*}
\vert \phi_t(R_1,R_2,\bar R_3, \bar R_4)  \rangle= a_{ts}\!\left[\begin{footnotesize}\begin{array}{cc}R_1&R_2 \\\bar R_3 &\bar R_4 \end{array}\end{footnotesize}\right]  
\vert \hat {\phi}_s(R_1,R_2,\bar R_3,\bar R_4) \rangle~. \label{dual}
\end{equation*}
The fusion matrix is determined by the $SU(N)$ quantum Racah coefficients. We have obtained these coefficients for a few representations in \cite{Zodinmawia:2011ud} and recently for symmetric representations in \cite{Nawata:2013ppa}.

\begin{figure}[htbp]
\centering
\includegraphics[scale=1]{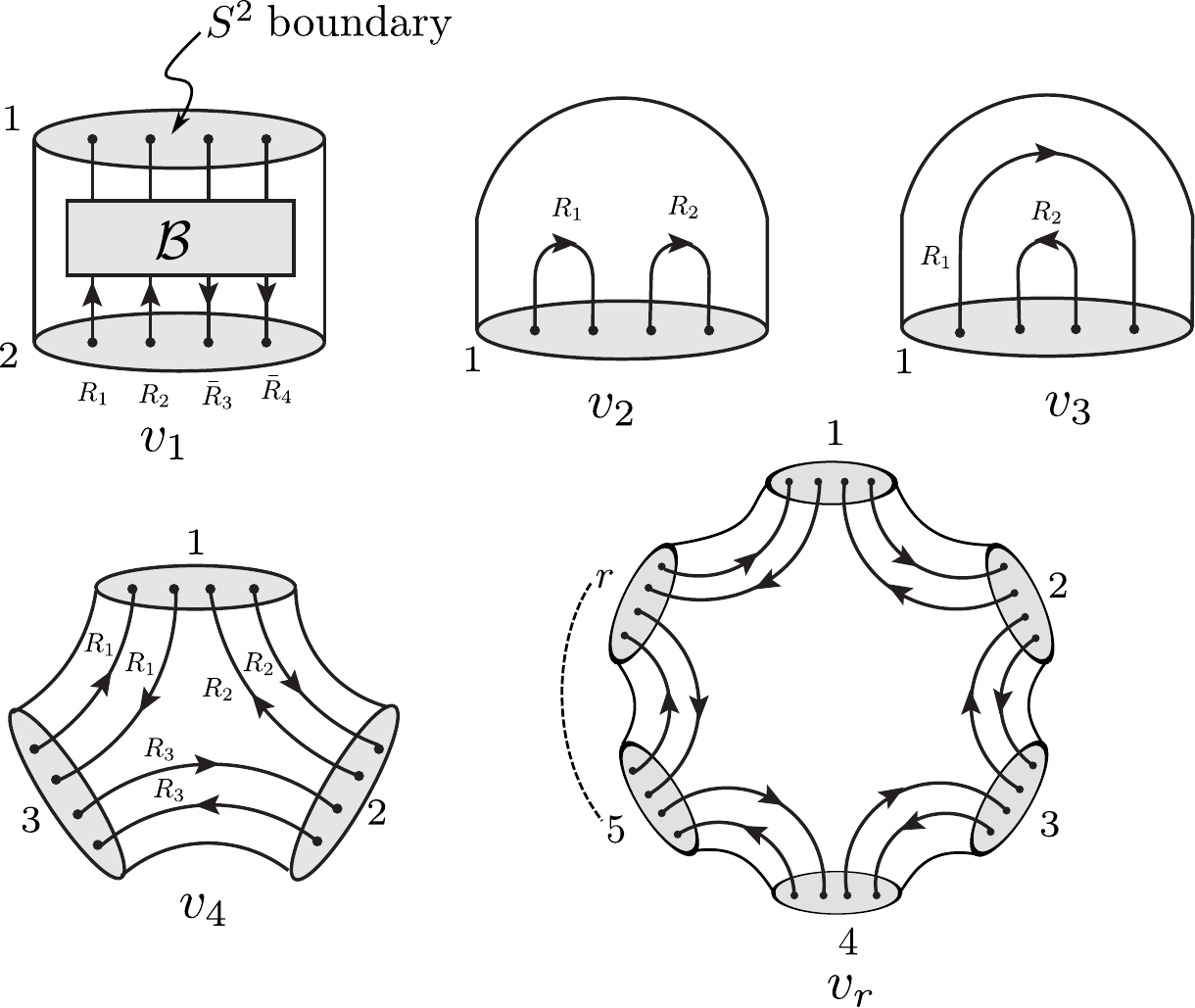}
\caption{Fundamental building blocks}
\label{figs:blocks1}
\end{figure}
\begin{figure}[htbp]
\centering
\includegraphics[scale=1]{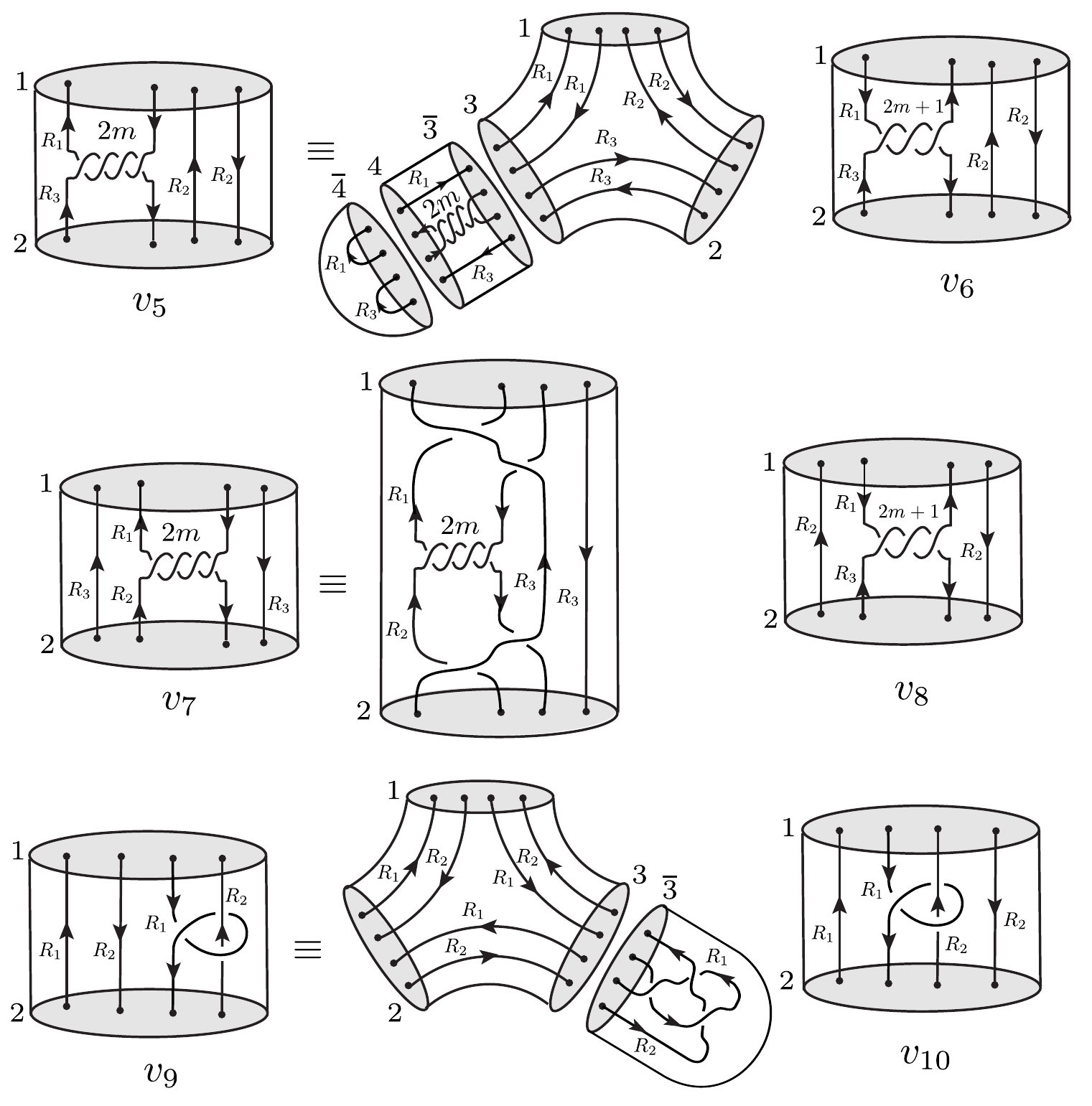}
\caption{Composite building blocks}
\label{figs:blocks2}
\end{figure}

In order to write the explicit polynomial form of $SU(N)$ invariants for many knots, we will require  
the states  corresponding to Chern-Simons functional integral over three-balls with several four-punctured $S^2$ boundaries.  Therefore, we will present these states  which will serve as the necessary building blocks 
for the knots and links in \S \ref{sec:knots} and \S\ref{sec:links}. 
Inside a  three-ball with two $S^2$ boundaries, we have a four-strand braid with braid word 
${\cal B}$ as shown by $v_1$ in Figure \ref{figs:blocks1}.  We shall call the 
state  as  $v_1$ whose form in terms of four-point conformal blocks of the WZNW 
model is
\begin{equation*}
v_1=\sum_{l\in(R_{1}\otimes R_{2})\cap(R_{3}\otimes R_{4})}\left\lbrace\mathcal{B}\,|\phi_l (R_{1},R_{2},\bar{R}_{3},\bar{R}_{4})\rangle\right\rbrace^{(1)}~{|\phi_l(R_{1},R_{2},\bar{R}_{3},\bar{R}_{4})\rangle}^{(2)}~,
\end{equation*}
where the superscripts outside the four-point conformal blocks denotes the boundaries (1) and (2) as indicated in Figure \ref{figs:blocks1}.  For the simplest three-ball with one $S^2$ boundary, the states
$v_2$ and $v_3$ will be 
\begin{equation*}
v_2={|\phi_{0}(R_{1},\bar{R}_{1},R_{2},R_{2})\rangle}^{(1)}~,~
v_3=|\hat{\phi}_{0}(R_{1},\bar{R}_{2},R_{2},\bar{R}_{1})\rangle^{(1)}~,
\end{equation*}
where the subscript $0$ in $\phi_0$ represents the singlet representation. This procedure can be generalized to three-ball with more than one $S^2$ boundary. 
For definiteness, we first write the
state $v_4$ for three $S^2$ boundaries and generalize to state $v_r$ for  $r$ $S^2$ boundaries:
\begin{eqnarray*}
v_4&=&\sum_{l\in(R_{1}\otimes\bar{R}_{1})\cap(R_{2}\otimes\bar{R}_{2})\cap(R_{3}\otimes\bar{R}_{3})}\frac{1}{\epsilon_{l}\sqrt{\dim_{q}l}}~
{|\phi_{l}(\bar{R}_{1},R_{1},\bar{R}_{2},R_{2})\rangle}^{(1)}\\
~~~&~&~~~~~~~~~~~~~~~~~~{|\phi_{l}(\bar{R}_{2},R_{2},\bar{R}_{3},R_{3})\rangle}^{(2)}{|\phi_{l}(\bar{R}_{3},R_{3},\bar{R}_{1},R_{1})\rangle}^{(3)}~,\\
v_{r}&=&\sum_{l}\frac{1}{\left(\epsilon_{l}\sqrt{\dim_{q}l}\right)^{r-2}}~|\phi_{l}(\bar{R}_{1},R_{1},\bar{R}_{2},R_{2})\rangle^{(1)}\ldots ~
|\phi_{l}(\bar{R}_{r},R_{r},\bar{R}_{1},R_{1})\rangle^{(r)}~.
\end{eqnarray*}
Here $\epsilon_l\equiv \epsilon_l^{R_1,\bar R_1}=\pm 1$ (See (3.1) in \cite{Zodinmawia:2011ud}).
Using these fundamental building blocks, we can obtain states for three-ball with two  $S^2$ boundaries in Figure \ref{figs:blocks2} which we call composite building blocks. 
For example, the state $v_5$ for the
two $S^2$ boundaries can be viewed as gluing of appropriate oppositely oriented boundaries of $v_1$, $v_2$ and $v_4$ as shown.  In the equivalent diagram for $v_5$, we have 
indicated $\bar 3$ on an $S^2$ boundary  which denotes that it is oppositely oriented to the $S^2$ boundary numbered by $3$. 
Gluing along two oppositely oriented $S^2$ boundaries amounts to taking inner product of the states corresponding to the boundaries. For example, gluing along the $S^2$ boundaries $3$ and $\bar{3}$ results in
\begin{equation*}
~^{(\bar 3)}{\langle \phi_{l}(R_3, \bar{R}_{3},R_1,\bar{R}_{1},)\vert}
{|\phi_{x}(\bar{R}_{3},R_{3},\bar{R}_{1},R_{1})\rangle}^{(3)}=\delta_{lx}~.
\eee   
Writing the states $v_1$, $v_2$ and $v_4$ and taking appropriate inner product, we obtain  the state $v_5$ as
\begin{eqnarray*}
v_{5}&=&\sum_{l,r}\frac{1}{\epsilon_{l} \sqrt{\dim_{q}l}}~\epsilon_{r}^{\bar{R}_1,R_3}\sqrt{\dim_{q}r}~(\lambda_{r}^{(-)}(\bar{R}_{1},R_{3})){}^{2m}~a_{lr}\!\left[\begin{footnotesize}\begin{array}{cc}
R_{1} & \bar{R}_{1}\\
R_{3} & \bar{R}_{3}
\end{array}\end{footnotesize}\right] \\
~&~&~\times {|\phi_{l}(\bar{R}_{1},R_{1},\bar{R}_{2},R_{2})\rangle}^{(1)}~{|\phi_{l}(R_{3},\bar{R}_{3},R_{2},\bar{R}_{2})\rangle}^{(2)}~.
\end{eqnarray*}
The state $v_6$  is similar to the state $v_5$, but it involves an odd number of braidings:
\begin{eqnarray*}
 v_{6}&=&\sum_{l,r}\frac{1}{\epsilon_{l}^{\bar{R}_1,R_3}\sqrt{\dim_{q}l}}~\epsilon_{r}^{R_1,R_3}\sqrt{\dim_{q}r}(\lambda_{r}^{(+)}(R_{1},R_{3}))^{-(2m+1)}~a_{rl}\!\left[\begin{footnotesize}\begin{array}{cc}
R_{1} & R_{3}\\
\bar{R}_{1} & \bar{R}_{3}
\end{array}\end{footnotesize}\right] \\
~&~&\times |\phi_{\bar{l}}(R_{1},\bar{R}_{3},\bar{R}_{2},R_{2})\rangle^{(1)}~|{\phi_{l}(R_{3},\bar{R}_{1},R_{2},\bar{R}_{2})\rangle}^{(2)}~. 
\end{eqnarray*}
The state $v_7$ can be obtained by gluing $v_1$ , $v_5$  and again a $v_1$ with appropriate braid words ${\cal B}$ in both $v_1$'s  as shown in the equivalent diagram: 
\begin{eqnarray*}
 v_{7}&=&\sum_{l,r,x,y}\frac{1}{\epsilon_{l}\sqrt{\dim_{q}l}}~\epsilon_{r}^{\bar{R}_1,R_3}\sqrt{\dim_{q}r}~(\lambda_{r}^{(-)}(\bar{R}_{1},R_{3}))^{2m}a_{lr}\!\left[\begin{footnotesize}\begin{array}{cc}
R_{1} & \bar{R}_{1}\\
R_{3} & \bar{R}_{3}
\end{array}\end{footnotesize}\right]\\
~&~&\times a_{lx}\!\left[\begin{footnotesize}\begin{array}{cc}
R_{1} & \bar{R}_{1}\\
\bar{R}_{3} & R_{3}
\end{array}\end{footnotesize}\right]a_{ly}\!\left[\begin{footnotesize}\begin{array}{cc}
\bar{R}_{2} & R_{2}\\
R_{3} & \bar{R}_{3}
\end{array}\end{footnotesize}\right]|{\phi_{x}(\bar{R}_{3},\bar{R}_{1},R_{1},R_{3})\rangle}^{(1)}~{|\phi_{y}(R_{3},R_{2},\bar{R}_{2},\bar{R}_{3})\rangle}^{(2)} ~.
 \end{eqnarray*}
The  state $v_8$ is almost the same as the state $v_7$ except for an odd number instead of an even number of braidings:
\begin{eqnarray*}
 v_{8}&=&\sum_{l,r,x,y}\frac{1}{\epsilon_{l}^{\bar{R}_1,R_3}\sqrt{\dim_{q}l}}~\epsilon_{r}^{R_1,R_3}\sqrt{\dim_{q}r}~(\lambda_{r}^{(+)}(R_{1},R_{3}))^{-(2m+1)}\!a_{rl}\left[\begin{footnotesize}\begin{array}{cc}
R_{1} & R_{3}\\
\bar{R}_{1} & \bar{R}_{3}
\end{array}\end{footnotesize}\right] \cr
&&\times a_{x\bar{l}}\!\left[\begin{footnotesize}\begin{array}{cc}
\bar{R}_{2} & R_{1}\\
\bar{R}_{3} & R_{2}
\end{array}\end{footnotesize}\right]a_{yl}\!\left[\begin{footnotesize}\begin{array}{cc}
R_{2} & R_{3}\\
\bar{R}_{1} & \bar{R}_{2}
\end{array}\end{footnotesize}\right]|{\phi_{x}(\bar{R}_{2},R_{1},\bar{R}_{3},R_{2})\rangle}^{(1)}~{|\phi_{y}(R_{2},R_{3},\bar{R}_{1},\bar{R}_{2})\rangle}^{(2)}~.
 \end{eqnarray*}
The equivalent diagram for $v_9$ in Figure \ref{figs:blocks2} determines the state as
\begin{eqnarray*}
 v_{9}&=&\sum_{l,x,y,z}\frac{1}{\epsilon_{l}^{\bar{R}_1,R_2}\sqrt{\dim_{q}l}}~\epsilon_{z}^{\bar R_1,R_2}\sqrt{\dim_{q}z}~a_{xl}\!\left[\begin{footnotesize}\begin{array}{cc}
R_{1} & \bar{R}_{1}\\
R_{2} & \bar{R}_{2}
\end{array}\end{footnotesize}\right]a_{yx}\!\left[\begin{footnotesize}\begin{array}{cc}
R_{2} & R_{1}\\
\bar{R}_{1} & \bar{R}_{2}
\end{array}\end{footnotesize}\right]\\
&&\times a_{zy}\!\left[\begin{footnotesize}\begin{array}{cc}
\bar{R}_{1} & R_{2}\\
R_{1} & \bar{R}_{2}
\end{array}\end{footnotesize}\right]\lambda_{x}^{(-)}(R_{1},\bar{R}_{1})~\lambda_{y}^{(+)}(R_{1},R_{2})~\lambda_{z}^{(-)}(\bar{R}_{1},R_{2})~{|\phi_{l}(\bar{R}_{1},R_{2},R_{1},\bar{R}_{2})\rangle}^{(1)} \\
&~&\times {|\phi_{\bar{l}}(R_{1},\bar{R}_{2},\bar{R}_{1},R_{2})\rangle}^{(2)}~. 
 \end{eqnarray*}
To get the state $v_{10}$, we could glue the state $v_1$ to the state $v_9$ with braid words ${\cal B}=b_2^{(+)} \{b_3^{(-)}\}^{-1}$:
\begin{eqnarray*}
v_{10}&=&\sum_{l,x,y,z}\frac{1}{\epsilon_{l}^{\bar{R}_1,R_2}\sqrt{\dim_{q}l}}~\epsilon_{z}^{\bar{R}_1,R_2}\sqrt{\dim_{q}z}~a_{xl}\!\left[\begin{footnotesize}\begin{array}{cc}
R_{1} & \bar{R}_{1}\\
R_{2} & \bar{R}_{2}
\end{array}\end{footnotesize}\right]a_{yx}\!\left[\begin{footnotesize}\begin{array}{cc}
R_{2} & R_{1}\\
\bar{R}_{1} & \bar{R}_{2}
\end{array}\end{footnotesize}\right]\\
~&~&\times a_{zy}\!\left[\begin{footnotesize}\begin{array}{cc}
\bar{R}_{1} & R_{2}\\
R_{1} & \bar{R}_{2}
\end{array}\end{footnotesize}\right]a_{sl}\!\left[\begin{footnotesize}\begin{array}{cc}
R_{1} & \bar{R}_{1}\\
R_{2} & \bar{R}_{2}
\end{array}\end{footnotesize}\right]a_{tl}\!\left[\begin{footnotesize}\begin{array}{cc}
R_{1} & \bar{R}_{1}\\
R_{2} & \bar{R}_{2}
\end{array}\end{footnotesize}\right]\lambda_{x}^{(-)}(R_{1},\bar{R}_{1})~\lambda_{y}^{(+)}(R_{1},R_{2})\\
~&~&\times \lambda_{z}^{(-)}(\bar{R}_{1},R_{2})~{|\phi_{s}(\bar{R}_{1},R_{1},\bar{R}_{2},R_{2})\rangle}^{(1)}~{|\phi_{t}(R_{1},\bar{R}_{1},R_{2},\bar{R}_{2})\rangle}^{(2)} ~.
\end{eqnarray*}
Our main aim is to 
redraw many knots and links in $S^3$ using these  building blocks so that the invariant involves 
only the multiplicity-free  Racah coefficients. 
For instance, see Figure~6 and Figure~7 in \cite{Ramadevi:1993hu}
where equivalent diagrams of  the knots $\bf 9_{42}$ 
and $\bf 10_{71}$  are drawn.
 
As an example, we will demonstrate the evaluation of the Chern-Simons invariant for 
the knot $\bf {10_{152}}$ by using these building blocks. 
\begin{figure}[htbp]
\centering
\includegraphics[scale=1]{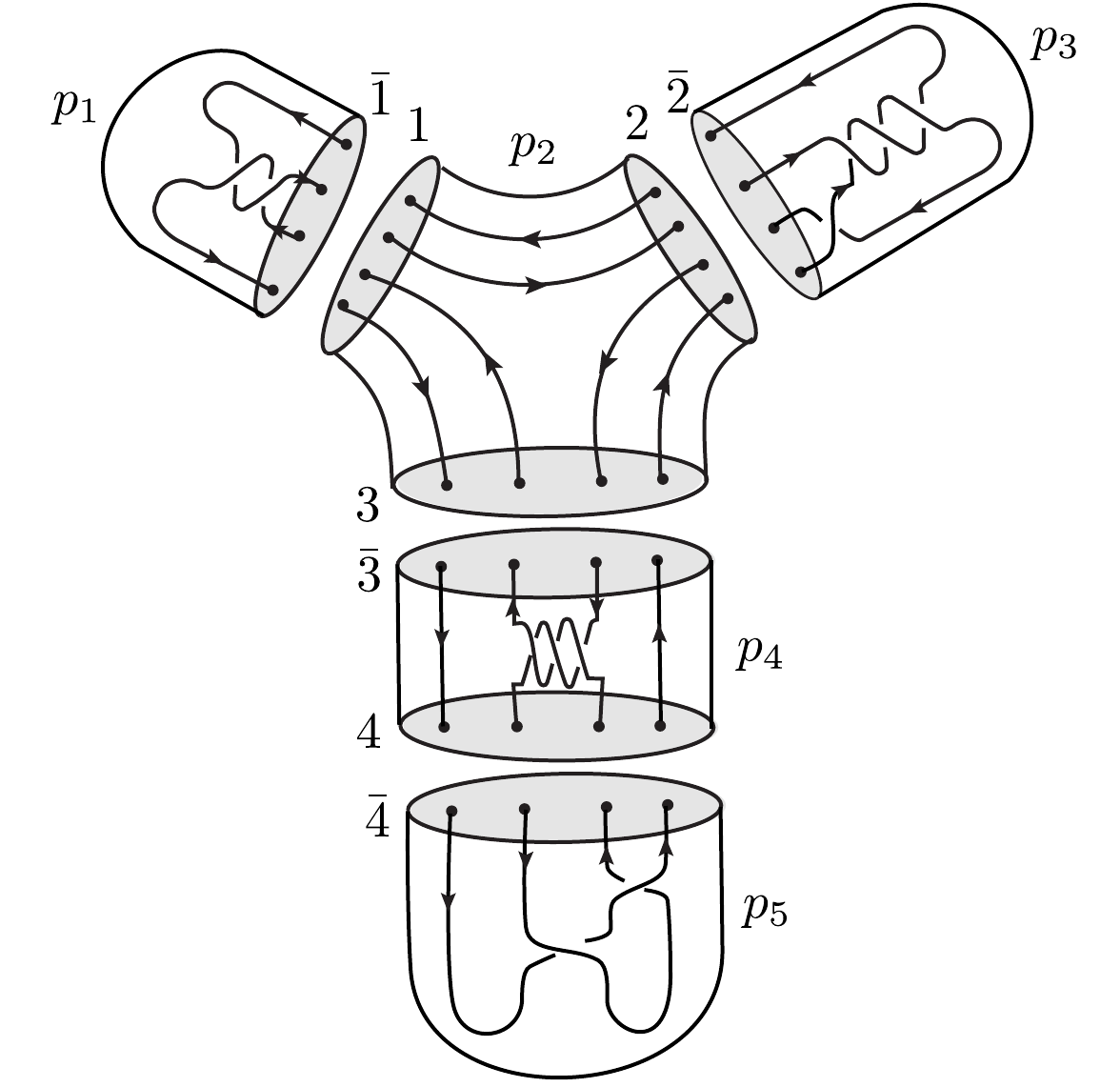}
\caption{The knot $\bf{10_{152}}$ in gluing of three-balls.}
\label{figs:glue}
\end{figure}
This knot can be viewed  as  gluing of five three-balls as shown in Figure \ref{figs:glue}. Using the states for the fundamental and composite building blocks , we can 
directly write the different states corresponding to the three-balls $\{p_i\}$ ($i=1,\cdots,5$) as follows:
\beaa
 p_{1}&=&\sum_{s_1,u}\epsilon_{u}^{R,\bar{R}}\sqrt{\dim_{q}u}~a_{s_1u}\!\left[\begin{footnotesize}\begin{array}{cc}
R & \bar{R}\\
R & \bar{R}
\end{array}\end{footnotesize}\right]~(\lambda_{u}^{(-)}(R,\bar{R}))^{2}~{|\phi_{s_1}(\bar{R},R,\bar{R},R)\rangle}^{(\bar{1})} ~,\\
p_{2}&=&\sum_l\frac{1}{\epsilon_{l}^{R,\bar{R}}\sqrt{\dim_{q}l}}~{|\phi_{l}(R,\bar{R},R,\bar{R})\rangle}^{(1)}~{|\phi_{l}(R,\bar{R},R,\bar{R})\rangle}^{(2)}~{|\phi_{l}(\bar{R},R,\bar{R},R)\rangle}^{(3)} ~,\\
 p_{3}&=&\sum_{s_2,v}\epsilon_{v}^{R,R}\sqrt{\dim_{q}v}~a_{s_2v}\!\left[\begin{footnotesize}\begin{array}{cc}
R & R\\
\bar{R} & \bar{R}
\end{array}\end{footnotesize}\right]~\lambda_{s_2}^{(-)}(R,\bar{R})~(\lambda_{v}^{(+)}(R,R))^{3}~{|\phi_{s_2}(\bar{R},R,\bar{R},R)\rangle}^{(\bar{2})} ~, \\
 p_{4}&=&\sum_{l_1,r,x,y}\frac{1}{\epsilon_{l_1}^{R,\bar{R}}\sqrt{\dim_{q}l_1}}~\epsilon_{r}^{R,R}\sqrt{\dim_{q}r}(\lambda_{r}^{(+)}(R,R))^{3}\!a_{rl_1}\left[\begin{footnotesize}\begin{array}{cc}
R & R\\
\bar{R} & \bar{R}
\end{array}\end{footnotesize}\right] \\
&&\times a_{xl_1}\!\left[\begin{footnotesize}\begin{array}{cc}
R & \bar{R}\\
R & \bar{R}
\end{array}\end{footnotesize}\right]a_{yl_1}\!\left[\begin{footnotesize}\begin{array}{cc}
\bar{R} & \bar{R}\\
R & R
\end{array}\end{footnotesize}\right]{|\phi_{x}(R,\bar{R},R,\bar{R})\rangle}^{(\bar{3})}~{|\phi_{y}(\bar{R},\bar{R},R,R)\rangle}^{(4)} ~,\\
 p_{5}&=&\sum_{s_3,z}\epsilon_{z}^{R,\bar{R}}\sqrt{\dim_{q}z}~a_{s_3z}\!\left[\begin{footnotesize}\begin{array}{cc}
R & R\\
\bar{R} & \bar{R}
\end{array}\end{footnotesize}\right]~\lambda_{s_3}^{(+)}(R,R)~\lambda_{z}^{(-)}(R,\bar{R})~{|\phi_{s_3}(R,R,\bar{R},\bar{R})\rangle} ^{(\bar{4})} ~.
\eeaa
One can obtain the  $SU(N)$ invariant after gluing all the three-balls together which amounts to taking appropriate inner products of 
the above five states:
\begin{flalign*}
V_{R}^{\{SU(N)\}}[{\bf 10_{152}}]= & \sum_{l,l_1,r,u,v,x,y,z}\frac{1}{\epsilon_{l}^{R,\bar{R}}\sqrt{\dim_{q}l}~\epsilon_{l_1}^{R,\bar{R}}\sqrt{\dim_{q}l_1}}~\epsilon_{z}^{R,\bar{R}}\sqrt{\dim_{q}z}&\\
   & \times \epsilon_{u}^{R,\bar{R}}\sqrt{\dim_{q}u}~\epsilon_{r}^{R,R}\sqrt{\dim_{q}r}~\epsilon_{v}^{R,R}\sqrt{\dim_{q}v}&\\
&\times a_{rl_1}\!\left[\begin{footnotesize}\begin{array}{cc}
R & R\\
\bar{R} & \bar{R}
\end{array}\end{footnotesize}\right]~a_{ll_1}\!\left[\begin{footnotesize}\begin{array}{cc}
R & \bar{R}\\
R & \bar{R}
\end{array}\end{footnotesize}\right]~a_{yl_1}\!\left[\begin{footnotesize}\begin{array}{cc}
R & R\\
\bar{R} & \bar{R}
\end{array}\end{footnotesize}\right]~a_{yz}\!\left[\begin{footnotesize}\begin{array}{cc}
R & R\\
\bar{R} & \bar{R}
\end{array}\end{footnotesize}\right]~a_{lu}\!\left[\begin{footnotesize}\begin{array}{cc}
R & \bar{R}\\
R & \bar{R}
\end{array}\end{footnotesize}\right]& \\
   & \times a_{vl}\!\left[\begin{footnotesize}\begin{array}{cc}
R & R\\
\bar{R} & \bar{R}
\end{array}\end{footnotesize}\right]~(\lambda_{r}^{(+)}(R,R))^{3}~\lambda_{y}^{(+)}(R,R)~\lambda_{z}^{(-)}(R,\bar{R})&\\
   & \times(\lambda_{u}^{(-)}(R,\bar{R}))^{2}~\lambda_{l}^{(-)}(R,\bar{R})~(\lambda_{v}^{(+)}(R,R))^{3}.\end{flalign*}
The framing number for the knot $\bf 10_{152}$ as drawn in Figure \ref{figs:glue} is $f=-11$, giving the $U(1)$ invariant \eqref{u1}:
\bee
V_{R}^{\{U(1)\}}[{\bf 10_{152}}]= q^{-11 \ell^2 \over 2N}~.
\end{equation*}
To adjust the framing number to zero, we introduce an additional twist with framing number $- f$ to the knot.  
This additional twist  leads to a multiplication by
a factor $q^{ -f C_R}$, giving the 
unreduced HOMFLY polynomial
\beaa
\overline P_R({\bf 10_{152}};a=q^N,q)= q^{11 C_R}V_{R}^{\{U(1)\}}[{\bf 10_{152}}] V_R^{\{SU(N)\}}[{\bf 10_{152}}] ~.
\eeaa
Note that the factor $q^{- f C_R}$ can be incorporated as a framing correction in the
vertical framing braiding eigenvalues:
\begin{equation*}
\hat {\lambda}^{(+)}_s(R,R)= q^{C_R}\lambda^{(+)}_s(R,R)~,~ \hat {\lambda}^{(-)}_s(R,\bar{R})= q^{C_R}\lambda^{(-)}_s(R,\bar{R})~,\label{sframe}
\end{equation*}
where $\hat {\lambda}$ denotes standard framing eigenvalues which will be used in the explicit computation of colored HOMFLY polynomials of knots.

Let us conclude this section by providing the definition of  a  reduced HOMFLY polynomial. The reduced colored HOMFLY polynomial of a knot $\cal K$ is expressed by
\beaa
 P_R({\cal K};a,q)=\overline P_R({\cal K};a,q)/\overline P_R(\bigcirc;a,q)~.
\eeaa
The unknot factor carrying the rank-$n$ symmetric representation is
\beaa
\overline P_{[n]}(\bigcirc;a,q)=\frac{q^{n/2}(a;q)_n}{a^{n/2}(q;q)_n}~,
\eeaa
where we denote the $q$-Pochhammer symbols by $(z;q)_{k}=\prod_{j=0}^{k-1} (1-zq^j)$.


\section{Colored HOMFLY polynomials for knots}\label{sec:knots}

In this section, we shall demonstrate computations of the colored HOMFLY polynomials of knots. 
The closed form expressions of the colored HOMFLY polynomials with all the symmetric representations are known for the $(2,2p+1)$-torus knots \cite{Fuji:2012pm} and the twist knots  \cite{Itoyama:2012fq,Nawata:2012pg,Kawagoe:2012bt}. In addition, we have verified the results in \cite{Itoyama:2012re,Itoyama:2012qt} for  the colored HOMFLY polynomials of the knots ${\bf 6_2}$, ${\bf 6_3}$, ${\bf 7_3}$ and ${\bf 7_5}$ up to 4 boxes. Hence, we present  the $[3]$-colored HOMFLY polynomials for the other seven-crossing knots in \S \ref{sec:seven}. (The $[2]$-colored HOMFLY polynomials are collected in \cite{Zodinmawia:2011ud}.) For each figure, we redraw the (left) diagram in the table of Rolfsen into the right diagram to which we apply the method in \S\ref{sec:CS}.

In \S \ref{sec:thick}, we shall compute the colored HOMFLY polynomials of thick knots \cite{Dunfield:2005si}. 
If all the generators of the HOMFLY homology of a given knot have the same $\delta$-grading, the knot is called homologically thin. (See more detail in \cite{Dunfield:2005si}.) Otherwise, it is called homologically thick. For a thick knot, the colored HOMFLY polynomial is a crucial information to obtain the homological invariant since it is not clear the homological invariant obey the exponential growth property. For the $(3,4)$-torus knot ($\bf 8_{19}$) and the knot $\bf 9_{42}$, the $[2]$-colored superpolynomials are given in {\cite{Gukov:2011ry}. In addition, the colored HOMFLY polynomials of the knot $\bf 10_{139}$ are given up to 4 boxes \cite{Itoyama:2012re}. The evaluation of the colored HOMFLY polynomials for these knots are beyond the scope of the method provided by \cite{Kawagoe:2012bt}. We have verified these 
results of the knots $\bf 9_{42}$ and $\bf 10_{139}$ using our approach.
Here we present  the invariants for 10-crossing thick knots except  the knot $\bf 10_{161}$ \footnote{Since the knot $\bf 10_{161}$ can be written as a three-strand knot, the colored HOMFLY polynomials can be obtained by the method in \cite{Itoyama:2012re}.}.

Before going into detail, let us fix the notation. In this paper, we use the skein relation 
\bee a^{1/2}P_{L_+}-a^{-1/2}P_{L_-}=(q^{1/2}-q^{-1/2})P_{L_0}~,
\eee for an uncolored HOMFLY polynomial $P({\cal K};a,q)=P_{[1]}({\cal K};a,q)$ with $P(\bigcirc;a,q)=1$.
To show the colored HOMFLY polynomials concisely, we use the following convention. 

{\bf Example}
\beaa
&&f(a,q)\left( \begin{array}{ccccccccccc}     9 & 10 & 11 & 12 \\ 5 & 6 & 7 & 8  \\ 1 & 2 & 3 & 4  \end{array} \right)\\
&=&f(a,q)\Big[(1+2q+3q^2+4q^3)+a(5+6q+7q^2+8q^3)+a^2(9+10q+11q^2+12q^3)\Big]\eeaa
In the matrix,  the $q$-degree is assigned to the horizontal axis and the $a$-degree is scaled along the vertical axis.

\subsection{Seven-crossing knots}\label{sec:seven}
\subsubsection{$7_4$ knot}
\begin{figure}[h]
\centering{\includegraphics[scale=1]{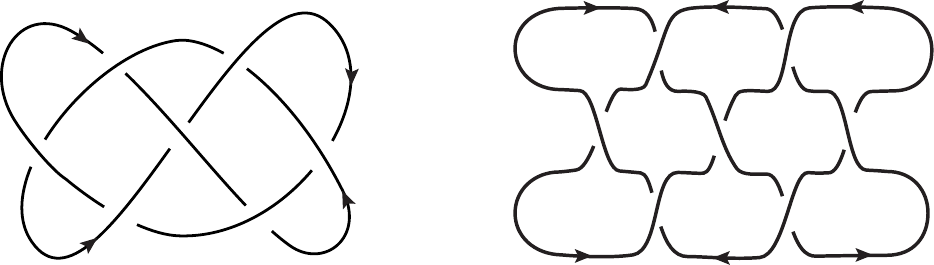}}
\caption{$\bf{7_{4}}$ knot}
\end{figure}
\beaa
P_{R}({\bf 7_{4}},;\,a,q) &= & \frac{1}{\dim_qR}\sum_{s,t,s^{\prime},u,v}\epsilon_{s}^{R,R}\,\sqrt{\dim_{q}s}\,\epsilon_{v}^{R,R}\,\sqrt{\dim_{q}v}\,\hat{\lambda}_{s}^{(+)}(R,\, R)\, a_{ts}\!\left[\begin{footnotesize}\begin{array}{cc}
\bar{R} & R\\
R & \bar{R}
\end{array}\end{footnotesize}\right]\\
  & &\times (\hat{\lambda}_{t}^{(-)}(R,\,\bar{R}))^{2}\, a_{ts^{\prime}}\!\left[\begin{footnotesize}\begin{array}{cc}
R & \bar{R}\\
\bar{R} & R
\end{array}\end{footnotesize}\right]\,\hat{\lambda}_{s^{\prime}}^{(+)}(\bar{R},\,\bar{R})\, a_{us^{\prime}}\!\left[\begin{footnotesize}\begin{array}{cc}
R & \bar{R}\\
\bar{R} & R
\end{array}\end{footnotesize}\right]\\
  & &\times (\hat{\lambda}_{u}^{(-)}(R,\,\bar{R}))^{2}\, a_{uv}\!\left[\begin{footnotesize}\begin{array}{cc}
\bar{R} & R\\
R & \bar{R}
\end{array}\end{footnotesize}\right]\,\hat\lambda_{v}^{(+)}(R,\, R).
\eeaa
\beaa
&&P_{[3]}({\bf 7_4}; a,q) =\tfrac{a^3}{ q^{3}} \times \\
&&\begin{tiny}
\left( \begin{array}{cccccccccccccccccccccccccc}  0 & 0 & 0 & 0 & 0 & 0 & 0 & 0 & 0 & 0 & 0 & 0 & 0 & 0 & 0 & 0 & 0 & 0 & 0 & 0 & 0 & 0 & 0 & 0 & -1 & 0 \\  0 & 0 & 0 & 0 & 0 & 0 & 0 & 0 & 0 & 0 & 0 & 0 & 0 & 0 & 0 & 0 & 0 & 0 & 0 & 1 & -1 & -1 & -2 & 1 & 1 & 1 \\  0 & 0 & 0 & 0 & 0 & 0 & 0 & 0 & 0 & 0 & 0 & 0 & 0 & 0 & 0 & 1 & 2 & 3 & -2 & -2 & -2 & 4 & 3 & 1 & -1 & -1 \\  0 & 0 & 0 & 0 & 0 & 0 & 0 & 0 & 0 & 0 & 0 & -1 & 0 & 2 & 4 & 3 & -5 & -5 & -3 & 6 & 4 & -1 & -3 & -2 & 1 & 0 \\  0 & 0 & 0 & 0 & 0 & 0 & 0 & -1 & -1 & -4 & -1 & 2 & 6 & 2 & -9 & -9 & -3 & 8 & 5 & -3 & -4 & -2 & 2 & 0 & 0 & 0 \\  0 & 0 & 0 & 0 & 1 & 0 & -1 & -7 & -2 & 5 & 12 & 4 & -11 & -11 & -1 & 11 & 7 & -4 & -4 & -2 & 3 & 0 & 0 & 0 & 0 & 0 \\  0 & 0 & 2 & 3 & 1 & -9 & -6 & 4 & 18 & 8 & -12 & -15 & -1 & 13 & 8 & -4 & -4 & -2 & 4 & 0 & 0 & 0 & 0 & 0 & 0 & 0 \\  1 & 2 & 3 & -6 & -9 & -3 & 15 & 14 & -8 & -17 & -6 & 12 & 9 & -4 & -4 & -2 & 3 & 0 & 0 & 0 & 0 & 0 & 0 & 0 & 0 & 0 \\  2 & 0 & -4 & -6 & 4 & 12 & 2 & -10 & -10 & 6 & 8 & 0 & -4 & -2 & 2 & 0 & 0 & 0 & 0 & 0 & 0 & 0 & 0 & 0 & 0 & 0 \\  1 & -2 & -1 & 2 & 3 & 0 & -6 & 0 & 3 & 2 & -1 & -2 & 1 & 0 & 0 & 0 & 0 & 0 & 0 & 0 & 0 & 0 & 0 & 0 & 0 & 0 \\ \end{array} \right)\end{tiny}\eeaa

\subsubsection{$7_6$ knot}
\begin{figure}[h]
\centering{\includegraphics[scale=1]{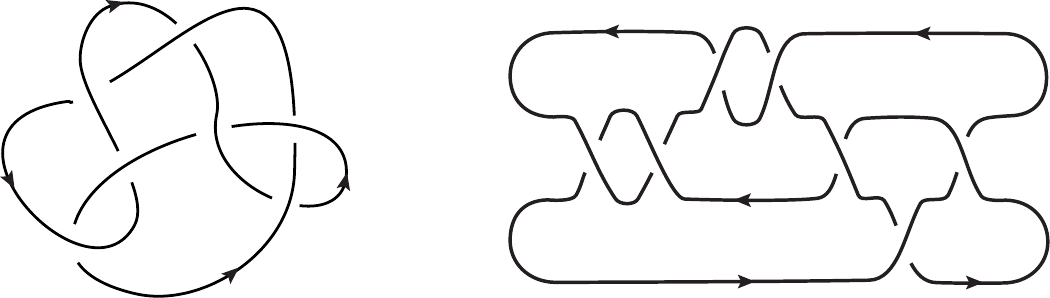}}
\caption{$\bf{7_{6}}$ knot}
\end{figure}
\begin{align*}
P_{R}({\bf 7_{6}};\,a,q)= & \frac{1}{\dim_qR}\sum_{s,t,s^{\prime},u,v}\epsilon_{s}^{R,\bar{R}}\,\sqrt{\dim_{q}s}\,\epsilon_{v}^{\bar{R},R}\,\sqrt{\dim_{q}v}\,(\hat{\lambda}_{s}^{(-)}(R,\,\bar{R}))^{-2}\, a_{ts}\!\left[\begin{footnotesize}\begin{array}{cc}
\bar{R} & R\\
\bar{R} & R
\end{array}\end{footnotesize}\right]&\cr
&\times (\hat{\lambda}_{t}^{(-)}(\bar{R},\, R))^{2}\, a_{ts^{\prime}}\!\left[\begin{footnotesize}\begin{array}{cc}
\bar{R} & R\\
\bar{R} & R
\end{array}\end{footnotesize}\right]\,(\hat{\lambda}_{s^{\prime}}^{(-)}(R,\,\bar{R}))^{-1}\, a_{us^{\prime}}\!\left[\begin{footnotesize}\begin{array}{cc}
\bar{R} & \bar{R}\\
R & R
\end{array}\end{footnotesize}\right]&\cr
   &\times (\hat{\lambda}_{u}^{(+)}(\bar{R},\,\bar{R}))^{-1}\, a_{uv}\!\left[\begin{footnotesize}\begin{array}{cc}
\bar{R} & \bar{R}\\
R & R
\end{array}\end{footnotesize}\right]\,(\hat{\lambda}_{v}^{(-)}(\bar{R},\, R))^{-1}.&
\end{align*}
\beaa
&&P_{[3]}({\bf 7_6}; a,q) =\tfrac{1}{a^9 q^{17}} \times \\
&&\begin{tiny}
\left( \begin{array}{cccccccccccccccccccccccccccc}  0 & 0 & 0 & 0 & 0 & 0 & 0 & 0 & 0 & 0 & 0 & 0 & 0 & 0 & 1 & -1 & 0 & 1 & 1 & -1 & -2 & 1 & 2 & 0 & -1 & -1 & 1 & 0 \\  0 & 0 & 0 & 0 & 0 & 0 & 0 & 0 & 0 & -1 & 1 & 0 & 0 & -3 & 1 & 3 & 0 & -5 & -3 & 4 & 5 & -2 & -4 & -2 & 3 & 1 & 0 & -1 \\  0 & 0 & 0 & 0 & 0 & 1 & -1 & -1 & 3 & 2 & -2 & -5 & 6 & 10 & -3 & -12 & -3 & 14 & 10 & -7 & -10 & -1 & 8 & 3 & -2 & -2 & 0 & 1 \\  0 & 0 & -1 & 2 & -2 & -3 & 2 & 4 & -5 & -13 & 5 & 19 & -3 & -26 & -14 & 24 & 21 & -11 & -23 & -4 & 14 & 6 & -5 & -4 & 0 & 2 & -1 & 0 \\  0 & 2 & -1 & -2 & 4 & 7 & -1 & -14 & 4 & 26 & 9 & -25 & -24 & 21 & 34 & 0 & -24 & -12 & 13 & 11 & -1 & -4 & -1 & 2 & 0 & 0 & 0 & 0 \\  -1 & -3 & 2 & 2 & -2 & -15 & -2 & 16 & 10 & -20 & -27 & 5 & 26 & 6 & -16 & -16 & 4 & 6 & 2 & -3 & -1 & 0 & 0 & 0 & 0 & 0 & 0 & 0 \\  2 & 2 & 2 & -6 & 1 & 12 & 10 & -6 & -15 & 3 & 16 & 8 & -4 & -8 & 2 & 2 & 2 & 0 & 0 & 0 & 0 & 0 & 0 & 0 & 0 & 0 & 0 & 0 \\  -1 & -3 & -2 & 3 & 1 & -3 & -9 & -1 & 3 & 3 & -2 & -3 & -1 & 0 & 0 & 0 & 0 & 0 & 0 & 0 & 0 & 0 & 0 & 0 & 0 & 0 & 0 & 0 \\  0 & 2 & 1 & 1 & -2 & 1 & 1 & 2 & 0 & 0 & 0 & 0 & 0 & 0 & 0 & 0 & 0 & 0 & 0 & 0 & 0 & 0 & 0 & 0 & 0 & 0 & 0 & 0 \\  0 & 0 & -1 & 0 & 0 & 0 & 0 & 0 & 0 & 0 & 0 & 0 & 0 & 0 & 0 & 0 & 0 & 0 & 0 & 0 & 0 & 0 & 0 & 0 & 0 & 0 & 0 & 0 \\ \end{array} \right)\end{tiny}\eeaa

\subsubsection{$7_7$ knot}
\begin{figure}[h]
\centering{\includegraphics[scale=1]{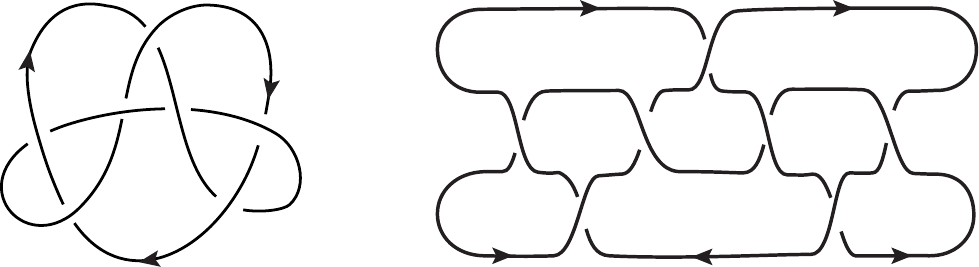}}
\caption{$\bf{7_{7}}$ knot}
\end{figure}
\begin{flalign*}
P_{R}({\bf 7_{7}};\,a,q) = & \frac{1}{\dim_qR}\sum_{s,t,s^{\prime},u,v,w,x}\epsilon_{s}^{R,R}\,\sqrt{\dim_{q}s}\,\epsilon_{x}^{R,R}\,\sqrt{\dim_{q}x}\,(\hat{\lambda}_{s}^{(+)}(R,\, R))\, a_{ts}\!\left[\begin{footnotesize}\begin{array}{cc}
\bar{R} & R\\
R & \bar{R}
\end{array}\end{footnotesize}\right]&\\
   &\times (\hat{\lambda}_{t}^{(-)}(\bar{R},\, R))\, a_{ts^{\prime}}\!\left[\begin{footnotesize}\begin{array}{cc}
R & \bar{R}\\
R & \bar{R}
\end{array}\end{footnotesize}\right]\,(\hat{\lambda}_{s^{\prime}}^{(-)}(\bar{R},\, R))^{-1}\, a_{us^{\prime}}\!\left[\begin{footnotesize}\begin{array}{cc}
R & R\\
\bar{R} & \bar{R}
\end{array}\end{footnotesize}\right]&\\
   &\times (\hat{\lambda}_{\bar{u}}^{(+)}(\bar{R},\,\bar{R}))^{-1}\, a_{uv}\!\left[\begin{footnotesize}\begin{array}{cc}
R & R\\
\bar{R} & \bar{R}
\end{array}\end{footnotesize}\right]\,(\hat{\lambda}_{v}^{(-)}(R,\,\bar{R}))^{-1}a_{wv}\!\left[\begin{footnotesize}\begin{array}{cc}
R & \bar{R}\\
R & \bar{R}
\end{array}\end{footnotesize}\right]&\\
   &\times \hat{\lambda}_{w}^{(-)}(R,\bar{R})\, a_{wx}\!\left[\begin{footnotesize}\begin{array}{cc}
\bar{R} & R\\
R & \bar{R}
\end{array}\end{footnotesize}\right]\,\hat{\lambda}_{x}^{(+)}(R,R).
\end{flalign*}
\beaa
&&P_{[3]}({\bf 7_7}; a,q) =\tfrac{1}{a^3 q^{13}} \times \\
&&\begin{tiny}
\left( \begin{array}{cccccccccccccccccccccccccccc}  0 & 0 & 0 & 0 & 0 & 0 & 0 & 0 & 0 & 0 & 0 & 0 & 0 & 0 & 0 & 0 & 0 & 0 & 0 & 0 & 0 & 0 & 0 & 0 & 0 & 1 & 0 & 0 \\  0 & 0 & 0 & 0 & 0 & 0 & 0 & 0 & 0 & 0 & 0 & 0 & 0 & 0 & 0 & 0 & 0 & 0 & 0 & 0 & -2 & 0 & 0 & 2 & -2 & -2 & -2 & 0 \\  0 & 0 & 0 & 0 & 0 & 0 & 0 & 0 & 0 & 0 & 0 & 0 & 0 & 0 & 0 & 1 & 2 & 0 & -4 & 2 & 6 & 8 & -3 & -4 & 0 & 5 & 4 & 1 \\  0 & 0 & 0 & 0 & 0 & 0 & 0 & 0 & 0 & 0 & 0 & -2 & 0 & 0 & 4 & -6 & -12 & -2 & 11 & 9 & -13 & -18 & -3 & 7 & 5 & -6 & -4 & -2 \\  0 & 0 & 0 & 0 & 0 & 0 & 0 & 1 & 2 & -4 & 0 & 6 & 13 & -8 & -19 & 2 & 30 & 22 & -16 & -25 & 4 & 20 & 13 & -7 & -4 & 1 & 4 & 1 \\  0 & 0 & 0 & 0 & -2 & 2 & 2 & -6 & -7 & 7 & 20 & -9 & -38 & -10 & 36 & 35 & -25 & -42 & -4 & 25 & 13 & -13 & -10 & 1 & 3 & 0 & -2 & 0 \\  0 & 1 & -2 & 2 & 3 & -3 & -9 & 4 & 24 & 4 & -37 & -23 & 29 & 51 & -11 & -43 & -14 & 26 & 19 & -9 & -10 & 2 & 4 & 1 & -2 & 1 & 0 & 0 \\  -1 & 1 & 2 & -2 & -6 & 0 & 16 & 7 & -22 & -22 & 11 & 38 & 3 & -27 & -17 & 12 & 16 & -3 & -7 & -1 & 2 & 1 & -1 & 0 & 0 & 0 & 0 & 0 \\  1 & -1 & -2 & -1 & 7 & 4 & -9 & -11 & 3 & 18 & 3 & -11 & -9 & 4 & 7 & -1 & -2 & -1 & 1 & 0 & 0 & 0 & 0 & 0 & 0 & 0 & 0 & 0 \\  0 & -1 & 2 & 1 & -2 & -3 & 0 & 6 & 0 & -3 & -2 & 1 & 2 & -1 & 0 & 0 & 0 & 0 & 0 & 0 & 0 & 0 & 0 & 0 & 0 & 0 & 0 & 0 \\ \end{array} \right)\end{tiny}\eeaa

\subsection{Thick knots}\label{sec:thick}

\subsubsection{$10_{124}$ knot}
Note that the knot $\bf 10_{124}$ is the $(3,5)$-torus knot.
\begin{figure}[h]
\centering{\includegraphics[scale=1]{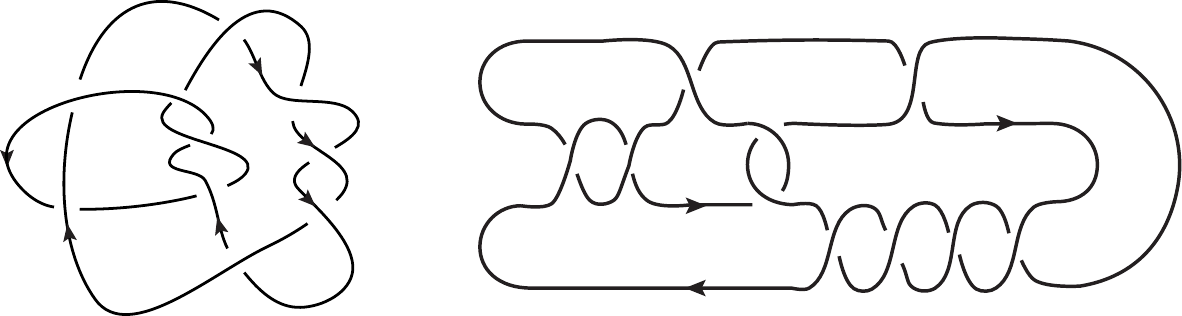}}
\caption{$\bf{10_{124}}$ knot}
\end{figure}
\beaa
P_R({\bf 10_{124}};a,q) &= & \tfrac{1}{\dim_qR}\sum_{l,r,x,y,z}\frac{1}{\epsilon_{l}^{R,\bar{R}}\sqrt{\dim_{q}l}}~\epsilon_{r}^{R,R}\sqrt{\dim_{q}r}~\epsilon_{x}^{R,R}\sqrt{\dim_{q}x} \\
& & \times \epsilon_{z}(R,R)\sqrt{\dim_{q}z}~a_{rl}\!\left[\begin{footnotesize}\begin{array}{cc}
R & R\\
\bar{R} & \bar{R}
\end{array}\end{footnotesize}\right]a_{xl}\!\left[\begin{footnotesize}\begin{array}{cc}
R & R\\
\bar{R} & \bar{R}
\end{array}\end{footnotesize}\right]a_{yl}\!\left[\begin{footnotesize}\begin{array}{cc}
R & \bar{R}\\
R & \bar{R}
\end{array}\end{footnotesize}\right]a_{zy}\!\left[\begin{footnotesize}\begin{array}{cc}
R & R\\
\bar{R} & \bar{R}
\end{array}\end{footnotesize}\right]\\
& &\times (\hat{\lambda}_{r}^{(+)}(R,R))^{-2}~(\hat{\lambda}_{x}^{(+)}(R,R))^{-5}~(\hat{\lambda}_{y}^{(-)}(R,\bar{R}))^{-1}~(\hat{\lambda}_{z}^{(+)}(R,R))^{-2}\eeaa

\begin{align*}
&P_{[2]}({\bf 10_{124}}; a,q)= \tfrac{1}{a^{12}q^{16}} \times\\
&\begin{footnotesize}
\left( \begin{array}{ccccccccccccccccccccccccc}  1 & 0 & 1 & 2 & 2 & 2 & 4 & 2 & 4 & 4 & 3 & 3 & 4 & 2 & 3 & 3 & 2 & 1 & 2 & 1 & 1 & 1 & 0 & 0 & 1 \\  -1 & -2 & -2 & -4 & -6 & -6 & -8 & -9 & -8 & -9 & -9 & -8 & -8 & -7 & -5 & -5 & -5 & -3 & -2 & -2 & -1 & -1 & -1 & 0 & 0 \\  0 & 2 & 2 & 4 & 5 & 7 & 7 & 9 & 8 & 9 & 7 & 7 & 6 & 6 & 4 & 3 & 2 & 2 & 1 & 1 & 0 & 0 & 0 & 0 & 0 \\  0 & 0 & -1 & -2 & -2 & -3 & -4 & -3 & -3 & -4 & -3 & -2 & -2 & -1 & -1 & -1 & 0 & 0 & 0 & 0 & 0 & 0 & 0 & 0 & 0 \\  0 & 0 & 0 & 0 & 1 & 0 & 1 & 1 & 0 & 0 & 1 & 0 & 0 & 0 & 0 & 0 & 0 & 0 & 0 & 0 & 0 & 0 & 0 & 0 & 0 \\ \end{array} \right)\end{footnotesize}
\end{align*}

\subsubsection{$10_{128}$ knot}
\begin{figure}[h]
\centering{\includegraphics[scale=1]{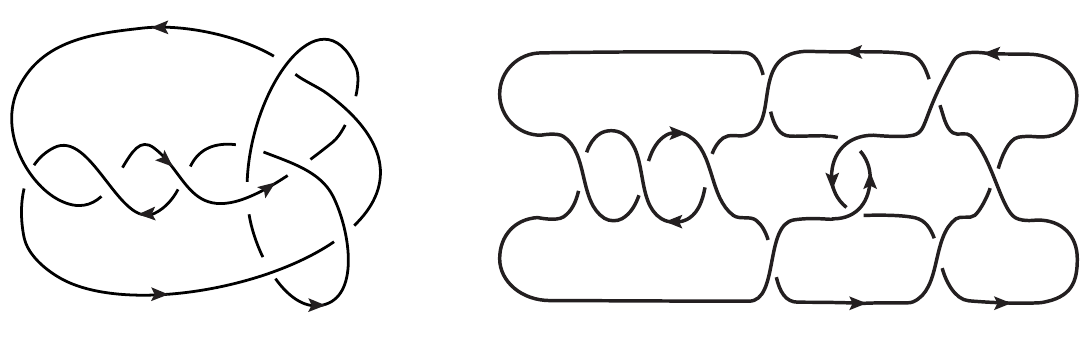}}
\caption{$\bf{10_{128}}$ knot}
\end{figure}
\begin{flalign*}
P_R({\bf 10_{128}};a,q) = & \sum_{l,r,x,y,x^{\prime},y^{\prime}}\frac{1}{\epsilon_{l}^{R,\bar{R}}\sqrt{\dim_{q}l}}~\epsilon_{r}^{R,\bar{R}}\sqrt{\dim_{q}r}~\epsilon_{x}^{R,\bar{R}}\sqrt{\dim_{q}x}&\\
 &\times \epsilon_{x^{\prime}}^{R,\bar{R}}\sqrt{\dim_{q}x^{\prime}}~a_{rl}\!\left[\begin{footnotesize}\begin{array}{cc}
R & \bar{R}\\
R & \bar{R}
\end{array}\end{footnotesize}\right]a_{yl}\!\left[\begin{footnotesize}\begin{array}{cc}
R & R\\
\bar{R} & \bar{R}
\end{array}\end{footnotesize}\right]a_{y^{\prime}l}\!\left[\begin{footnotesize}\begin{array}{cc}
R & R\\
\bar{R} & \bar{R}
\end{array}\end{footnotesize}\right]a_{yx}\!\left[\begin{footnotesize}\begin{array}{cc}
R & R\\
\bar{R} & \bar{R}
\end{array}\end{footnotesize}\right]&\\
 & \times a_{y^{\prime}x^{\prime}}\!\left[\begin{footnotesize}\begin{array}{cc}
R & R\\
\bar{R} & \bar{R}
\end{array}\end{footnotesize}\right]~(\hat{\lambda}_{r}^{(-)}(R,\bar{R}))^{-2}~(\hat{\lambda}_{y}^{(+)}(R,R))^{-2}~(\hat{\lambda}_{x}^{(-)}(R,\bar{R}))^{-1}~& \\
&\times (\hat{\lambda}_{y^{\prime}}^{(+)}(R,R))^{-2}~ (\hat{\lambda}_{x^{\prime}}^{(-)}(R,\bar{R}))^{-3}\end{flalign*}

\beaa
&&P_{[2]}({\bf 10_{128}}; a,q)=\\
&& \tfrac{1}{a^{12}q^{14}} \times\left( \begin{array}{ccccccccccccccccccccc}  0 & 0 & 1 & -1 & 0 & 2 & 0 & -1 & 1 & 1 & 0 & 0 & 0 & 0 & 1 & -1 & 0 & 2 & -1 & -1 & 1 \\  0 & 1 & -1 & -1 & 2 & 1 & -2 & -1 & 3 & 1 & -1 & 1 & 1 & 1 & 0 & 0 & 3 & 0 & -2 & 1 & 1 \\  1 & -1 & -1 & 2 & 1 & -2 & -2 & 1 & 1 & -3 & -2 & -1 & -1 & -2 & -1 & 1 & -1 & -2 & 0 & 0 & 0 \\  -1 & -2 & 0 & 0 & -3 & -4 & 1 & 1 & -3 & 0 & 0 & -1 & 0 & 0 & 1 & 0 & -1 & 0 & 0 & 0 & 0 \\  0 & 2 & 2 & 2 & 0 & 3 & 4 & 1 & 1 & 2 & 0 & 1 & 1 & 1 & 0 & 0 & 0 & 0 & 0 & 0 & 0 \\  0 & 0 & -1 & -2 & -1 & 0 & -1 & -1 & -1 & -1 & 0 & 0 & 0 & 0 & 0 & 0 & 0 & 0 & 0 & 0 & 0 \\  0 & 0 & 0 & 0 & 1 & 0 & 0 & 0 & 0 & 0 & 0 & 0 & 0 & 0 & 0 & 0 & 0 & 0 & 0 & 0 & 0 \\ \end{array} \right)\eeaa

\subsubsection{$10_{132}$ knot}
\begin{figure}[h]
\centering{\includegraphics[scale=1]{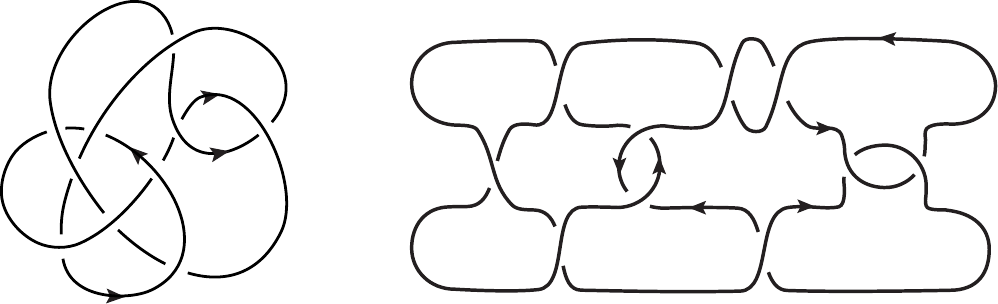}}
\caption{$\bf{10_{132}}$ knot}
\end{figure}
\begin{flalign*}
P_R({\bf 10_{132}};a,q) = & \sum_{l,r,x,y,x^{\prime},y^{\prime}}\frac{1}{\epsilon_{l}^{R,\bar{R}}\sqrt{\dim_{q}l}}~\epsilon_{r}^{R,R}\sqrt{\dim_{q}r}~\epsilon_{x}^{R,R}\sqrt{\dim_{q}x} & \\
 & \times \epsilon_{x^{\prime}}^{R,\bar{R}}\sqrt{\dim_{q}x^{\prime}}~a_{rl}\!\left[\begin{footnotesize}\begin{array}{cc}
R & R\\
\bar{R} & \bar{R}
\end{array}\end{footnotesize}\right]a_{yl}\!\left[\begin{footnotesize}\begin{array}{cc}
R & \bar{R}\\
R & \bar{R}
\end{array}\end{footnotesize}\right]a_{y^{\prime}l}\!\left[\begin{footnotesize}\begin{array}{cc}
R & R\\
\bar{R} & \bar{R}
\end{array}\end{footnotesize}\right]a_{xy}\!\left[\begin{footnotesize}\begin{array}{cc}
R & R\\
\bar{R} & \bar{R}
\end{array}\end{footnotesize}\right]&\\
   & \times a_{y^{\prime}x^{\prime}}\!\left[\begin{footnotesize}\begin{array}{cc}
R & R\\
\bar{R} & \bar{R}
\end{array}\end{footnotesize}\right]~(\hat{\lambda}_{r}^{(+)}(R,R))^{2}~(\hat{\lambda}_{y}^{(-)}(R,\bar{R}))^{3}~(\hat{\lambda}_{x}^{(+)}(R,R))^{2} & \\
 & \times (\hat{\lambda}_{y^{\prime}}^{(+)}(R,R))^{-2}~(\hat{\lambda}_{x^{\prime}}^{(-)}(R,\bar{R}))^{-1}\end{flalign*}

\beaa
&&P_{[2]}({\bf 10_{132}}; a,q)=\tfrac{a}{q^{4}} 
\left( \begin{array}{ccccccccccccccc}  0 & 0 & 0 & 0 & 0 & 0 & 0 & 1 & 0 & 1 & 1 & 0 & 0 & 1 & 0 \\  0 & 0 & 0 & 0 & -1 & -1 & 0 & -1 & -2 & -1 & -2 & -2 & 0 & -1 & -1 \\  0 & 0 & 1 & -1 & -1 & 2 & 1 & -1 & 2 & 1 & 1 & 2 & 1 & 0 & 1 \\  0 & 1 & 0 & -2 & 2 & 2 & -2 & 0 & 1 & -1 & 0 & 0 & -1 & 0 & 0 \\  1 & 0 & -2 & 1 & 2 & -2 & -1 & 1 & 0 & 0 & 0 & 0 & 0 & 0 & 0 \\  0 & -1 & 0 & 2 & -1 & -1 & 1 & 0 & 0 & 0 & 0 & 0 & 0 & 0 & 0 \\ \end{array} \right)
\eeaa

\subsubsection{$10_{136}$ knot}
\begin{figure}[h]
\centering{\includegraphics[scale=1]{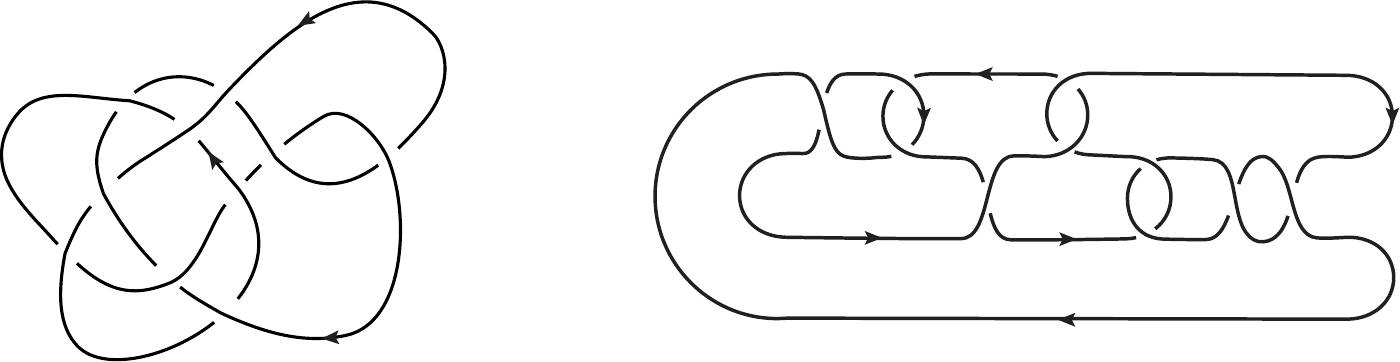}}
\caption{$\bf{10_{136}}$ knot}
\end{figure}
\begin{flalign*}
P_R({\bf 10_{136}};a,q) = & \sum_{l,l_{1},r,x,y,z}\frac{1}{\epsilon_{l}^{R,\bar{R}}\sqrt{\dim_{q}l}}~\epsilon_{x}^{R,R}\sqrt{\dim_{q}x}~\epsilon_{r}^{R,\bar{R}}\sqrt{\dim_{q}r} &\\
 &\times \epsilon_{z}^{R,\bar{R}}\sqrt{\dim_{q}u}~a_{xl}\!\left[\begin{footnotesize}\begin{array}{cc}
R & R\\
\bar{R} & \bar{R}
\end{array}\end{footnotesize}\right]a_{rl_{1}}\!\left[\begin{footnotesize}\begin{array}{cc}
R & \bar{R}\\
R & \bar{R}
\end{array}\end{footnotesize}\right]a_{ll_{1}}\!\left[\begin{footnotesize}\begin{array}{cc}
R & \bar{R}\\
R & \bar{R}
\end{array}\end{footnotesize}\right]a_{yl}\!\left[\begin{footnotesize}\begin{array}{cc}
R & R\\
\bar{R} & \bar{R}
\end{array}\end{footnotesize}\right] &\\
 & \times a_{yz}\!\left[\begin{footnotesize}\begin{array}{cc}
R & R\\
\bar{R} & \bar{R}
\end{array}\end{footnotesize}\right]~(\hat{\lambda}_{x}^{(+)}(R,R))^{-2}~(\hat{\lambda}_{r}^{(-)}(R,\bar{R}))^{-2}~(\hat{\lambda}_{l_{1}}^{(-)}(R,\bar{R}))^{2} &\\
&\times (\hat{\lambda}_{y}^{(+)}(R,R))^{2} ~(\hat{\lambda}_{z}^{(-)}(R,\bar{R}))^{-3}\end{flalign*}

\beaa
&&P_{[2]}({\bf 10_{136}}; a,q)=\tfrac{1}{a^4q^6}\left( \begin{array}{cccccccccccccc}  0 & 0 & 0 & 0 & 0 & 0 & 1 & -1 & 0 & 2 & -1 & -1 & 1 & 0 \\  0 & 0 & 0 & -1 & 0 & 1 & -2 & -1 & 2 & -2 & -2 & 2 & 0 & -1 \\  0 & 1 & 0 & 0 & 2 & 2 & 0 & 2 & 0 & 0 & 3 & 0 & -1 & 1 \\  -1 & 0 & -2 & -3 & 1 & -1 & -4 & -2 & 0 & 0 & -1 & -1 & 0 & 0 \\  2 & 0 & 0 & 4 & 4 & -1 & 0 & 2 & 2 & 0 & 0 & 0 & 0 & 0 \\  -1 & -2 & 1 & 0 & -3 & -1 & 0 & 0 & 0 & 0 & 0 & 0 & 0 & 0 \\  0 & 1 & 0 & 0 & 0 & 0 & 0 & 0 & 0 & 0 & 0 & 0 & 0 & 0 \\ \end{array} \right)
\eeaa

\subsubsection{$10_{145}$ knot}

\begin{figure}[h]
\centering{\includegraphics[scale=1]{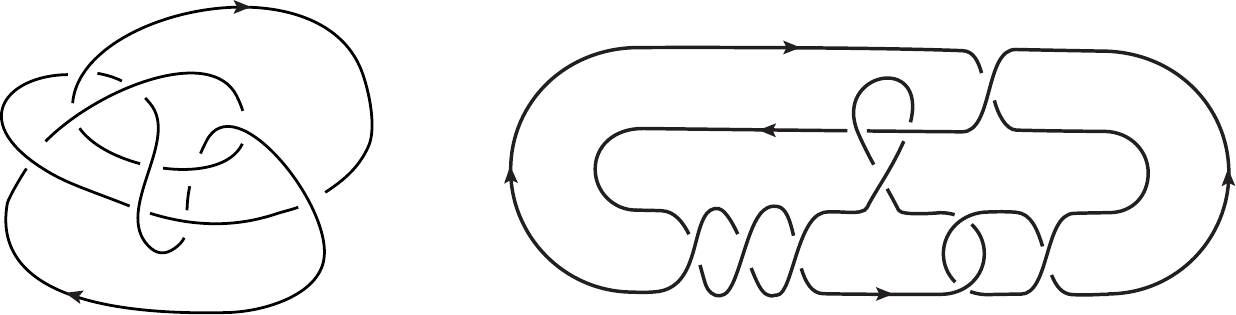}}
\caption{$\bf{10_{145}}$ knot}
\end{figure}

\begin{flalign*}
P_{R}({\bf 10_{145}};a,q)= & \sum_{l,r,u,v,x,y,z}\frac{1}{\epsilon_{l}^{R,R}\sqrt{\dim_{q}l}}~\epsilon_{z}^{R,R}\sqrt{\dim_{q}z}~\epsilon_{u}^{R,\bar{R}}\sqrt{\dim_{q}u}&\\
   & \times\,\epsilon_{r}^{R,R}\sqrt{\dim_{q}r}~a_{lx}\!\left[\begin{footnotesize}\begin{array}{cc}
R & R\\
\bar{R} & \bar{R}
\end{array}\end{footnotesize}\right]~a_{yx}\!\left[\begin{footnotesize}\begin{array}{cc}
R & \bar{R}\\
R & \bar{R}
\end{array}\end{footnotesize}\right]~a_{zy}\!\left[\begin{footnotesize}\begin{array}{cc}
R & R\\
\bar{R} & \bar{R}
\end{array}\end{footnotesize}\right]~a_{lu}\!\left[\begin{footnotesize}\begin{array}{cc}
R & R\\
\bar{R} & \bar{R}
\end{array}\end{footnotesize}\right]&\\
   & \times a_{lv}\!\left[\begin{footnotesize}\begin{array}{cc}
R & R\\
\bar{R} & \bar{R}
\end{array}\end{footnotesize}\right]~a_{rv}\!\left[\begin{footnotesize}\begin{array}{cc}
R & R\\
\bar{R} & \bar{R}
\end{array}\end{footnotesize}\right]~(\hat{\lambda}_{x}^{(-)}(R,\bar{R}))^{-1}~\hat{\lambda}_{y}^{(-)}(R,\bar{R})~ \hat{\lambda}_{z}^{(+)}(R,R)&\\
   & \times(\hat{\lambda}_{u}^{(-)}(R,\bar{R}))^{3}~(\hat{\lambda}_{v}^{(-)}(R,\bar{R}))^2~(\hat{\lambda}_{r}^{(+)}(R,R))^{2}\end{flalign*}

\beaa
&&P_{[2]}({\bf 10_{145}}; a,q)=\tfrac{a^4}{q^4}\left( \begin{array}{ccccccccccccccc}  0 & 0 & 0 & 0 & 0 & 0 & 0 & 0 & 0 & 0 & 0 & 0 & 0 & 1 & 0 \\  0 & 0 & 0 & 0 & 0 & 0 & 0 & 0 & 0 & 0 & -1 & 0 & 1 & -1 & -1 \\  0 & 0 & 0 & 0 & 0 & 0 & 0 & 0 & 1 & 0 & 0 & 2 & 0 & -1 & 1 \\  0 & 0 & 0 & 0 & 0 & -1 & -2 & 1 & 0 & -3 & 1 & 0 & -2 & 0 & 0 \\  0 & 0 & 0 & 1 & 0 & -2 & 2 & 2 & -2 & 2 & 2 & -1 & 0 & 1 & 0 \\  0 & 0 & 1 & 0 & -3 & 1 & 2 & -3 & 0 & 1 & -2 & -1 & 0 & 0 & 0 \\  1 & 0 & 0 & -1 & 0 & 2 & -1 & 0 & 2 & 0 & 0 & 0 & 1 & 0 & 0 \\ \end{array} \right)
\eeaa

\subsubsection{$10_{152}$ knot}
\begin{figure}[h]
\centering{\includegraphics[scale=1]{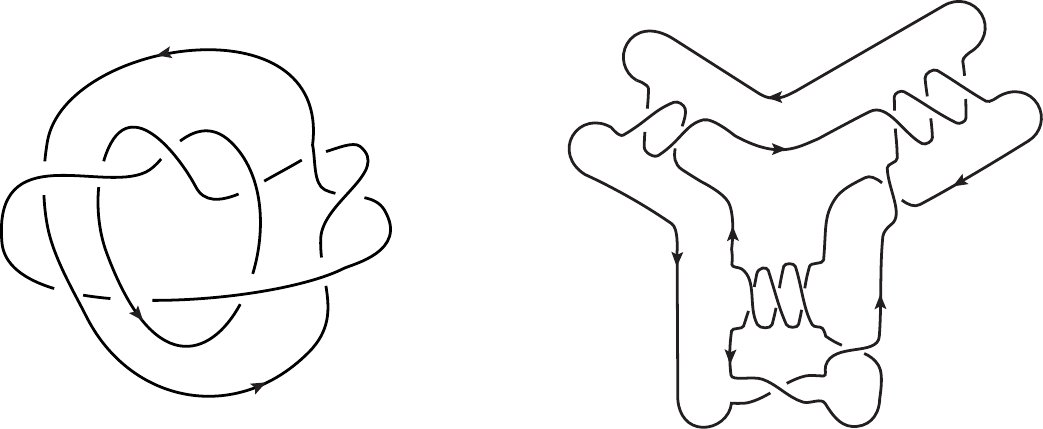}}
\caption{$\bf{10_{152}}$ knot}
\end{figure}

\begin{flalign*}
P_{R}({\bf 10_{152}};a,q)= & \sum_{l,l_1,r,u,v,x,y,z}\frac{1}{\epsilon_{l}^{R,\bar{R}}\sqrt{\dim_{q}l}~\epsilon_{l_1}^{R,\bar{R}}\sqrt{\dim_{q}l_1}}~\epsilon_{z}^{R,\bar{R}}\sqrt{\dim_{q}z}&\\
   & \times \epsilon_{u}^{R,\bar{R}}\sqrt{\dim_{q}u}~\epsilon_{r}^{R,R}\sqrt{\dim_{q}r}~\epsilon_{v}^{R,R}\sqrt{\dim_{q}v}&\\
&\times a_{rl_1}\!\left[\begin{footnotesize}\begin{array}{cc}
R & R\\
\bar{R} & \bar{R}
\end{array}\end{footnotesize}\right]~a_{ll_1}\!\left[\begin{footnotesize}\begin{array}{cc}
R & \bar{R}\\
R & \bar{R}
\end{array}\end{footnotesize}\right]~a_{yl_1}\!\left[\begin{footnotesize}\begin{array}{cc}
R & R\\
\bar{R} & \bar{R}
\end{array}\end{footnotesize}\right]~a_{yz}\!\left[\begin{footnotesize}\begin{array}{cc}
R & R\\
\bar{R} & \bar{R}
\end{array}\end{footnotesize}\right]~a_{lu}\!\left[\begin{footnotesize}\begin{array}{cc}
R & \bar{R}\\
R & \bar{R}
\end{array}\end{footnotesize}\right]& \\
   & \times a_{vl}\!\left[\begin{footnotesize}\begin{array}{cc}
R & R\\
\bar{R} & \bar{R}
\end{array}\end{footnotesize}\right]~(\hat{\lambda}_{r}^{(+)}(R,R))^{3}~\hat{\lambda}_{y}^{(+)}(R,R)~\hat{\lambda}_{z}^{(-)}(R,\bar{R})&\\
   & \times(\hat{\lambda}_{u}^{(-)}(R,\bar{R}))^{2}~\hat{\lambda}_{l}^{(-)}(R,\bar{R})~(\hat{\lambda}_{v}^{(+)}(R,R))^{3}\end{flalign*}

\beaa
&&P_{[2]}({\bf 10_{152}}; a,q)=\tfrac{1}{a^{12}q^{16}}\\
&&\begin{footnotesize}\left( \begin{array}{ccccccccccccccccccccccccc}  1 & 0 & 1 & 3 & 2 & 2 & 6 & 2 & 6 & 5 & 2 & 7 & 6 & 0 & 5 & 6 & 2 & 0 & 3 & 2 & 1 & 1 & 0 & 0 & 1 \\  -1 & -3 & -2 & -5 & -9 & -8 & -11 & -11 & -13 & -16 & -9 & -11 & -17 & -10 & -4 & -8 & -10 & -4 & -1 & -3 & -2 & -1 & -1 & 0 & 0 \\  0 & 3 & 3 & 6 & 7 & 10 & 14 & 12 & 9 & 20 & 13 & 6 & 12 & 13 & 6 & 4 & 3 & 4 & 2 & 1 & 0 & 0 & 0 & 0 & 0 \\  0 & 0 & -2 & -4 & -3 & -3 & -9 & -8 & -1 & -7 & -10 & -3 & -2 & -3 & -3 & -2 & 0 & 0 & 0 & 0 & 0 & 0 & 0 & 0 & 0 \\  0 & 0 & 0 & 0 & 3 & -1 & 0 & 5 & 0 & -2 & 3 & 1 & 0 & 0 & 0 & 0 & 0 & 0 & 0 & 0 & 0 & 0 & 0 & 0 & 0 \\ \end{array} \right)
\end{footnotesize}\eeaa

\subsubsection{$10_{153}$ knot}
\begin{figure}[h]
\centering{\includegraphics[scale=1]{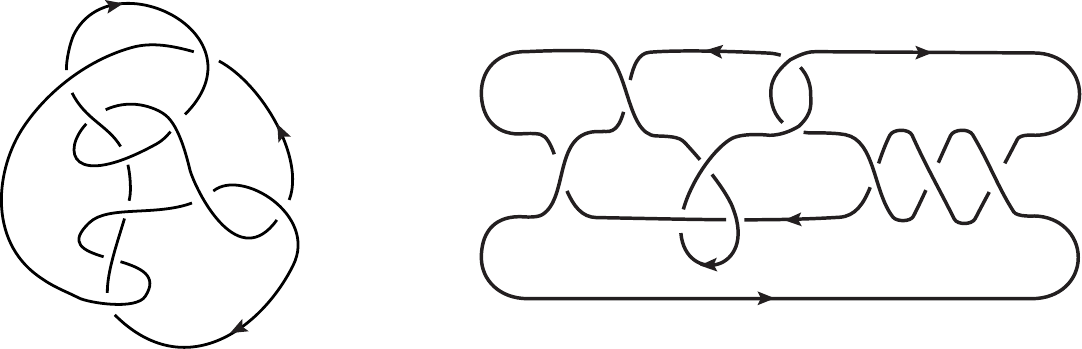}}
\caption{$\bf{10_{153}}$ knot}
\end{figure}

\begin{flalign*}
P_{R}({\bf 10_{153}};a,q) = & \sum_{l,l_1,v,u,x,y,z,r} \frac{1}{\epsilon_{l}^{R,\bar{R}}\sqrt{\dim_{q}l}~\epsilon_{l_1}^{R,\bar{R}}\sqrt{\dim_{q}l_1}}~\epsilon_{z}^{R,\bar{R}}\sqrt{\dim_{q}z}\\
 & \times \epsilon_{v}^{R,\bar{R}}\sqrt{\dim_{q}v}~\epsilon_{u}^{R,R}\sqrt{\dim_{q}u}~\epsilon_{r}^{R,R}\sqrt{\dim_{q}r}&\\
   & \times a_{lx}\!\left[\begin{footnotesize}\begin{array}{cc}
R & \bar{R}\\
R & \bar{R}
\end{array}\end{footnotesize}\right]~a_{yx}\!\left[\begin{footnotesize}\begin{array}{cc}
R & R\\
\bar{R} & \bar{R}
\end{array}\end{footnotesize}\right]~a_{yz}\!\left[\begin{footnotesize}\begin{array}{cc}
R & R\\
\bar{R} & \bar{R}
\end{array}\end{footnotesize}\right]~a_{ll_1}\!\left[\begin{footnotesize}\begin{array}{cc}
R & \bar{R}\\
R & \bar{R}
\end{array}\end{footnotesize}\right]~a_{lv}\!\left[\begin{footnotesize}\begin{array}{cc}
R & \bar{R}\\
R & \bar{R}
\end{array}\end{footnotesize}\right]&\\
   & \times a_{rl_1}\!\left[\begin{footnotesize}\begin{array}{cc}
R & R\\
\bar{R} & \bar{R}
\end{array}\end{footnotesize}\right]~a_{ul_1}\!\left[\begin{footnotesize}\begin{array}{cc}
R & R\\
\bar{R} & \bar{R}
\end{array}\end{footnotesize}\right]~(\hat{\lambda}_{x}^{(-)}(R,\bar{R}))^{-1}~(\hat{\lambda}_{y}^{(+)}(R,R))^{-1}~(\hat{\lambda}_{z}^{(-)}(R,\bar{R}))^{-1}&\\
  &\times (\hat{\lambda}_{r}^{(+)}(R,R))^{2}~(\hat{\lambda}_{v}^{(-)}(R,\bar{R}))^{-2}~(\hat{\lambda}_{u}^{(+)}(R,R))^{3}\end{flalign*}

\beaa
&&P_{[2]}({\bf 10_{153}}; a,q)=\tfrac{1}{a^4q^9}\\
&&\begin{small}
\left( \begin{array}{cccccccccccccccccccc}  0 & 0 & 0 & 0 & 0 & 0 & 1 & 0 & 1 & 1 & 1 & 1 & 1 & 0 & 1 & 1 & 0 & 0 & 1 & 0 \\  0 & 0 & 0 & -1 & -1 & 0 & -2 & -5 & -2 & -2 & -6 & -3 & -2 & -4 & -2 & -2 & -2 & 0 & -1 & -1 \\  0 & 1 & -1 & 0 & 3 & 2 & -1 & 5 & 5 & 3 & 5 & 6 & 2 & 3 & 3 & 2 & 2 & 1 & 0 & 1 \\  1 & 0 & -1 & 3 & 2 & -3 & 1 & 2 & -4 & 0 & 1 & -4 & -1 & 0 & -2 & 0 & 0 & -1 & 0 & 0 \\  -1 & -1 & 0 & -1 & -2 & -2 & 0 & -2 & -1 & 1 & -1 & -1 & 0 & 0 & 0 & 0 & 0 & 0 & 0 & 0 \\  0 & 0 & 1 & 1 & -1 & 0 & 1 & 0 & 0 & 0 & 0 & 0 & 0 & 0 & 0 & 0 & 0 & 0 & 0 & 0 \\  0 & 0 & 1 & -1 & -1 & 2 & 0 & -1 & 1 & 0 & 0 & 0 & 0 & 0 & 0 & 0 & 0 & 0 & 0 & 0 \\ \end{array} \right)\end{small}
\eeaa

\subsubsection{$10_{154}$ knot}
\begin{figure}[h]
\centering{\includegraphics[scale=1]{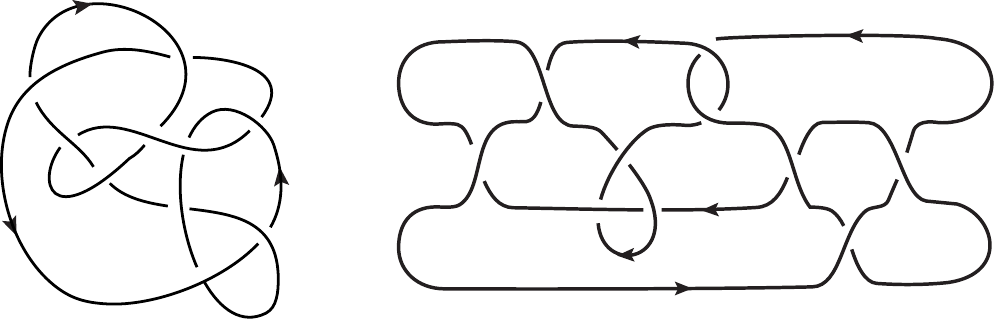}}
\caption{$\bf{10_{154}}$ knot}
\end{figure}

\begin{flalign*}
P_{R}({\bf 10_{154}};a,q)= & \sum_{l,l_1,r,s,u,v,w,x,y,z}\frac{1}{\epsilon_{l}^{R,\bar{R}}\sqrt{\dim_{q}l}~\epsilon_{l_1}^{R,\bar{R}}\sqrt{\dim_{q}l_1}}~\epsilon_{z}^{R,\bar{R}}\sqrt{\dim_{q}z}\\
  & \times \epsilon_{s}^{R,\bar{R}}\sqrt{\dim_{q}s}~\epsilon_{w}^{R,\bar{R}}\sqrt{\dim_{q}w}~\epsilon_{r}^{R,\bar{R}}\sqrt{\dim_{q}r}&\\
   & \times a_{lx}\!\left[\begin{footnotesize}\begin{array}{cc}
R & \bar{R}\\
R & \bar{R}
\end{array}\end{footnotesize}\right]~ a_{yx}\!\left[\begin{footnotesize}\begin{array}{cc}
R & R\\
\bar{R} & \bar{R}
\end{array}\end{footnotesize}\right]~a_{yz}\!\left[\begin{footnotesize}\begin{array}{cc}
R & R\\
\bar{R} & \bar{R}
\end{array}\end{footnotesize}\right]~a_{ll_1}\!\left[\begin{footnotesize}\begin{array}{cc}
R & \bar{R}\\
R & \bar{R}
\end{array}\end{footnotesize}\right]~a_{ls}\!\left[\begin{footnotesize}\begin{array}{cc}
R & \bar{R}\\
R & \bar{R}
\end{array}\end{footnotesize}\right]& \\
&\times a_{rl_1}\!\left[\begin{footnotesize}\begin{array}{cc}
R & \bar{R}\\
R & \bar{R}
\end{array}\end{footnotesize}\right]~a_{l_1u}\!\left[\begin{footnotesize}\begin{array}{cc}
R & \bar{R}\\
R & \bar{R}
\end{array}\end{footnotesize}\right]~a_{vu}\!\left[\begin{footnotesize}\begin{array}{cc}
R & R\\
\bar{R} & \bar{R}
\end{array}\end{footnotesize}\right]
a_{vw}\!\left[\begin{footnotesize}\begin{array}{cc}
R & R\\
\bar{R} & \bar{R}
\end{array}\end{footnotesize}\right]~(\hat{\lambda}_{x}^{(-)}(R,\bar{R}))^{-1}\\
  & \times (\hat{\lambda}_{y}^{(+)}(R,R))^{-1}~(\hat{\lambda}_{z}^{(-)}(R,\bar{R}))^{-1}(\hat{\lambda}_{r}^{(-)}(R,\bar{R}))^{-2}~(\hat{\lambda}_{s}^{(-)}(R,\bar{R}))^{-2}&\\
 & \times (\hat{\lambda}_{u}^{(-)}(R,\bar{R}))^{-1}~(\hat{\lambda}_{v}^{(+)}(R,R))^{-1}~(\hat{\lambda}_{w}^{(-)}(R,\bar{R}))^{-1}\end{flalign*}

\beaa
&&P_{[2]}({\bf 10_{154}}; a,q)=\tfrac{a^6}{q^6}\begin{small}
\left( \begin{array}{ccccccccccccccccccc}  0 & 0 & 0 & 0 & 0 & 0 & 0 & 0 & 0 & 0 & 0 & 0 & 0 & 0 & 0 & 0 & 1 & 0 & 0 \\  0 & 0 & 0 & 0 & 0 & 0 & 0 & 0 & 0 & 0 & 0 & 0 & 0 & -2 & 0 & 2 & -2 & -2 & 0 \\  0 & 0 & 0 & 0 & 0 & 0 & 0 & 0 & 0 & 0 & 0 & -1 & -4 & 2 & 4 & -3 & -2 & 3 & 1 \\  0 & 0 & 0 & 0 & 0 & 0 & 0 & 2 & 3 & 1 & 0 & 1 & 4 & 5 & -1 & -2 & 3 & 1 & -1 \\  0 & 0 & 0 & 0 & -1 & -2 & 0 & 3 & -2 & -6 & -1 & 2 & -1 & -2 & -2 & 0 & 1 & 0 & -1 \\  0 & 0 & 0 & 0 & -3 & -1 & 5 & -2 & -7 & 1 & 2 & -3 & -1 & -2 & -3 & 1 & -1 & -2 & 0 \\  1 & 0 & 0 & 0 & 0 & 3 & 2 & -3 & 2 & 4 & -1 & 1 & 3 & -1 & 2 & 2 & 0 & 0 & 1 \\ \end{array} \right)
\end{small}\eeaa


\section{Colored HOMFLY invariants for links}\label{sec:links}
In this section, we shall compute the colored HOMFLY invariants of links. First of all, we should emphasize that the invariants are no longer polynomials, but \emph{rational functions} with respect to the variables $(a,q)$. In addition, the colored HOMFLY invariants of links are crucially dependent of the orientation of each component of a link. For each link in this section, we choose the orientation presented in Knot Atlas \cite{KnotAtlas}. In \cite{Gukov:2013}, the cyclotomic expansions of the colored HOMFLY invariants of the twist links including the Whitehead link $\bf 5_1^2$ and the link $\bf 7_3^2$, and the Borromean rings $\bf 6_3^3$ are given. Therefore, we treat two-component links with six and seven crossings\footnote{We would like to thank Andrey Morozov for pointing out our mistake in the previous version: The colored HOMFLY invariants of the links ${\bf 7_4^2}$ and ${\bf 7_5^2}$ are not symmetric under the exchange of the two colors.}
 in \S\ref{sec:two}. We have not succeeded in computing the invariants of the link $\bf 7_6^2$ by this method\footnote{Since the link $\bf 7_6^2$ can be written as a three-strand link, the colored HOMFLY invariants can be obtained by the method in \cite{Itoyama:2012re}.}. In \S\ref{sec:three}, we consider three-component links including the links $\bf 6_1^3$ and $\bf 6_3^3$.

In \S\ref{sec:two},  every unreduced colored HOMFLY invariant ${\overline P}_{([n_1],[ n_2])}({\cal L};a,q)$ contains the unknot factor  ${\overline P}_{[n_{\max}] }(\bigcirc ;a,q)$  colored by the highest rank $n_{\max}=\max(n_1,n_2)$. Furthermore, one can observe that it includes the factor $(a;q)_{n_{\max}}/(q;q)_{n_1}(q;q)_{n_2}$. If we normalize by 
\bea\label{Laurent2}
\frac{(q;q)_{n_1}(q;q)_{n_2}}{(a;q)_{n_{\max}}}{\overline P}_{([n_1],[n_2])}({\cal L};a,q)~,
\eea
then it becomes a Laurent polynomial with respect to the variables $(a,q)$. Interestingly, they satisfy the exponential growth property (the property which special polynomials satisfy) \cite{DuninBarkowski:2011yx,Zhu:2012tm,Fuji:2012pi}
\beaa
&&\lim_{q\to1}\frac{(q;q)_{kn_1}(q;q)_{kn_2}}{(a;q)_{kn_{\max}}}{\overline P}_{([k n_1],[k n_2])}({\cal L};a,q)=\left[\lim_{q\to1}\frac{(q;q)_{n_1}(q;q)_{n_2}}{(a;q)_{n_{\max}}}{\overline P}_{([n_1],[n_2])}({\cal L};a,q)\right]^k~.\eeaa
where ${\rm gcd}(n_1,n_2)=1$ and $k\in \bZ_{\ge0}$. In fact, the forms of the Laurent polynomials \eqref{Laurent2} strongly suggest the interpretation at homological level.
For instance, it is easy to see that the difference between the $([1],[3])$-color invariant 
and the $([1],[4])$-color invariant in the matrix form expressions below is just a shift in $q$-degree. In higher ranks, though the cancellation between coefficients make this shift obscure, it is not difficult that only a shift in $q$-degree is involved if you increase the rank of the larger color. The homological interpretations of link invariants will be given in the separate paper \cite{Gukov:2013}.

Similarly, for a three-component link, a colored HOMFLY invariant ${\overline P}_{([n_1],[n_2],[ n_3])}({\cal L};a,q)$ contains the unknot factor  ${\overline P}_{[n_{\max}] }(\bigcirc ;a,q)$  colored by the highest rank $n_{\max}=\max(n_1,n_2,n_3)$. In addition, it  also includes the factor $(a;q)_{n_{\max}}/(q;q)_{n_1}(q;q)_{n_2}(q;q)_{n_3}$. If we normalize by 
\beaa\label{Laurent}
\frac{(q;q)_{n_1}(q;q)_{n_2}(q;q)_{n_3}}{(a;q)_{n_{\max}}}{\overline P}_{([n_1],[n_2],[n_3])}({\cal L};a,q)~,
\eeaa
then it becomes a Laurent polynomial, which obeys
the exponential growth property
\beaa
&&\lim_{q\to1}\frac{(q;q)_{kn_1}(q;q)_{kn_2}(q;q)_{kn_3}}{(a;q)_{kn_{\max}}}{\overline P}_{([k n_1],[k n_2],[k n_3])}({\cal L};a,q)\\
&=&\left[\lim_{q\to1}\frac{(q;q)_{n_1}(q;q)_{n_2}(q;q)_{n_3}}{(a;q)_{n_{\max}}}{\overline P}_{([n_1],[n_2],[ n_3])}({\cal L};a,q)\right]^k
\eeaa
where ${\rm gcd}(n_1,n_2,n_3)=1$.

\subsection{Two-component links}\label{sec:two}

\subsubsection{$6_2^2$ link}

\begin{figure}[h]
\centering{\includegraphics[scale=1]{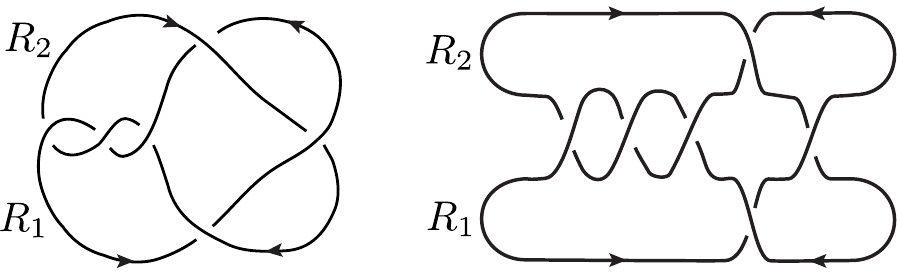}}
\caption{$\bf{6_2^2}$ link}
\end{figure}
\begin{flalign*}
\bar{P}_{(R_{1},R_{2})}({\bf 6_{2}^2};\,a,q) = & q^{\frac{3{\ell}^{(1)}{\ell}^{(2)}}{N}}\sum_{s,t,s^{\prime}}\epsilon_{s}^{R_{1},R_{2}}\,\sqrt{\dim_{q}s}\,\epsilon_{s^{\prime}}^{\bar{R}_{1},\bar{R}_{2}}\,\sqrt{\dim_{q}s^{\prime}}\,(\lambda_{s}^{(+)}(R_{1},\, R_{2}))^{-3}\,&\\
   &\times  a_{t\bar{s}}\!\left[\begin{footnotesize}\begin{array}{cc}
R_{2} & \bar{R}_{1}\\
\bar{R}_{2} & R_{1}
\end{array}\end{footnotesize}\right]~(\lambda_{\bar{t}}^{(-)}(\bar{R}_{1},\, R_{2}))^{-2}\, a_{t\bar{s^{\prime}}}\!\left[\begin{footnotesize}\begin{array}{cc}
\bar{R}_{1} & R_{2}\\
R_{1} & \bar{R}_{2}
\end{array}\end{footnotesize}\right]\,(\lambda_{s^{\prime}}^{(+)}(\bar{R}_{1},\,\bar{R}_{2}))^{-1}.\end{flalign*}
The colored HOMFLY invariants of the link $\bf 6_2^2$ is symmetric under interchanging the two colors.

\begin{itemize}
\item{$\overline P_{([1],[1])}({\bf 6_2^2}; a,q) =$}
\beaa
\tfrac{(1-a) }{a^{4}q(1-q)^2 }\left( \begin{array}{ccccc}  -1 & 2 & -2 & 2 & -1 \\  -1 & 2 & -3 & 2 & -1 \\  0 & 1 & -1 & 1 & 0 \\ \end{array} \right)
\eeaa

\item{$\overline P_{([1],[2])}({\bf 6_2^2}; a,q) =\overline P_{([2],[1])}({\bf 6_2^2}; a,q) =$}
\beaa
\tfrac{(1-a) (1-a q) }{a^{9/2}q^{7/2}(1-q)^2 (1-q^2) } \left(   \begin{array}{cccccccc}  0 & -1 & 1 & 1 & -2 & 1 & 1 & -1 \\  -1 & 0 & 2 & -2 & -1 & 2 & -1 & 0 \\  0 & 1 & 0 & -1 & 1 & 0 & 0 & 0 \\ \end{array} \right)
\eeaa

\item{$\overline P_{([1],[3])}({\bf 6_2^2}; a,q) =\overline P_{([3],[1])}({\bf 6_2^2}; a,q) =$}
\beaa
\tfrac{(1-a) (1-a q) (1-a q^2)}{a^5q^6(1-q)^2 (1-q^2) (1-q^3)}\left( \begin{array}{ccccccccccc}  0 & 0 & -1 & 1 & 0 & 1 & -2 & 1 & 0 & 1 & -1 \\  -1 & 0 & 0 & 2 & -2 & 0 & -1 & 2 & -1 & 0 & 0 \\  0 & 1 & 0 & 0 & -1 & 1 & 0 & 0 & 0 & 0 & 0 \\ \end{array} \right)
\eeaa
\item{$\overline P_{([1],[4])}({\bf 6_2^2}; a,q) =\overline P_{([4],[1])}({\bf 6_2^2}; a,q) =$}
\beaa
\tfrac{ (1-a) (1-a q) (1-a q^2) (1-a q^3)}{a^{{11}/{2}} q^{17/2}(1-q)^2 (1-q^2) \left(1-q^3\right)(1-q^4) }\left( \begin{array}{cccccccccccccc}  0 & 0 & 0 & -1 & 1 & 0 & 0 & 1 & -2 & 1 & 0 & 0 & 1 & -1 \\  -1 & 0 & 0 & 0 & 2 & -2 & 0 & 0 & -1 & 2 & -1 & 0 & 0 & 0 \\  0 & 1 & 0 & 0 & 0 & -1 & 1 & 0 & 0 & 0 & 0 & 0 & 0 & 0 \\ \end{array} \right)
\eeaa

\item{$\overline P_{([2],[2])}({\bf 6_2^2}; a,q) =$}
\beaa
\tfrac{(1-a) (1-a q)}{a^{8}q^{7}(1-q)^2 (1-q^2)^2 }\begin{small}\left( \begin{array}{ccccccccccccccc}  0 & 0 & 1 & -2 & 0 & 3 & -4 & 1 & 5 & -5 & -2 & 5 & -1 & -2 & 1 \\  0 & 1 & -2 & 0 & 5 & -6 & -2 & 10 & -5 & -7 & 7 & 2 & -4 & 0 & 1 \\  1 & -2 & 0 & 5 & -5 & -3 & 10 & -3 & -7 & 5 & 2 & -2 & 0 & 0 & 0 \\  -1 & 0 & 2 & -3 & -2 & 5 & -1 & -4 & 2 & 1 & -1 & 0 & 0 & 0 & 0 \\  0 & 1 & -1 & 0 & 2 & -1 & -1 & 1 & 0 & 0 & 0 & 0 & 0 & 0 & 0 \\ \end{array} \right)\end{small}
\eeaa

\item{$\overline P_{([2],[3])}({\bf 6_2^2}; a,q) =\overline P_{([3],[2])}({\bf 6_2^2}; a,q) =$}
\beaa
&&\tfrac{(1-a) (1-a q) (1-a q^2)}{a^{{17}/{2}}q^{25/2}(1-q)^2 (1-q^2)^2 (1-q^3) }\times\\
&&\begin{small}
\left( \begin{array}{ccccccccccccccccccccc}  0 & 0 & 0 & 0 & 1 & -1 & -2 & 2 & 2 & -2 & -3 & 2 & 5 & -2 & -5 & 1 & 3 & 1 & -2 & -1 & 1 \\  0 & 0 & 1 & 0 & -3 & 0 & 5 & 1 & -7 & -3 & 8 & 5 & -7 & -6 & 4 & 5 & -1 & -3 & 0 & 1 & 0 \\  1 & 0 & -2 & -1 & 3 & 4 & -4 & -6 & 4 & 7 & -1 & -7 & -1 & 5 & 1 & -2 & 0 & 0 & 0 & 0 & 0 \\  -1 & -1 & 1 & 2 & -1 & -4 & 0 & 4 & 1 & -3 & -2 & 2 & 1 & -1 & 0 & 0 & 0 & 0 & 0 & 0 & 0 \\  0 & 1 & 0 & -1 & 0 & 1 & 1 & -1 & -1 & 1 & 0 & 0 & 0 & 0 & 0 & 0 & 0 & 0 & 0 & 0 & 0 \\ \end{array} \right)
\end{small}\eeaa

\item{$\overline P_{([2],[4])}({\bf 6_2^2}; a,q) =\overline P_{([4],[2])}({\bf 6_2^2}; a,q) =$}
\beaa
&&\tfrac{(1-a) (1-a q) (1-a q^2)(1-a q^3)}{a^{9}q^{18} (1-q)^2 (1-q^2)^2 (1-q^3) (1-q^4) }\times\\
&&\begin{footnotesize}
\left( \begin{array}{ccccccccccccccccccccccccccc}  0 & 0 & 0 & 0 & 0 & 0 & 1 & -1 & -1 & 0 & 1 & 2 & -2 & -1 & -1 & 1 & 4 & -1 & -2 & -2 & 0 & 3 & 0 & 0 & -1 & -1 & 1 \\  0 & 0 & 0 & 1 & 0 & -1 & -2 & 0 & 4 & 1 & -1 & -5 & -2 & 5 & 3 & 2 & -5 & -5 & 3 & 2 & 2 & 0 & -3 & 0 & 1 & 0 & 0 \\  1 & 0 & 0 & -2 & -1 & 2 & 2 & 3 & -4 & -4 & 1 & 2 & 6 & -2 & -4 & -1 & -1 & 4 & 1 & -2 & 0 & 0 & 0 & 0 & 0 & 0 & 0 \\  -1 & -1 & 0 & 1 & 2 & -1 & -2 & -2 & 0 & 4 & 0 & -1 & -1 & -2 & 2 & 1 & -1 & 0 & 0 & 0 & 0 & 0 & 0 & 0 & 0 & 0 & 0 \\  0 & 1 & 0 & 0 & -1 & 0 & 1 & 0 & 1 & -1 & -1 & 1 & 0 & 0 & 0 & 0 & 0 & 0 & 0 & 0 & 0 & 0 & 0 & 0 & 0 & 0 & 0 \\ \end{array} \right)
\end{footnotesize}\eeaa

\end{itemize}

\subsubsection{$6_3^2$ link}
\begin{figure}[h]
\centering{\includegraphics[scale=1]{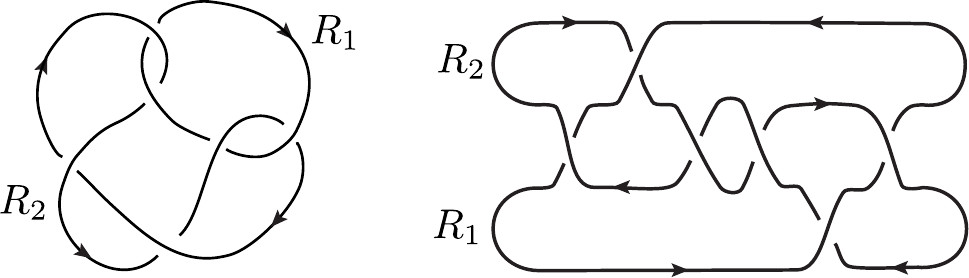}}
\caption{$\bf{6_3^2}$ link}
\end{figure}

\begin{flalign*}
\bar{P}_{(R_{1},R_{2})}({\bf 6_{3}^2};\,a,q) = & q^{(-2C_{R_{2}}-\tfrac{2{\ell}^{(1)}{\ell}^{(2)}}{N})}\sum_{s,t,s^{\prime},u,v}\epsilon_{s}^{R_{1},R_{2}}\,\sqrt{\dim_{q}s}\,\epsilon_{v}^{\bar{R}_{1},\bar{R}_{2}}\,\sqrt{\dim_{q}v}\,\lambda_{s}^{(+)}(R_{1},\, R_{2})&\\
   &\times a_{ts}\!\left[\begin{footnotesize}\begin{array}{cc}
\bar{R}_{1} & R_{2}\\
R_{1} & \bar{R}_{2}
\end{array}\end{footnotesize}\right]\,\lambda_{\bar{t}}^{(-)}(R_{1},\,\bar{R}_{2})\, a_{ts^{\prime}}\!\left[\begin{footnotesize}\begin{array}{cc}
\bar{R}_{1} & R_{2}\\
\bar{R}_{2} & R_{1}
\end{array}\end{footnotesize}\right]\,(\lambda_{s^{\prime}}^{(-)}(R_{2},\,\bar{R}_{2}))^{-2}&\\
   &\times a_{us^{\prime}}\!\left[\begin{footnotesize}\begin{array}{cc}
\bar{R}_{1} & R_{2}\\
\bar{R}_{2} & R_{1}
\end{array}\end{footnotesize}\right]\,\lambda_{u}^{(-)}(\bar{R}_{1},\, R_{2})\, a_{uv}\!\left[\begin{footnotesize}\begin{array}{cc}
R_{2} & \bar{R}_{1}\\
\bar{R}_{2} & R_{1}
\end{array}\end{footnotesize}\right]\,\lambda_{v}^{(+)}(\bar{R}_{1},\,\bar{R}_{2}).
\end{flalign*}
The colored HOMFLY invariants of the link $\bf 6_3^2$ is symmetric under interchanging the two colors.

\begin{itemize}

\item{$\overline P_{([1],[1])}({\bf 6_3^2}; a,q) =$}
\beaa
\tfrac{(1-a)  }{a q(1-q)^2 }\left( \begin{array}{ccccc}  0 & 0 & -1 & 0 & 0 \\  0 & 2 & -3 & 2 & 0 \\  -1 & 3 & -4 & 3 & -1 \\  0 & 1 & -2 & 1 & 0 \\ \end{array} \right)
\eeaa

\item{$\overline P_{([1],[2])}({\bf 6_3^2}; a,q) =\overline P_{([2],[1])}({\bf 6_3^2}; a,q) =$}
\beaa
\tfrac{(1-a) (1-a q) }{a^{3/2} q^{1/2}(1-q)^2 (1-q^2)}\left( \begin{array}{ccccccc}  0 & 0 & 0 & 0 & -1 & 0 & 0 \\  0 & 0 & 2 & -2 & -1 & 2 & 0 \\  -1 & 2 & 0 & -3 & 2 & 1 & -1 \\  0 & 1 & -1 & -1 & 1 & 0 & 0 \\ \end{array} \right)
\eeaa

\item{$\overline P_{([1],[3])}({\bf 6_3^2}; a,q) =\overline P_{([3],[1])}({\bf 6_3^2}; a,q) =$}
\beaa
\tfrac{(1-a) (1-a q) (1-a q^2)}{a^2 (1-q)^2 (1-q^2) (1-q^3)}\left( \begin{array}{ccccccccc}  0 & 0 & 0 & 0 & 0 & 0 & -1 & 0 & 0 \\  0 & 0 & 0 & 2 & -2 & 0 & -1 & 2 & 0 \\  -1 & 2 & -1 & 1 & -3 & 2 & 0 & 1 & -1 \\  0 & 1 & -1 & 0 & -1 & 1 & 0 & 0 & 0 \\ \end{array} \right)
\eeaa
\item{$\overline P_{([1],[4])}({\bf 6_3^2}; a,q) =\overline P_{([4],[1])}({\bf 6_3^2}; a,q) =$}
\beaa
\tfrac{q^{1/2}(1-a) (1-a q) (1-a q^2) (1-a q^3)}{a^{{5}/{2}}  (1-q)^2 (1-q^2) \left(1-q^3\right)(1-q^4) }
\left( \begin{array}{ccccccccccc}  0 & 0 & 0 & 0 & 0 & 0 & 0 & 0 & -1 & 0 & 0 \\  0 & 0 & 0 & 0 & 2 & -2 & 0 & 0 & -1 & 2 & 0 \\  -1 & 2 & -1 & 0 & 1 & -3 & 2 & 0 & 0 & 1 & -1 \\  0 & 1 & -1 & 0 & 0 & -1 & 1 & 0 & 0 & 0 & 0 \\ \end{array} \right)
\eeaa

\item{$\overline P_{([2],[2])}({\bf 6_3^2}; a,q) =$}
\beaa
 \tfrac{(1-a) (1-a q)}{a^{2}q^{3}(1-q)^2 (1-q^2)^2}\begin{small}\left( \begin{array}{ccccccccccccc}  0 & 0 & 0 & 0 & 0 & 0 & 0 & 0 & 0 & 0 & 1 & 0 & 0 \\  0 & 0 & 0 & 0 & 0 & 0 & 0 & -2 & 0 & 3 & -1 & -2 & 0 \\  0 & 0 & 0 & 0 & 1 & 2 & -6 & -1 & 10 & -3 & -6 & 3 & 1 \\  0 & 0 & -2 & 3 & 5 & -12 & -2 & 17 & -5 & -10 & 6 & 2 & -2 \\  1 & -3 & 1 & 9 & -11 & -8 & 19 & -2 & -13 & 7 & 2 & -3 & 1 \\  -1 & 1 & 4 & -5 & -5 & 9 & 1 & -7 & 2 & 2 & -1 & 0 & 0 \\  0 & 1 & -2 & -1 & 4 & -1 & -2 & 1 & 0 & 0 & 0 & 0 & 0 \\ \end{array} \right)\end{small}
\eeaa

\item{$\overline P_{([2],[3])}({\bf 6_3^2}; a,q) =\overline P_{([3],[2])}({\bf 6_3^2}; a,q) =$}
\beaa
&& \tfrac{(1-a) (1-a q) (1-a q^2)}{a^{5/2}q^{5/2}(1-q)^2 (1-q^2)^2 (1-q^3) }\times\\
&&\begin{small}
\left( \begin{array}{ccccccccccccccccc}  0 & 0 & 0 & 0 & 0 & 0 & 0 & 0 & 0 & 0 & 0 & 0 & 0 & 0 & 1 & 0 & 0 \\  0 & 0 & 0 & 0 & 0 & 0 & 0 & 0 & 0 & 0 & -2 & 0 & 2 & 1 & -1 & -2 & 0 \\  0 & 0 & 0 & 0 & 0 & 0 & 1 & 2 & -4 & -4 & 4 & 6 & 0 & -6 & -2 & 3 & 1 \\  0 & 0 & 0 & -2 & 2 & 5 & -3 & -9 & -1 & 12 & 6 & -9 & -7 & 3 & 5 & 0 & -2 \\  1 & -2 & -2 & 6 & 4 & -8 & -10 & 6 & 15 & -2 & -12 & -2 & 6 & 3 & -3 & -1 & 1 \\  -1 & 0 & 4 & 1 & -7 & -4 & 7 & 7 & -4 & -7 & 1 & 4 & 0 & -1 & 0 & 0 & 0 \\  0 & 1 & -1 & -2 & 1 & 2 & 1 & -2 & -1 & 1 & 0 & 0 & 0 & 0 & 0 & 0 & 0 \\ \end{array} \right)\end{small}
\eeaa

\item{$\overline P_{([2],[4])}({\bf 6_3^2}; a,q) =\overline P_{([4],[2]])}({\bf 6_3^2}; a,q) =$}
\beaa
&& \tfrac{(1-a) (1-a q) (1-a q^2)(1-a q^3)}{a^{3}q^{2}(1-q)^2 (1-q^2)^2 (1-q^3) (1-q^4) }\times\\
&&\begin{small}
\left( \begin{array}{ccccccccccccccccccccc}  0 & 0 & 0 & 0 & 0 & 0 & 0 & 0 & 0 & 0 & 0 & 0 & 0 & 0 & 0 & 0 & 0 & 0 & 1 & 0 & 0 \\  0 & 0 & 0 & 0 & 0 & 0 & 0 & 0 & 0 & 0 & 0 & 0 & 0 & -2 & 0 & 2 & 0 & 1 & -1 & -2 & 0 \\  0 & 0 & 0 & 0 & 0 & 0 & 0 & 0 & 1 & 2 & -4 & -2 & 1 & 1 & 6 & -1 & -3 & -2 & -2 & 3 & 1 \\  0 & 0 & 0 & 0 & -2 & 2 & 4 & -3 & -1 & -5 & -2 & 9 & 3 & 1 & -5 & -6 & 2 & 2 & 3 & 0 & -2 \\  1 & -2 & -1 & 3 & 1 & 3 & -5 & -7 & 2 & 3 & 9 & 0 & -7 & -3 & -2 & 5 & 2 & -1 & -1 & -1 & 1 \\  -1 & 0 & 3 & 1 & -2 & -4 & -2 & 4 & 4 & 2 & -3 & -5 & 0 & 2 & 2 & 0 & -1 & 0 & 0 & 0 & 0 \\  0 & 1 & -1 & -1 & 0 & 0 & 2 & 0 & 0 & -1 & -1 & 1 & 0 & 0 & 0 & 0 & 0 & 0 & 0 & 0 & 0 \\ \end{array} \right)
\end{small}\eeaa

\item{$\overline P_{([3],[3])}({\bf 6_3^2}; a,q)=$}
\beaa
&&  \tfrac{(1-a) (1-a q) (1-a q^2)}{a^{3}q^{7}(1-q)^2 (1-q^2)^2 (1-q^3)^2 }\times\\
&&\begin{tiny}
\left( \begin{array}{cccccccccccccccccccccccccccc}  0 & 0 & 0 & 0 & 0 & 0 & 0 & 0 & 0 & 0 & 0 & 0 & 0 & 0 & 0 & 0 & 0 & 0 & 0 & 0 & 0 & 0 & 0 & 0 & 0 & -1 & 0 & 0 \\  0 & 0 & 0 & 0 & 0 & 0 & 0 & 0 & 0 & 0 & 0 & 0 & 0 & 0 & 0 & 0 & 0 & 0 & 0 & 0 & 2 & 0 & 0 & -3 & 1 & 1 & 2 & 0 \\  0 & 0 & 0 & 0 & 0 & 0 & 0 & 0 & 0 & 0 & 0 & 0 & 0 & 0 & 0 & -1 & -2 & 0 & 6 & 2 & -3 & -10 & 1 & 5 & 5 & -2 & -3 & -1 \\  0 & 0 & 0 & 0 & 0 & 0 & 0 & 0 & 0 & 0 & 0 & 2 & 0 & -1 & -9 & 1 & 14 & 10 & -9 & -24 & 1 & 17 & 11 & -6 & -11 & 1 & 2 & 2 \\  0 & 0 & 0 & 0 & 0 & 0 & 0 & -1 & -2 & 6 & 5 & -4 & -22 & -2 & 30 & 24 & -19 & -43 & -1 & 33 & 19 & -13 & -20 & 3 & 7 & 4 & -3 & -1 \\  0 & 0 & 0 & 0 & 2 & -3 & -5 & 5 & 15 & 2 & -35 & -18 & 36 & 48 & -14 & -64 & -13 & 44 & 32 & -17 & -29 & 4 & 12 & 3 & -5 & -2 & 2 & 0 \\  0 & -1 & 3 & -1 & -6 & -1 & 14 & 14 & -26 & -32 & 16 & 54 & 12 & -58 & -33 & 31 & 41 & -7 & -31 & 0 & 12 & 4 & -5 & -2 & 3 & -1 & 0 & 0 \\  1 & -1 & -3 & 0 & 7 & 8 & -12 & -18 & 3 & 28 & 15 & -27 & -24 & 8 & 26 & 4 & -17 & -5 & 5 & 5 & -2 & -2 & 1 & 0 & 0 & 0 & 0 & 0 \\  -1 & 1 & 2 & 2 & -6 & -6 & 4 & 11 & 4 & -13 & -7 & 4 & 10 & 0 & -6 & -1 & 1 & 2 & -1 & 0 & 0 & 0 & 0 & 0 & 0 & 0 & 0 & 0 \\  0 & 1 & -2 & -1 & 2 & 3 & 0 & -6 & 0 & 3 & 2 & -1 & -2 & 1 & 0 & 0 & 0 & 0 & 0 & 0 & 0 & 0 & 0 & 0 & 0 & 0 & 0 & 0 \\ \end{array} \right)
\end{tiny}\eeaa

\end{itemize}

\subsubsection{$7_1^2$ link}
\begin{figure}[h]
\centering{\includegraphics[scale=1]{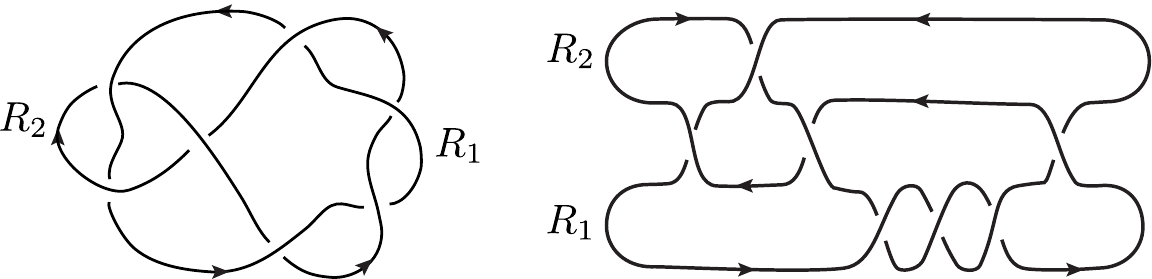}}
\caption{$\bf{7_1^2}$ link}
\end{figure}
\begin{flalign*}
\bar{P}_{(R_{1},R_{2})}({\bf 7_{1}^2};\,a,q) = & q^{\left(-C_{R_{2}}+\tfrac{{\ell}^{(1)}{\ell}^{(2)}}{N}\right)}\sum_{s,t,s^{\prime},u,v}\epsilon_{s}^{R_{1},R_{2}}\,\sqrt{\dim_{q}s}\,\epsilon_{v}^{\bar{R}_{1},R_{2}}\,\sqrt{\dim_{q}v}\,(\lambda_{s}^{(+)}(R_{1},\, R_{2}))&\\
  &\times a_{ts}\!\left[\begin{footnotesize}\begin{array}{cc}
\bar{R}_{1} & R_{2}\\
R_{1} & \bar{R}_{2}
\end{array}\end{footnotesize}\right]\,\lambda_{\bar{t}}^{(-)}(R_{1},\,\bar{R}_{2})\, a_{ts^{\prime}}\!\left[\begin{footnotesize}\begin{array}{cc}
\bar{R}_{1} & R_{2}\\
\bar{R}_{2} & R_{1}
\end{array}\end{footnotesize}\right]\,(\lambda_{s^{\prime}}^{(-)}(R_{2},\,\bar{R}_{2}))^{-1}&\\
   &\times a_{us^{\prime}}\!\left[\begin{footnotesize}\begin{array}{cc}
\bar{R}_{1} & \bar{R}_{2}\\
R_{2} & R_{1}
\end{array}\end{footnotesize}\right]\,\lambda_{u}^{(+)}(\bar{R}_{1},\,\bar{R}_{2})^{-3}\, a_{uv}\!\left[\begin{footnotesize}\begin{array}{cc}
\bar{R}_{2} & \bar{R}_{1}\\
R_{2} & R_{1}
\end{array}\end{footnotesize}\right]\,(\lambda_{v}^{(-)}(\bar{R}_{1},\, R_{2}))^{-1}.
\end{flalign*}
The colored HOMFLY invariants of the link $\bf 7_1^2$ is symmetric under interchanging the two colors.
\begin{itemize}

\item{$\overline P_{([1],[1])}({\bf 7_1^2}; a,q)=$}
\beaa
\tfrac{(1-a) }{a^{3}q^{2}(1-q)^2  }
\left( \begin{array}{ccccccc}  0 & -1 & 1 & -1 & 1 & -1 & 0 \\  1 & -2 & 3 & -3 & 3 & -2 & 1 \\  0 & -1 & 2 & -2 & 2 & -1 & 0 \\ \end{array} \right)
\eeaa

\item{$\overline P_{([1],[2])}({\bf 7_1^2}; a,q)=\overline P_{([2],[1])}({\bf 7_1^2}; a,q)=$}
\beaa
\tfrac{(1-a) (1-a q) }{a^{7/2}q^{7/2}(1-q)^2 (1-q^2) }\left( \begin{array}{cccccccccc}  0 & 0 & -1 & 0 & 1 & -1 & 0 & 1 & -1 & 0 \\  1 & -1 & -1 & 3 & -1 & -2 & 3 & -1 & -1 & 1 \\  0 & -1 & 1 & 1 & -2 & 1 & 1 & -1 & 0 & 0 \\ \end{array} \right)
\eeaa

\item{$\overline P_{([1],[3])}({\bf 7_1^2}; a,q)=\overline P_{([3],[1])}({\bf 7_1^2}; a,q)=$}
\beaa
\tfrac{(1-a) (1-a q) (1-a q^2)}{a^{4}q^{5}(1-q)^2 (1-q^2) (1-q^3)}\left( \begin{array}{ccccccccccccc}  0 & 0 & 0 & -1 & 0 & 0 & 1 & -1 & 0 & 0 & 1 & -1 & 0 \\  1 & -1 & 0 & -1 & 3 & -1 & 0 & -2 & 3 & -1 & 0 & -1 & 1 \\  0 & -1 & 1 & 0 & 1 & -2 & 1 & 0 & 1 & -1 & 0 & 0 & 0 \\ \end{array} \right)
\eeaa

\item{$\overline P_{([1],[4])}({\bf 7_1^2}; a,q)=\overline P_{([4],[1])}({\bf 7_1^2}; a,q)=$}
\beaa
 \tfrac{(1-a) (1-a q) (1-a q^2) (1-a q^3)}{a^{9/2} q^{13/2} (1-q)^2 (1-q^2) \left(1-q^3\right)(1-q^4) }
\begin{small}
\left( \begin{array}{cccccccccccccccc}  0 & 0 & 0 & 0 & -1 & 0 & 0 & 0 & 1 & -1 & 0 & 0 & 0 & 1 & -1 & 0 \\  1 & -1 & 0 & 0 & -1 & 3 & -1 & 0 & 0 & -2 & 3 & -1 & 0 & 0 & -1 & 1 \\  0 & -1 & 1 & 0 & 0 & 1 & -2 & 1 & 0 & 0 & 1 & -1 & 0 & 0 & 0 & 0 \\ \end{array} \right)
\end{small}
\eeaa

\item{$\overline P_{([2],[2])}({\bf 7_1^2}; a,q)=$}
\beaa
\tfrac{(1-a) (1-a q) }{a^{6}q^{8}(1-q)^2 (1-q^2)^2 }\begin{footnotesize}\left( \begin{array}{ccccccccccccccccccc}  0 & 0 & 0 & 0 & 0 & 1 & -1 & 0 & 1 & -1 & 1 & 1 & -2 & 0 & 2 & -1 & -1 & 1 & 0 \\  0 & 0 & -1 & 1 & 1 & -3 & 1 & 1 & -3 & 3 & 1 & -6 & 2 & 5 & -4 & -2 & 3 & 0 & -1 \\  1 & -2 & 1 & 3 & -5 & 2 & 3 & -7 & 6 & 5 & -12 & 2 & 11 & -7 & -5 & 6 & 0 & -2 & 1 \\  -1 & 1 & 2 & -4 & 2 & 3 & -8 & 5 & 7 & -11 & -1 & 10 & -3 & -5 & 3 & 1 & -1 & 0 & 0 \\  0 & 1 & -2 & 0 & 3 & -4 & 1 & 5 & -5 & -2 & 5 & -1 & -2 & 1 & 0 & 0 & 0 & 0 & 0 \\ \end{array} \right)\end{footnotesize}\eeaa

\item{$\overline P_{([2],[3])}({\bf 7_1^2}; a,q)=\overline P_{([3],[2])}({\bf 7_1^2}; a,q)=$}
\beaa
&&\tfrac{(1-a) (1-a q) (1-a q^2)}{a^{13/2}q^{23/2}(1-q)^2 (1-q^2)^2 (1-q^3) }\times\\
&&\begin{footnotesize}
\left( \begin{array}{ccccccccccccccccccccccccc}  0 & 0 & 0 & 0 & 0 & 0 & 0 & 1 & 0 & -1 & 0 & 1 & 0 & -1 & 0 & 2 & 0 & -2 & 0 & 1 & 1 & -1 & -1 & 1 & 0 \\  0 & 0 & 0 & -1 & 0 & 2 & 0 & -3 & -1 & 3 & 1 & -4 & -1 & 5 & 1 & -5 & -3 & 4 & 4 & -3 & -3 & 1 & 2 & 0 & -1 \\  1 & -1 & -2 & 3 & 3 & -4 & -4 & 4 & 6 & -6 & -7 & 8 & 8 & -6 & -10 & 3 & 11 & -1 & -8 & -1 & 4 & 2 & -2 & -1 & 1 \\  -1 & 0 & 3 & 0 & -5 & 0 & 7 & 0 & -9 & -1 & 11 & 3 & -11 & -5 & 8 & 6 & -4 & -5 & 1 & 3 & 0 & -1 & 0 & 0 & 0 \\  0 & 1 & -1 & -2 & 2 & 2 & -2 & -3 & 2 & 5 & -2 & -5 & 1 & 3 & 1 & -2 & -1 & 1 & 0 & 0 & 0 & 0 & 0 & 0 & 0 \\ \end{array} \right)\end{footnotesize}\eeaa

\item{$\overline P_{([2],[4])}({\bf 7_1^2}; a,q)=\overline P_{([4],[2])}({\bf 7_1^2}; a,q)=$}
\beaa
&& \tfrac{(1-a) (1-a q) (1-a q^2)(1-a q^3)}{a^{7}q^{15}(1-q)^2 (1-q^2)^2 (1-q^3)(1-q^4)  }\times\\
&&\begin{tiny}
\left( \begin{array}{ccccccccccccccccccccccccccccccc}  0 & 0 & 0 & 0 & 0 & 0 & 0 & 0 & 0 & 1 & 0 & 0 & -1 & 0 & 1 & 0 & 0 & -1 & 0 & 1 & 1 & 0 & -2 & 0 & 1 & 0 & 1 & -1 & -1 & 1 & 0 \\  0 & 0 & 0 & 0 & -1 & 0 & 1 & 1 & 0 & -3 & -1 & 1 & 2 & 1 & -3 & -2 & 1 & 4 & 1 & -4 & -2 & -1 & 3 & 4 & -3 & -2 & 0 & 0 & 2 & 0 & -1 \\  1 & -1 & -1 & 0 & 2 & 3 & -3 & -3 & -1 & 3 & 5 & -2 & -5 & -4 & 6 & 7 & -2 & -4 & -7 & 2 & 9 & 0 & -2 & -4 & -2 & 4 & 1 & 0 & -1 & -1 & 1 \\  -1 & 0 & 2 & 1 & -1 & -4 & 0 & 4 & 3 & -1 & -7 & -1 & 4 & 6 & 2 & -9 & -4 & 3 & 4 & 4 & -3 & -4 & 0 & 1 & 2 & 0 & -1 & 0 & 0 & 0 & 0 \\  0 & 1 & -1 & -1 & 0 & 1 & 2 & -2 & -1 & -1 & 1 & 4 & -1 & -2 & -2 & 0 & 3 & 0 & 0 & -1 & -1 & 1 & 0 & 0 & 0 & 0 & 0 & 0 & 0 & 0 & 0 \\ \end{array} \right)\end{tiny}\eeaa

\end{itemize}

\subsubsection{$7_2^2$ link}

\begin{figure}[h]
\centering{\includegraphics[scale=1]{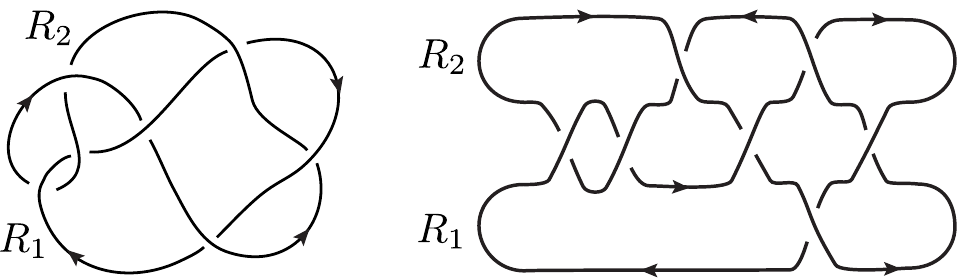}}
\caption{$\bf{ 7_2^2}$ link}
\end{figure}
\begin{flalign*}
\bar{P}_{(R_{1},R_{2})}({\bf 7_{2}^2};\,a,q)=&q^{\left(-C_{R_{2}}+\tfrac{{\ell}^{(1)}{\ell}^{(2)}}{N}\right)}\sum_{s,t,s^{\prime},u,v}\epsilon_{s}^{\bar{R}_{1},R_{2}}\,\sqrt{\dim_{q}s}\,\epsilon_{v}^{R_{1},R_{2}}\,\sqrt{\dim_{q}v}\,(\lambda_{s}^{(-)}(\bar{R}{}_{1},\, R_{2}))^{2}& \\
		& \times a_{ts}\left[\begin{array}{cc}
R_{1} & \bar{R}_{1}\\
R_{2} & \bar{R}_{2}
\end{array}\right]\,(\lambda_{t}^{(-)}(R_{2},\,\bar{R}_{2}))^{-1}\, a_{s^{\prime}t}\left[\begin{array}{cc}
R_{1} & R_{2}\\
\bar{R}_{2} & \bar{R}_{1}
\end{array}\right]\,(\lambda_{s^{\prime}}^{(+)}(R_{1},\, R_{2}))^{-1} & \\
		& \times a_{s^{\prime}u}\left[\begin{array}{cc}
R_{1} & R_{2}\\
\bar{R}_{1} & \bar{R}{}_{2}
\end{array}\right]\,(\lambda_{u}^{(-)}(\bar{R}_{1},\, R_{2}))^{-2}\, a_{vu}\left[\begin{array}{cc}
R_{1} & R_{2}\\
\bar{R}_{1} & \bar{R}_{2}
\end{array}\right]\,(\lambda_{v}^{(+)}(R_{1},\, R_{2}))^{-1} \end{flalign*}
The colored HOMFLY invariants of the link $\bf 7_2^2$ is symmetric under interchanging the two colors.
\begin{itemize}

\item{$\overline P_{([1],[1])}({\bf 7_2^2}; a,q)=$}
\beaa
\tfrac{(1-a) }{a^3 q(1-q)^2}\left( \begin{array}{ccccc}  0 & -1 & 2 & -1 & 0 \\  1 & -4 & 5 & -4 & 1 \\  1 & -3 & 5 & -3 & 1 \\  0 & -1 & 2 & -1 & 0 \\ \end{array} \right)
\eeaa

\item{$\overline P_{([1],[2])}({\bf 7_2^2}; a,q)=\overline P_{([2],[1])}({\bf 7_2^2}; a,q)=$}
\beaa
 \tfrac{(1-a) (1-a q) }{a^{7/2}q^{5/2} (1-q)^2 (1-q^2) }\left( \begin{array}{cccccccc}  0 & 0 & 0 & -1 & 1 & 1 & -1 & 0 \\  0 & 1 & -3 & 0 & 4 & -3 & -1 & 1 \\  1 & -1 & -2 & 4 & 0 & -2 & 1 & 0 \\  0 & -1 & 1 & 1 & -1 & 0 & 0 & 0 \\ \end{array} \right)
\eeaa

\item{$\overline P_{([1],[3])}({\bf 7_2^2}; a,q)=\overline P_{([3],[1])}({\bf 7_2^2}; a,q)=$}
\beaa
 \tfrac{(1-a) (1-a q) (1-a q^2)}{a^4 q^4 (1-q)^2 (1-q^2) (1-q^3)}\left( \begin{array}{ccccccccccc}  0 & 0 & 0 & 0 & 0 & -1 & 1 & 0 & 1 & -1 & 0 \\  0 & 0 & 1 & -3 & 1 & -1 & 4 & -3 & 0 & -1 & 1 \\  1 & -1 & 0 & -2 & 4 & -1 & 1 & -2 & 1 & 0 & 0 \\  0 & -1 & 1 & 0 & 1 & -1 & 0 & 0 & 0 & 0 & 0 \\ \end{array} \right)
\eeaa

\item{$\overline P_{([1],[4])}({\bf 7_2^2}; a,q)=\overline P_{([4],[1])}({\bf 7_2^2}; a,q)=$}
\beaa
\tfrac{(1-a) (1-a q) (1-a q^2) (1-a q^3)}{a^{{9}/{2}} q^{{11}/{2}} (1-q)^2 (1-q^2) \left(1-q^3\right)(1-q^4) }
\begin{small}
\left( \begin{array}{cccccccccccccc}  0 & 0 & 0 & 0 & 0 & 0 & 0 & -1 & 1 & 0 & 0 & 1 & -1 & 0 \\  0 & 0 & 0 & 1 & -3 & 1 & 0 & -1 & 4 & -3 & 0 & 0 & -1 & 1 \\  1 & -1 & 0 & 0 & -2 & 4 & -1 & 0 & 1 & -2 & 1 & 0 & 0 & 0 \\  0 & -1 & 1 & 0 & 0 & 1 & -1 & 0 & 0 & 0 & 0 & 0 & 0 & 0 \\ \end{array} \right)
\end{small}
\eeaa

\item{$\overline P_{([2],[2])}({\bf 7_2^2}; a,q)=$}
\beaa
&& \tfrac{(1-a) (1-a q)}{a^{6}q^{6}(1-q)^2 (1-q^2)^2 }\times\\
&&\begin{footnotesize}
\left( \begin{array}{ccccccccccccccc}  0 & 0 & 0 & 0 & 0 & 0 & 0 & 1 & -2 & -1 & 4 & -1 & -2 & 1 & 0 \\  0 & 0 & 0 & 0 & -1 & 2 & 3 & -8 & -1 & 12 & -4 & -8 & 4 & 2 & -1 \\  0 & 0 & 1 & -4 & 2 & 12 & -17 & -8 & 29 & -6 & -20 & 11 & 5 & -4 & 0 \\  0 & 1 & -4 & 1 & 13 & -15 & -15 & 29 & 2 & -25 & 7 & 9 & -5 & -1 & 1 \\  1 & -3 & 1 & 10 & -11 & -11 & 21 & 2 & -16 & 4 & 5 & -2 & 0 & 0 & 0 \\  -1 & 1 & 4 & -6 & -5 & 11 & 1 & -8 & 2 & 2 & -1 & 0 & 0 & 0 & 0 \\  0 & 1 & -2 & -1 & 4 & -1 & -2 & 1 & 0 & 0 & 0 & 0 & 0 & 0 & 0 \\ \end{array} \right)\end{footnotesize}
\eeaa

\item{$\overline P_{([2],[3])}({\bf 7_2^2}; a,q)=\overline P_{([3],[2])}({\bf 7_2^2}; a,q)=$}
\beaa
&& \tfrac{(1-a) (1-a q) (1-a q^2)}{a^{{13}/{2}}q^{{19}/2}(1-q)^2 (1-q^2)^2 (1-q^3) }\times\\
&&\begin{footnotesize}
\left( \begin{array}{ccccccccccccccccccccc}  0 & 0 & 0 & 0 & 0 & 0 & 0 & 0 & 0 & 0 & 0 & 1 & -1 & -2 & 1 & 2 & 1 & -2 & -1 & 1 & 0 \\  0 & 0 & 0 & 0 & 0 & 0 & 0 & -1 & 1 & 4 & -2 & -7 & -1 & 9 & 5 & -7 & -6 & 2 & 4 & 0 & -1 \\  0 & 0 & 0 & 0 & 1 & -3 & -2 & 10 & 4 & -14 & -12 & 13 & 21 & -7 & -19 & 0 & 10 & 4 & -4 & -2 & 1 \\  0 & 0 & 1 & -2 & -4 & 5 & 11 & -6 & -21 & -1 & 25 & 10 & -19 & -14 & 8 & 10 & -2 & -4 & 0 & 1 & 0 \\  1 & -1 & -3 & 2 & 7 & 2 & -13 & -9 & 14 & 13 & -5 & -13 & -1 & 8 & 1 & -2 & 0 & 0 & 0 & 0 & 0 \\  -1 & 0 & 3 & 2 & -5 & -6 & 4 & 8 & 0 & -6 & -2 & 3 & 1 & -1 & 0 & 0 & 0 & 0 & 0 & 0 & 0 \\  0 & 1 & -1 & -2 & 1 & 2 & 1 & -2 & -1 & 1 & 0 & 0 & 0 & 0 & 0 & 0 & 0 & 0 & 0 & 0 & 0 \\ \end{array} \right)\end{footnotesize}\eeaa

\item{$\overline P_{([2],[4])}({\bf 7_2^2}; a,q)=\overline P_{([4],[2])}({\bf 7_2^2}; a,q)=$}
\beaa
&& \tfrac{(1-a) (1-a q) (1-a q^2)(1-a q^3)}{a^{7}q^{13}(1-q)^2 (1-q^2)^2 (1-q^3) (1-q^4) }\times\\
&&\begin{tiny}
\left( \begin{array}{ccccccccccccccccccccccccccc}  0 & 0 & 0 & 0 & 0 & 0 & 0 & 0 & 0 & 0 & 0 & 0 & 0 & 0 & 0 & 1 & -1 & -1 & 0 & 0 & 2 & 0 & 0 & -1 & -1 & 1 & 0 \\  0 & 0 & 0 & 0 & 0 & 0 & 0 & 0 & 0 & 0 & -1 & 1 & 3 & -1 & -2 & -4 & -1 & 6 & 3 & 1 & -4 & -5 & 1 & 2 & 2 & 0 & -1 \\  0 & 0 & 0 & 0 & 0 & 0 & 1 & -3 & -1 & 6 & 2 & 2 & -10 & -8 & 6 & 6 & 13 & -4 & -13 & -2 & 0 & 7 & 3 & -2 & -2 & -1 & 1 \\  0 & 0 & 0 & 1 & -2 & -2 & 1 & 3 & 8 & -5 & -11 & -5 & 0 & 17 & 7 & -7 & -8 & -9 & 5 & 7 & 1 & -1 & -3 & 0 & 1 & 0 & 0 \\  1 & -1 & -1 & -1 & 1 & 6 & 0 & -2 & -8 & -6 & 10 & 6 & 4 & -3 & -9 & -1 & 2 & 4 & 1 & -2 & 0 & 0 & 0 & 0 & 0 & 0 & 0 \\  -1 & 0 & 2 & 1 & 1 & -4 & -4 & 2 & 2 & 5 & 0 & -4 & -1 & -1 & 2 & 1 & -1 & 0 & 0 & 0 & 0 & 0 & 0 & 0 & 0 & 0 & 0 \\  0 & 1 & -1 & -1 & 0 & 0 & 2 & 0 & 0 & -1 & -1 & 1 & 0 & 0 & 0 & 0 & 0 & 0 & 0 & 0 & 0 & 0 & 0 & 0 & 0 & 0 & 0 \\ \end{array} \right)\end{tiny}\eeaa

\end{itemize}

\subsubsection{$7_4^2$ link}
\begin{figure}[h]
\centering{\includegraphics[scale=1]{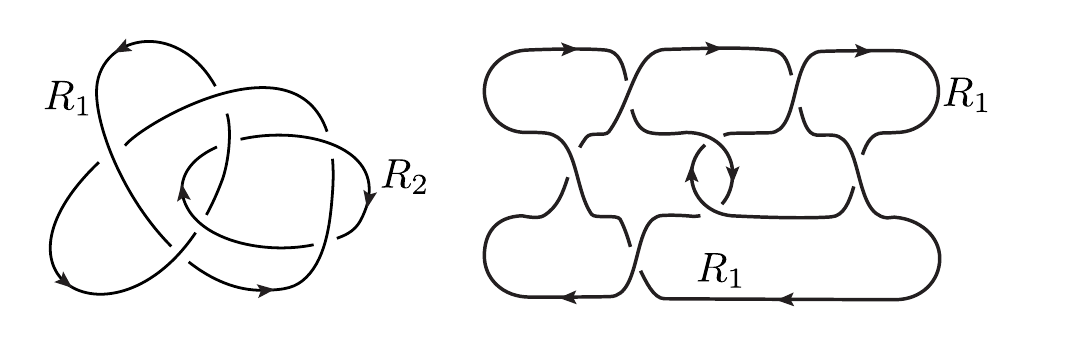}}
\caption{$\bf{7_4^2}$ link}
\end{figure}
\begin{flalign*}
\bar{P}_{R_{1},R_{2}}(7_{4}^{2};a,q) = & q^{-3C_{R_1}}\sum_{l,r,u,v,x,y}\frac{1}{\epsilon_{l}^{R,\bar{R}}\sqrt{\dim_{q}l}}~\epsilon_{r}^{R_{1},R_{2}}\sqrt{\dim_{q}r}~\epsilon_{u}^{R_{1},\bar{R}_{1}}&\\
   & \times \sqrt{\dim_{q}u}~\epsilon_{v}^{R_{1},R_{2}}\sqrt{\dim_{q}v}~ a_{rl}\!\left[\begin{footnotesize}\begin{array}{cc}
R_{1} & R_{2}\\
\bar{R}_{2} & \bar{R}_{1}
\end{array}\end{footnotesize}\right]a_{xl}\!\left[\begin{footnotesize}\begin{array}{cc}
R_{1} & R_{1}\\
\bar{R}_{1} & \bar{R}_{1}
\end{array}\end{footnotesize}\right]~a_{ly}\!\left[\begin{footnotesize}\begin{array}{cc}
R_{1} & \bar{R}_{1}\\
R_{2} & \bar{R}_{2}
\end{array}\end{footnotesize}\right]&\\
   & \times a_{xu}\!\left[\begin{footnotesize}\begin{array}{cc}
R_{1} & R_{1}\\
\bar{R}_{1} & \bar{R}_{1}
\end{array}\end{footnotesize}\right]a_{vy}\!\left[\begin{footnotesize}\begin{array}{cc}
R_{1} & R_{2}\\
\bar{R}_{1} & \bar{R}_{2}
\end{array}\end{footnotesize}\right]~(\lambda_{r}^{(+)}(R_{1},R_{2}))^{-2}~(\lambda_{x}^{(+)}(R_{1},R_{1}))^{-2}&\\
 &\times(\lambda_{u}^{(-)}(R_{1},\bar{R}_{1}))^{-1}~\lambda_{y}^{(-)}(\bar{R}_{1},R_{2})~\lambda_{v}^{(+)}(R_{1},R_{2})\end{flalign*}

\begin{itemize}

\item{$\overline P_{([1],[1])}({\bf 7_4^2}; a,q)=$}
\beaa
 \tfrac{(1-a)}{a^{3}q^{2}(1-q)^2}\left( \begin{array}{ccccccc}  0 & -1 & 1 & -2 & 1 & -1 & 0 \\  1 & -2 & 4 & -3 & 4 & -2 & 1 \\  0 & -1 & 2 & -3 & 2 & -1 & 0 \\ \end{array} \right)
\eeaa

\item{$\overline P_{([1],[2])}({\bf 7_4^2}; a,q)=$}
\beaa
 \tfrac{(1-a)(1-a q) }{a^{7/2}q^{5/2}(1-q)^2(1-q^2)}\left( \begin{array}{ccccccccc}  0 & 0 & -1 & 0 & 0 & -1 & 1 & -1 & 0 \\  1 & -1 & 0 & 2 & -1 & 2 & 0 & -1 & 1 \\  0 & -1 & 1 & 0 & -1 & 1 & -1 & 0 & 0 \\ \end{array} \right)
\eeaa

\item{$\overline P_{([1],[3])}({\bf 7_4^2}; a,q)=$}
\beaa
 \tfrac{(1-a)(1-a q) (1-a q^2)  }{a^{4} q^{3}(1-q)^2(1-q^2)(1-q^3)}
\left( \begin{array}{ccccccccccc}  0 & 0 & 0 & -1 & 0 & -1 & 1 & -1 & 1 & -1 & 0 \\  1 & -1 & 1 & -2 & 3 & -1 & 3 & -2 & 1 & -1 & 1 \\  0 & -1 & 1 & -1 & 2 & -2 & 1 & -1 & 0 & 0 & 0 \\ \end{array} \right)
\eeaa

\item{$\overline P_{([1],[4])}({\bf 7_4^2}; a,q)=$}
\beaa
 \tfrac{(1-a)(1-a q) (1-a q^2) \left(1-a q^3\right) }{a^{9/2} q^{7/2} (1-q)^2(1-q^2)(1-q^3)(1-q^4)  }
\left( \begin{array}{ccccccccccccc}  0 & 0 & 0 & 0 & -1 & 0 & -1 & 0 & 1 & -1 & 1 & -1 & 0 \\  1 & -1 & 1 & -1 & -1 & 3 & -1 & 3 & -1 & -1 & 1 & -1 & 1 \\  0 & -1 & 1 & -1 & 1 & 1 & -2 & 1 & -1 & 0 & 0 & 0 & 0 \\ \end{array} \right)
\eeaa

\item{$\overline P_{([2],[1])}({\bf 7_4^2}; a,q)=$}
\beaa
&& \tfrac{(1-a) (1-a q)}{a^{11/2}q^{11/2}(1-q)^2 (1-q^2) }\begin{footnotesize}\left( \begin{array}{ccccccccccccc}  0 & 0 & -1 & 0 & 0 & -2 & 1 & 0 & -2 & 0 & 1 & -1 & 0 \\  1 & -1 & 1 & 3 & -3 & 3 & 4 & -3 & 0 & 4 & -1 & -1 & 1 \\  -1 & 0 & 2 & -4 & -1 & 4 & -4 & -2 & 2 & 0 & -1 & 0 & 0 \\  0 & 1 & -1 & -1 & 3 & -1 & -1 & 1 & 0 & 0 & 0 & 0 & 0   \\ \end{array} \right)\end{footnotesize}
\eeaa

\item{$\overline P_{([2],[2])}({\bf 7_4^2}; a,q)=$}
\beaa
&& \tfrac{(1-a) (1-a q)}{a^{6}q^{8}(1-q)^2 (1-q^2)^2 }\times\\
&&\begin{footnotesize}\left( \begin{array}{ccccccccccccccccccc}  0 & 0 & 0 & 0 & 0 & 1 & -1 & 1 & 2 & -2 & 1 & 3 & -3 & 0 & 3 & -1 & -1 & 1 & 0 \\  0 & 0 & -1 & 1 & 0 & -4 & 2 & 0 & -7 & 3 & 2 & -9 & 1 & 6 & -5 & -3 & 3 & 0 & -1 \\  1 & -2 & 2 & 3 & -6 & 5 & 5 & -10 & 10 & 9 & -15 & 3 & 15 & -8 & -5 & 7 & 0 & -2 & 1 \\  -1 & 1 & 1 & -5 & 4 & 3 & -12 & 6 & 9 & -15 & -2 & 12 & -4 & -6 & 3 & 1 & -1 & 0 & 0 \\  0 & 1 & -2 & 1 & 3 & -6 & 2 & 7 & -7 & -2 & 6 & -1 & -2 & 1 & 0 & 0 & 0 & 0 & 0 \\ \end{array} \right)\end{footnotesize}
\eeaa

\item{$\overline P_{([2],[3])}({\bf 7_4^2}; a,q)=$}
\beaa
&& \tfrac{(1-a) (1-a q) (1-a q^2)}{a^{{13}/{2}}q^{{19}/{2}}(1-q)^2 (1-q^2)^2 (1-q^3) }\times\\
&&\begin{footnotesize}
\left( \begin{array}{ccccccccccccccccccccccc}  0 & 0 & 0 & 0 & 0 & 0 & 0 & 1 & 0 & 0 & 1 & 0 & 0 & 1 & 1 & 0 & -2 & 1 & 2 & -1 & -1 & 1 & 0 \\  0 & 0 & 0 & -1 & 0 & 1 & -1 & -1 & -1 & -2 & -2 & 0 & 2 & -3 & -5 & 1 & 4 & -1 & -4 & 0 & 2 & 0 & -1 \\  1 & -1 & -1 & 3 & 0 & -1 & 2 & 0 & -2 & 1 & 9 & 3 & -9 & -2 & 11 & 3 & -6 & -3 & 4 & 3 & -2 & -1 & 1 \\  -1 & 0 & 2 & -1 & -2 & 2 & -1 & -4 & 2 & 6 & -3 & -10 & 1 & 9 & -1 & -7 & 0 & 3 & 0 & -1 & 0 & 0 & 0 \\  0 & 1 & -1 & -1 & 2 & -1 & -1 & 2 & 1 & -1 & -3 & 2 & 3 & -2 & -1 & 1 & 0 & 0 & 0 & 0 & 0 & 0 & 0 \\ \end{array} \right)
\end{footnotesize}\eeaa

\item{$\overline P_{([2],[4])}({\bf 7_4^2}; a,q)=$}
\beaa
&& \tfrac{(1-a) (1-a q) (1-a q^2) (1-a q^3) }{a^7q^{11}(1-q)^2 (1-q^2)^2 (1-q^3) (1-q^4)}\times\\
&&\begin{footnotesize}
\left( \begin{array}{ccccccccccccccccccccccccccc}  0 & 0 & 0 & 0 & 0 & 0 & 0 & 0 & 0 & 1 & 0 & 1 & 0 & 0 & 0 & 0 & 2 & 0 & -1 & 1 & -1 & 0 & 2 & -1 & -1 & 1 & 0 \\  0 & 0 & 0 & 0 & -1 & 0 & 0 & 0 & 1 & -1 & -2 & -4 & -1 & 2 & -1 & 0 & -2 & -4 & 2 & 2 & -2 & 0 & -1 & -1 & 2 & 0 & -1 \\  1 & -1 & 0 & 0 & 0 & 2 & 1 & 2 & -4 & -3 & 4 & 3 & 6 & 0 & -4 & 0 & 3 & 3 & 1 & -3 & -1 & 2 & 1 & 1 & -1 & -1 & 1 \\  -1 & 0 & 1 & 0 & 1 & -1 & -1 & -3 & -1 & 6 & 0 & -2 & -2 & -4 & 1 & 3 & 0 & -2 & -2 & 0 & 2 & 0 & -1 & 0 & 0 & 0 & 0 \\  0 & 1 & -1 & 0 & 0 & -1 & 1 & 0 & 2 & -1 & -2 & 2 & -1 & 0 & 2 & -1 & -1 & 1 & 0 & 0 & 0 & 0 & 0 & 0 & 0 & 0 & 0 \\ \end{array} \right)
\end{footnotesize}\eeaa

\item{$\overline P_{([3],[1])}({\bf 7_4^2}; a,q)=$}
\beaa
&& \tfrac{(1-a) (1-a q) (1-a q^2) }{a^8q^{12}(1-q)^2 (1-q^2) (1-q^3) }\times\\
&&\begin{footnotesize}\left( \begin{array}{cccccccccccccccccccccc}  0 & 0 & 0 & 0 & -1 & 0 & -1 & 0 & -2 & 1 & -1 & -1 & -2 & 1 & 0 & 0 & -2 & 0 & 0 & 1 & -1 & 0 \\  0 & 1 & -1 & 2 & 0 & 3 & -2 & 5 & 2 & 3 & -4 & 4 & 3 & 4 & -3 & 0 & 0 & 4 & -1 & 0 & -1 & 1 \\  -1 & 0 & 0 & 0 & -4 & -1 & 1 & 0 & -6 & -4 & 1 & 3 & -3 & -3 & -2 & 2 & 0 & 0 & -1 & 0 & 0 & 0 \\  1 & 0 & 0 & -2 & 4 & 1 & 1 & -4 & 3 & 2 & 2 & -2 & 0 & 0 & 1 & 0 & 0 & 0 & 0 & 0 & 0 & 0 \\  0 & -1 & 1 & 0 & 1 & -3 & 1 & 0 & 1 & -1 & 0 & 0 & 0 & 0 & 0 & 0 & 0 & 0 & 0 & 0 & 0 & 0 \\ \end{array} \right)
\end{footnotesize}\eeaa

\item{$\overline P_{([3],[2])}({\bf 7_4^2}; a,q)=$}
\beaa
&& \tfrac{(1-a) (1-a q) (1-a q^2)}{a^{17/2}q^{31/2}(1-q)^2 (1-q^2)^2 (1-q^3) }\times\\
&&\begin{tiny}
\left( \begin{array}{cccccccccccccccccccccccccccccc}  0 & 0 & 0 & 0 & 0 & 0 & 0 & 0 & 1 & 0 & 0 & 1 & 1 & 0 & 0 & 2 & 2 & -2 & -1 & 2 & 3 & -1 & -2 & 0 & 2 & 1 & -1 & -1 & 1 & 0 \\  0 & 0 & 0 & 0 & -1 & 0 & 1 & -2 & -2 & 0 & -1 & -5 & -4 & 3 & 1 & -9 & -7 & 4 & 6 & -4 & -9 & -2 & 6 & 3 & -4 & -4 & 1 & 2 & 0 & -1 \\  0 & 1 & -1 & 0 & 3 & -1 & -1 & 4 & 5 & -3 & -4 & 13 & 13 & -7 & -12 & 10 & 20 & 2 & -15 & -5 & 11 & 12 & -4 & -8 & 0 & 5 & 2 & -2 & -1 & 1 \\  -1 & 0 & 2 & -2 & -4 & 2 & 6 & -5 & -14 & 4 & 16 & -6 & -24 & -5 & 20 & 9 & -18 & -14 & 5 & 11 & 1 & -7 & -4 & 2 & 2 & 0 & -1 & 0 & 0 & 0 \\  1 & 0 & -2 & 0 & 5 & 1 & -9 & -1 & 13 & 6 & -13 & -10 & 13 & 13 & -5 & -11 & 0 & 7 & 3 & -2 & -2 & 0 & 1 & 0 & 0 & 0 & 0 & 0 & 0 & 0 \\  0 & -1 & 1 & 2 & -3 & -2 & 3 & 4 & -3 & -7 & 3 & 6 & -1 & -4 & -1 & 2 & 1 & -1 & 0 & 0 & 0 & 0 & 0 & 0 & 0 & 0 & 0 & 0 & 0 & 0 \\ \end{array} \right)
\end{tiny}\eeaa

\item{$\overline P_{([4],[1])}({\bf 7_4^2}; a,q)=$}
\beaa
&& \tfrac{(1-a) (1-a q) (1-a q^2) (1-a q^3) }{a^{21/2}q^{43/2}(1-q)^2 (1-q^2) (1-q^3) (1-q^4)}\times\\
&&\begin{tiny}
\left( \begin{array}{cccccccccccccccccccccccccccccccccc}  0 & 0 & 0 & 0 & 0 & 0 & 0 & -1 & 0 & -1 & -1 & 0 & -2 & 0 & -1 & -2 & -1 & -1 & 0 & 0 & -2 & -1 & -2 & 1 & 0 & 0 & 0 & -2 & 0 & 0 & 0 & 1 & -1 & 0 \\  0 & 0 & 0 & 1 & -1 & 2 & 1 & 0 & 4 & 0 & 5 & 4 & 1 & 3 & 1 & 5 & 7 & 2 & 2 & -3 & 3 & 4 & 3 & 4 & -3 & 0 & 0 & 0 & 4 & -1 & 0 & 0 & -1 & 1 \\  0 & -1 & 0 & 0 & -2 & 0 & -5 & -1 & 0 & -5 & -3 & -9 & -4 & 0 & -1 & -2 & -9 & -6 & -2 & 0 & 4 & -4 & -2 & -3 & -2 & 2 & 0 & 0 & 0 & -1 & 0 & 0 & 0 & 0 \\  1 & 0 & 1 & -1 & 1 & 3 & 2 & 4 & -3 & 2 & 3 & 5 & 7 & -1 & -1 & 0 & 1 & 6 & 1 & 1 & -1 & -1 & 1 & 0 & 1 & 0 & 0 & 0 & 0 & 0 & 0 & 0 & 0 & 0 \\  -1 & 0 & 0 & 0 & 2 & -4 & -1 & -1 & -1 & 4 & -3 & -1 & -2 & -2 & 2 & 0 & 0 & 0 & -1 & 0 & 0 & 0 & 0 & 0 & 0 & 0 & 0 & 0 & 0 & 0 & 0 & 0 & 0 & 0 \\  0 & 1 & -1 & 0 & 0 & -1 & 3 & -1 & 0 & 0 & -1 & 1 & 0 & 0 & 0 & 0 & 0 & 0 & 0 & 0 & 0 & 0 & 0 & 0 & 0 & 0 & 0 & 0 & 0 & 0 & 0 & 0 & 0 & 0 \\ \end{array} \right)
\end{tiny}\eeaa

\end{itemize}

\subsubsection{$7_5^2$ link}
\begin{figure}[h]
\centering{\includegraphics[scale=1]{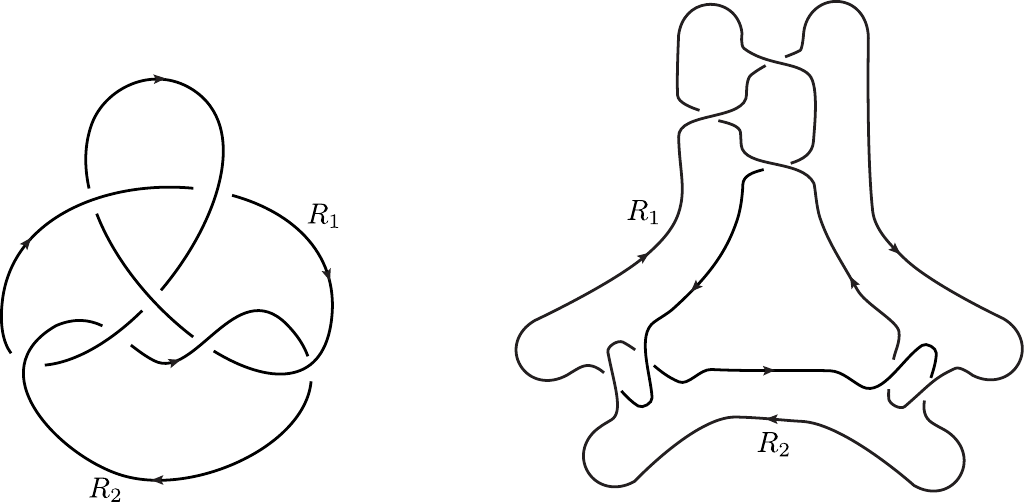}}
\caption{$\bf{7_5^2}$ link}
\end{figure}
\begin{flalign*}
\bar{P}_{R_1,R_2}({\bf 7_5^2};\, a,q) = &q^{3C_{R_1}+\frac{2{\ell}^{1}{\ell}^{2}}{N}} \sum_{l,u,v,x,y,z}\frac{1}{\epsilon_{l}^{R_1,\bar{R}_1}\sqrt{\dim_{q}l}}\epsilon_{y}^{R_{1},\bar{R}_{1}}\sqrt{\dim_{q}y}~\epsilon_{u}^{\bar{R}_{1},R_{2}}&\\
   & \times \sqrt{\dim_{q}u}~\epsilon_{v}^{\bar{R}_{1},R_{2}}\sqrt{\dim_{q}v}~a_{lx}\!\left[\begin{footnotesize}\begin{array}{cc}
R_{1} & \bar{R}_{1}\\
R_{1} & \bar{R}_{1}
\end{array}\end{footnotesize}\right]~a_{zx}\!\left[\begin{footnotesize}\begin{array}{cc}
R_{1} & R_{1}\\
\bar{R}_{1} & \bar{R}_{1}
\end{array}\end{footnotesize}\right]~a_{zy}\!\left[\begin{footnotesize}\begin{array}{cc}
R_{1} & R_{1}\\
\bar{R}_{1} & \bar{R}_{1}
\end{array}\end{footnotesize}\right]&\\
   & \times a_{lu}\!\left[\begin{footnotesize}\begin{array}{cc}
R_{1} & \bar{R}_{1}\\
R_2 & \bar{R}_{2}
\end{array}\end{footnotesize}\right]~a_{lv}\!\left[\begin{footnotesize}\begin{array}{cc}
R_{1} & \bar{R}_{1}\\
{R}_{2} & \bar{R}_{2}
\end{array}\end{footnotesize}\right]~\lambda_{x}^{(-)}(R_{1},\bar{R}_{1})~\lambda_{z}^{(+)}(R_1,R_1)~\lambda_{y}^{(-)}(R_{1},\bar{R}_{1})&\\
&\times (\lambda_{u}^{(-)}(\bar{R}_{1},R_{2}))^{2}~(\lambda_{v}^{(-)}(\bar{R}_{1},R_{2}))^{2}\end{flalign*}

\begin{itemize}

\item{$\overline P_{([1],[1])}({\bf 7_5^2}; a,q)=$}
\beaa
\tfrac{a(1-a)}{q(1-q)^2}\left( \begin{array}{ccccc}  0 & 0 & 1 & 0 & 0 \\  0 & -3 & 3 & -3 & 0 \\  2 & -4 & 6 & -4 & 2 \\  1 & -3 & 4 & -3 & 1 \\ \end{array} \right)
\eeaa

\item{$\overline P_{([1],[2])}({\bf 7_5^2}; a,q)=$}
\beaa
\tfrac{a^{1/2}(1-a)(1-a q)  }{q^{1/2}(1-q)^2(1-q^2)}\left( \begin{array}{ccccccc}  0 & 0 & 0 & 0 & 1 & 0 & 0 \\  0 & 0 & -2 & 1 & 1 & -3 & 0 \\  1 & -1 & -1 & 4 & -2 & -1 & 2 \\  1 & -2 & 1 & 1 & -2 & 1 & 0 \\ \end{array} \right)
\eeaa

\item{$\overline P_{([1],[3])}({\bf 7_5^2}; a,q)=$}
\beaa
\tfrac{(1-a)(1-a q) (1-a q^2)  }{(1-q)^2(1-q^2)(1-q^3)}
\left( \begin{array}{ccccccccc}  0 & 0 & 0 & 0 & 0 & 0 & 1 & 0 & 0 \\  0 & 0 & 0 & -2 & 2 & -1 & 1 & -3 & 0 \\  1 & -2 & 2 & -2 & 4 & -3 & 1 & -1 & 2 \\  1 & -2 & 2 & -2 & 2 & -2 & 1 & 0 & 0 \\ \end{array} \right)
\eeaa
\item{$\overline P_{([1],[4])}({\bf 7_5^2}; a,q)=$}
\beaa
\tfrac{q^{{1}/{2}}(1-a)(1-a q) (1-a q^2) (1-a q^3) }{a^{{1}/{2}} (1-q)^2(1-q^2)(1-q^3)(1-q^4)  }
\left( \begin{array}{ccccccccccc}  0 & 0 & 0 & 0 & 0 & 0 & 0 & 0 & 1 & 0 & 0 \\  0 & 0 & 0 & 0 & -2 & 2 & 0 & -1 & 1 & -3 & 0 \\  1 & -2 & 1 & 1 & -2 & 4 & -3 & 0 & 1 & -1 & 2 \\  1 & -2 & 2 & -1 & -1 & 2 & -2 & 1 & 0 & 0 & 0 \\ \end{array} \right)
\eeaa

\item{$\overline P_{([2],[1])}({\bf 7_5^2}; a,q)=$}
\beaa
\begin{footnotesize}\tfrac{a^{5/2}(1-a)(1-a q)   }{q^{5/2}(1-q)^2(1-q^2)}
\left( \begin{array}{ccccccccccc}  0 & 0 & 0 & 0 & 0 & 0 & 0 & 0 & -1 & 0 & 0 \\  0 & 0 & 0 & 0 & 0 & 1 & 3 & -2 & 0 & 3 & 0 \\  0 & 0 & 0 & -3 & 0 & 4 & -5 & -4 & 4 & -2 & -2 \\  0 & 2 & -3 & -1 & 7 & -2 & -5 & 6 & 0 & -2 & 2 \\  1 & -1 & -2 & 3 & 1 & -4 & 2 & 1 & -2 & 1 & 0 \\ \end{array} \right)
\end{footnotesize}
\eeaa

\item{$\overline P_{([2],[2])}({\bf 7_5^2}; a,q)=$}
\beaa
\tfrac{a^{2}(1-a) (1-a q) }{q^{2}(1-q)^2 (1-q^2)^2 }\begin{footnotesize}\left( \begin{array}{ccccccccccccccc}  0 & 0 & 0 & 0 & 0 & 0 & 0 & 0 & 0 & 0 & 0 & 0 & 1 & 0 & 0 \\  0 & 0 & 0 & 0 & 0 & 0 & 0 & 0 & 0 & -3 & -1 & 3 & -2 & -3 & 0 \\  0 & 0 & 0 & 0 & 0 & 0 & 3 & 4 & -9 & 1 & 15 & -4 & -6 & 7 & 2 \\  0 & 0 & 0 & -1 & -4 & 7 & 6 & -22 & 0 & 24 & -14 & -14 & 11 & 0 & -5 \\  0 & 1 & -1 & -7 & 11 & 12 & -28 & -1 & 34 & -16 & -16 & 18 & 0 & -6 & 3 \\  2 & -1 & -9 & 9 & 15 & -25 & -6 & 31 & -12 & -16 & 15 & 0 & -5 & 2 & 0 \\  1 & -3 & 0 & 9 & -8 & -8 & 15 & -2 & -10 & 7 & 1 & -3 & 1 & 0 & 0 \\ \end{array} \right)
\end{footnotesize}
\eeaa

\item{$\overline P_{([2],[3])}({\bf 7_5^2}; a,q)=$}
\beaa
&&\tfrac{a^{{3}/{2}}(1-a) (1-a q) (1-a q^2)}{q^{{3}/{2}}(1-q)^2 (1-q^2)^2 (1-q^3) }\times\\
&&\begin{footnotesize}
\left( \begin{array}{ccccccccccccccccccc}  0 & 0 & 0 & 0 & 0 & 0 & 0 & 0 & 0 & 0 & 0 & 0 & 0 & 0 & 0 & 0 & 1 & 0 & 0 \\  0 & 0 & 0 & 0 & 0 & 0 & 0 & 0 & 0 & 0 & 0 & 0 & -2 & -1 & 1 & 1 & -2 & -3 & 0 \\  0 & 0 & 0 & 0 & 0 & 0 & 0 & 0 & 1 & 4 & -3 & -5 & 4 & 8 & 2 & -6 & -1 & 7 & 2 \\  0 & 0 & 0 & 0 & 0 & -3 & 1 & 7 & -1 & -12 & -5 & 13 & 8 & -13 & -10 & 5 & 6 & -3 & -5 \\  0 & 0 & 2 & -4 & -5 & 10 & 9 & -10 & -16 & 7 & 22 & -5 & -17 & 2 & 11 & 2 & -6 & -1 & 3 \\  1 & 1 & -5 & -3 & 11 & 6 & -15 & -9 & 14 & 11 & -11 & -10 & 8 & 6 & -5 & -2 & 2 & 0 & 0 \\  1 & -2 & -2 & 6 & 1 & -7 & -1 & 5 & 3 & -5 & -2 & 5 & -1 & -2 & 1 & 0 & 0 & 0 & 0 \\ \end{array} \right)
\end{footnotesize}\eeaa

\item{$\overline P_{([2],[4])}({\bf 7_5^2}; a,q)=$}
\beaa
&& \tfrac{a(1-a) (1-a q) (1-a q^2) (1-a q^3) }{q(1-q)^2 (1-q^2)^2 (1-q^3) (1-q^4)}\times\\
&&\begin{footnotesize}
\left( \begin{array}{ccccccccccccccccccccccc}  0 & 0 & 0 & 0 & 0 & 0 & 0 & 0 & 0 & 0 & 0 & 0 & 0 & 0 & 0 & 0 & 0 & 0 & 0 & 0 & 1 & 0 & 0 \\  0 & 0 & 0 & 0 & 0 & 0 & 0 & 0 & 0 & 0 & 0 & 0 & 0 & 0 & 0 & -2 & 0 & 1 & -1 & 1 & -2 & -3 & 0 \\  0 & 0 & 0 & 0 & 0 & 0 & 0 & 0 & 0 & 0 & 1 & 2 & -2 & -1 & -1 & 3 & 6 & -2 & 0 & -1 & -1 & 7 & 2 \\  0 & 0 & 0 & 0 & 0 & 0 & -2 & 1 & 3 & 0 & -1 & -7 & 0 & 5 & 1 & 3 & -9 & -5 & 4 & 0 & 3 & -3 & -5 \\  0 & 0 & 1 & -1 & -3 & 0 & 6 & 3 & -6 & -2 & -3 & 2 & 11 & -4 & -4 & -2 & 0 & 8 & -1 & -2 & -1 & -1 & 3 \\  1 & 0 & -3 & 0 & 2 & 3 & 2 & -8 & -1 & 4 & -1 & 6 & -4 & -5 & 4 & 0 & 2 & -2 & -2 & 2 & 0 & 0 & 0 \\  1 & -2 & -1 & 4 & -1 & -1 & 0 & -2 & 2 & 0 & 1 & 0 & -3 & 3 & 0 & -2 & 1 & 0 & 0 & 0 & 0 & 0 & 0 \\ \end{array} \right)
\end{footnotesize}\eeaa

\item{$\overline P_{([3],[1])}({\bf 7_5^2}; a,q)=$}
\beaa
&& \tfrac{a^{4}(1-a) (1-a q) (1-a q^2)  }{q^4(1-q)^2 (1-q^2) (1-q^3)}\times\\
&&\begin{footnotesize}
\left( \begin{array}{cccccccccccccccccccc}  0 & 0 & 0 & 0 & 0 & 0 & 0 & 0 & 0 & 0 & 0 & 0 & 0 & 0 & 0 & 0 & 0 & 1 & 0 & 0 \\  0 & 0 & 0 & 0 & 0 & 0 & 0 & 0 & 0 & 0 & 0 & 0 & -1 & -1 & -3 & 2 & -1 & 0 & -3 & 0 \\  0 & 0 & 0 & 0 & 0 & 0 & 0 & 0 & 1 & 3 & 1 & 0 & -3 & 6 & 3 & 4 & -3 & 2 & 2 & 2 \\  0 & 0 & 0 & 0 & 0 & -3 & 0 & 1 & 3 & -4 & -7 & -1 & 3 & 1 & -6 & -4 & 1 & -1 & -1 & -2 \\  0 & 0 & 2 & -3 & 0 & -1 & 7 & -1 & -2 & -5 & 6 & 3 & 1 & -5 & 4 & 0 & 2 & -2 & 2 & 0 \\  1 & -1 & 0 & -2 & 3 & 0 & 1 & -4 & 1 & 1 & 2 & -3 & 1 & -1 & 2 & -2 & 1 & 0 & 0 & 0 \\ \end{array} \right)
\end{footnotesize}\eeaa

\item{$\overline P_{([3],[2])}({\bf 7_5^2}; a,q)=$}
\beaa
&& \tfrac{a^{7/2}(1-a) (1-a q) (1-a q^2)  }{q^{7/2}(1-q)^2 (1-q^2)^2 (1-q^3) }\times\\
&&\begin{tiny}
\left( \begin{array}{cccccccccccccccccccccccccc}  0 & 0 & 0 & 0 & 0 & 0 & 0 & 0 & 0 & 0 & 0 & 0 & 0 & 0 & 0 & 0 & 0 & 0 & 0 & 0 & 0 & 0 & 0 & -1 & 0 & 0 \\  0 & 0 & 0 & 0 & 0 & 0 & 0 & 0 & 0 & 0 & 0 & 0 & 0 & 0 & 0 & 0 & 0 & 0 & 1 & 3 & 1 & -2 & 0 & 2 & 3 & 0 \\  0 & 0 & 0 & 0 & 0 & 0 & 0 & 0 & 0 & 0 & 0 & 0 & 0 & 0 & -3 & -4 & -2 & 7 & 2 & -11 & -10 & 1 & 6 & -2 & -7 & -2 \\  0 & 0 & 0 & 0 & 0 & 0 & 0 & 0 & 0 & 0 & 3 & 5 & -2 & -11 & -1 & 22 & 15 & -13 & -18 & 9 & 23 & 5 & -9 & 0 & 5 & 5 \\  0 & 0 & 0 & 0 & 0 & 0 & -1 & -4 & 4 & 10 & -1 & -23 & -15 & 26 & 28 & -19 & -38 & -2 & 30 & 5 & -21 & -9 & 5 & 4 & -4 & -3 \\  0 & 0 & 0 & 1 & -1 & -6 & 0 & 16 & 10 & -23 & -28 & 19 & 44 & -2 & -44 & -13 & 35 & 17 & -18 & -12 & 9 & 8 & -4 & -3 & 3 & 0 \\  0 & 2 & 0 & -6 & -4 & 10 & 15 & -8 & -29 & -2 & 34 & 17 & -29 & -25 & 19 & 21 & -9 & -14 & 4 & 8 & -3 & -3 & 2 & 0 & 0 & 0 \\  1 & -1 & -3 & 1 & 6 & 3 & -9 & -9 & 8 & 14 & -4 & -15 & 1 & 11 & 1 & -7 & -1 & 5 & -1 & -2 & 1 & 0 & 0 & 0 & 0 & 0 \\ \end{array} \right)
\end{tiny}\eeaa

\item{$\overline P_{([4],[1])}({\bf 7_5^2}; a,q)=$}
\beaa
&& \tfrac{a^{11/2}(1-a) (1-a q) (1-a q^2) (1-a q^3) }{q^{11/2}(1-q)^2 (1-q^2) (1-q^3) (1-q^4)}\times\\
&&\begin{tiny}
\left( \begin{array}{cccccccccccccccccccccccccccccccc}  0 & 0 & 0 & 0 & 0 & 0 & 0 & 0 & 0 & 0 & 0 & 0 & 0 & 0 & 0 & 0 & 0 & 0 & 0 & 0 & 0 & 0 & 0 & 0 & 0 & 0 & 0 & 0 & 0 & -1 & 0 & 0 \\  0 & 0 & 0 & 0 & 0 & 0 & 0 & 0 & 0 & 0 & 0 & 0 & 0 & 0 & 0 & 0 & 0 & 0 & 0 & 0 & 0 & 0 & 1 & 1 & 1 & 3 & -2 & 1 & 1 & 0 & 3 & 0 \\  0 & 0 & 0 & 0 & 0 & 0 & 0 & 0 & 0 & 0 & 0 & 0 & 0 & 0 & 0 & 0 & -1 & -1 & -4 & -1 & -1 & -1 & 2 & -7 & -4 & -3 & -5 & 3 & -3 & -2 & -2 & -2 \\  0 & 0 & 0 & 0 & 0 & 0 & 0 & 0 & 0 & 0 & 0 & 1 & 3 & 1 & 3 & -3 & 0 & 3 & 5 & 11 & 1 & 1 & 2 & 0 & 11 & 4 & 4 & 2 & 0 & 4 & 1 & 2 \\  0 & 0 & 0 & 0 & 0 & 0 & 0 & -3 & 0 & 1 & 0 & 4 & -5 & -6 & -4 & -3 & 6 & 0 & -4 & -8 & -8 & 1 & -2 & 0 & -4 & -7 & 1 & -4 & 0 & -1 & -2 & 0 \\  0 & 0 & 0 & 2 & -3 & 0 & 0 & -1 & 7 & -1 & -1 & -2 & -5 & 6 & 3 & 5 & -2 & -4 & 2 & 0 & 6 & 1 & -3 & 4 & -2 & 2 & 2 & -2 & 2 & 0 & 0 & 0 \\  1 & -1 & 0 & 0 & -2 & 3 & 0 & 0 & 1 & -4 & 1 & 0 & 2 & 1 & -2 & 0 & -2 & 2 & 1 & -2 & 2 & -2 & 0 & 2 & -2 & 1 & 0 & 0 & 0 & 0 & 0 & 0 \\ \end{array} \right)
\end{tiny}\eeaa

\end{itemize}

\subsection{Three-component links}\label{sec:three}

\subsubsection{$6_1^3$ link}

\begin{figure}[h]
\centering{\includegraphics[scale=.8]{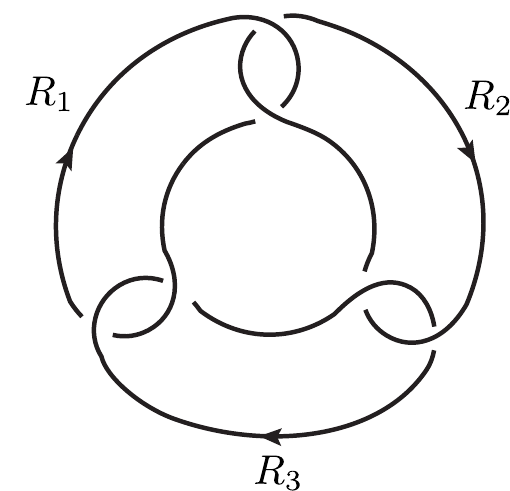}}
\caption{${\bf 6_1^3}$ link}
\end{figure}
\begin{eqnarray*}
\bar{P}_{(R_{1},R_{2},R_{3})}({\bf 6_{1}^{3}};\, a,q) & = &q^{-\frac{\ell^{(1)}\ell^{(2)}}{N}-\frac{\ell^{(2)}\ell^{(3)}}{N}-\frac{\ell^{(1)}\ell^{(3)}}{N}} \sum_{l,x,y,z}\frac{1}{\epsilon_{l}^{R_{1},\bar{R}_{1}}\sqrt{\dim_{q}l}}\epsilon_{x}^{\bar{R}_{1},R_{2}}\sqrt{\dim_{q}x}\,\\
 &  & \times \epsilon_{y}^{\bar{R}_{2},R_{3}}\sqrt{\dim_{q}y}\,\epsilon_{z}^{\bar{R}_{1},R_{3}}\sqrt{\dim_{q}z}\, a_{lx}\!\left[\begin{footnotesize}\begin{array}{cc}
R_{1} & \bar{R}_{1}\\
R_{2} & \bar{R}_{2}
\end{array}\end{footnotesize}\right]\, a_{ly}\!\left[\begin{footnotesize}\begin{array}{cc}
R_{2} & \bar{R}_{2}\\
R_{3} & \bar{R}_{3}
\end{array}\end{footnotesize}\right]\,\\
 &  & \times a_{lz}\!\left[\begin{footnotesize}\begin{array}{cc}
R_{1} & \bar{R}_{1}\\
R_{3} & \bar{R}_{3}
\end{array}\end{footnotesize}\right](\lambda_{x}^{(-)}(\bar{R}_{1},R_{2}))^{2}\,(\lambda_{y}^{(-)}(\bar{R}_{2},R_{3}))^{2}\,(\lambda_{z}^{(-)}(\bar{R}_{1},R_{3}))^{2}\end{eqnarray*}

The colored HOMFLY invariants of the link $\bf 6_{1}^{3}$ are symmetric under permutations over the representations $(R_1,R_2,R_3)$.
\begin{itemize}
\item{$\bar{P}_{([1],[1],[1])}({\bf 6_{1}^{3}}; a,q)=$}
\beaa
\tfrac{a^{1/2}(1-a)}{q^{1/2} (1-q)^3}\begin{small}
\left( \begin{array}{ccccc}  0 & 0 & 1 & 0 & 0 \\  0 & -3 & 4 & -3 & 0 \\  2 & -5 & 7 & -5 & 2 \\  1 & -4 & 6 & -4 & 1 \\ \end{array} \right)\end{small}
\eeaa

\item{$\bar{P}_{([1],[1],[2])}({\bf 6_{1}^{3}}; a,q)=$}
\beaa
\tfrac{(1-a) (1-a q) }{ (1-q)^3  (1-q^2)}
\begin{small}\left( \begin{array}{ccccccc}  0 & 0 & 0 & 0 & 1 & 0 & 0 \\  0 & 0 & -2 & 1 & 2 & -3 & 0 \\  1 & -1 & -2 & 5 & -2 & -2 & 2 \\  1 & -3 & 2 & 2 & -3 & 1 & 0 \\ \end{array} \right)\end{small}
\eeaa
\item{$\bar{P}_{([1],[1],[3])}({\bf 6_{1}^{3}}; a,q)=$}
\beaa
\tfrac{q^{1/2} (1-a) (1-a q) (1-a q^2)}{a^{1/2} (1-q)^3(1-q^2) (1-q^3)}
\begin{small}\left( \begin{array}{ccccccccc}  0 & 0 & 0 & 0 & 0 & 0 & 1 & 0 & 0 \\  0 & 0 & 0 & -2 & 2 & -1 & 2 & -3 & 0 \\  1 & -2 & 2 & -3 & 5 & -3 & 1 & -2 & 2 \\  1 & -3 & 3 & -2 & 3 & -3 & 1 & 0 & 0 \\ \end{array} \right)\end{small}
\eeaa

\item{$\bar{P}_{([1],[2],[2])}({\bf 6_{1}^{3}}; a,q)=$}
\beaa
\tfrac{(1-a) (1-a q)}{(1-q)^3 (1-q^2)^2}
\begin{small}\left( \begin{array}{ccccccccccc}  0 & 0 & 0 & 0 & 0 & 0 & 0 & 0 & -1 & 0 & 0 \\  0 & 0 & 0 & 0 & 0 & 1 & 2 & -3 & 0 & 3 & 0 \\  0 & 0 & 0 & -3 & 3 & 4 & -8 & -1 & 6 & -2 & -2 \\  0 & 2 & -5 & 1 & 10 & -9 & -5 & 10 & -2 & -3 & 2 \\  1 & -2 & -2 & 7 & -2 & -7 & 6 & 1 & -3 & 1 & 0 \\ \end{array} \right)\end{small}
\eeaa

\item{$\bar{P}_{([1],[2],[3])}({\bf 6_{1}^{3}}; a,q)=$}
\beaa
\tfrac{q^{1/2} (1-a) (1-a q) (1-a q^2) }{a^{1/2} (1-q)^3 (1-q^2)^2 (1-q^3)}
\begin{small}\left( \begin{array}{cccccccccccccc}  0 & 0 & 0 & 0 & 0 & 0 & 0 & 0 & 0 & 0 & 0 & -1 & 0 & 0 \\  0 & 0 & 0 & 0 & 0 & 0 & 0 & 1 & 1 & -1 & -1 & 0 & 3 & 0 \\  0 & 0 & 0 & 0 & -2 & 1 & 3 & 0 & -5 & -2 & 4 & 2 & -2 & -2 \\  0 & 1 & -1 & -3 & 3 & 5 & -3 & -6 & 1 & 6 & 0 & -3 & -1 & 2 \\  1 & -2 & -1 & 3 & 2 & -3 & -3 & 2 & 3 & -1 & -2 & 1 & 0 & 0 \\ \end{array} \right)\end{small}
\eeaa

\item{$\bar{P}_{([1],[3],[3])}({\bf 6_{1}^{3}}; a,q)=$}
\beaa
&&\tfrac{q^{1/2}(1-a) (1-a q) (1-a q^2) }{ a^{1/2}(1-q)^3  (1-q^2)^2 (1-q^3)^2}\\
&&\begin{small}\left( \begin{array}{cccccccccccccccccccc}  0 & 0 & 0 & 0 & 0 & 0 & 0 & 0 & 0 & 0 & 0 & 0 & 0 & 0 & 0 & 0 & 0 & 1 & 0 & 0 \\  0 & 0 & 0 & 0 & 0 & 0 & 0 & 0 & 0 & 0 & 0 & 0 & -1 & 0 & -2 & 3 & -1 & 0 & -3 & 0 \\  0 & 0 & 0 & 0 & 0 & 0 & 0 & 0 & 1 & 2 & -3 & -1 & -2 & 8 & 0 & 0 & -5 & 2 & 2 & 2 \\  0 & 0 & 0 & 0 & 0 & -3 & 3 & 4 & -1 & -7 & -6 & 11 & 4 & -2 & -9 & 0 & 5 & 0 & -1 & -2 \\  0 & 0 & 2 & -5 & 1 & 3 & 7 & -7 & -10 & 4 & 11 & 2 & -9 & -5 & 7 & 1 & 0 & -3 & 2 & 0 \\  1 & -2 & 0 & -1 & 6 & -1 & -5 & -3 & 4 & 6 & -3 & -5 & 2 & 1 & 2 & -3 & 1 & 0 & 0 & 0 \\ \end{array} \right)\end{small}
\eeaa

\item{$\bar{P}_{([2],[2],[2])}({\bf 6_{1}^{3}}; a,q)=$}
\beaa
\tfrac{a(1-a) (1-a q) }{q(1-q)^3 (1-q^2)^3}
\begin{footnotesize}\left( \begin{array}{ccccccccccccccc}  0 & 0 & 0 & 0 & 0 & 0 & 0 & 0 & 0 & 0 & 0 & 0 & 1 & 0 & 0 \\  0 & 0 & 0 & 0 & 0 & 0 & 0 & 0 & 0 & -3 & 0 & 4 & -2 & -3 & 0 \\  0 & 0 & 0 & 0 & 0 & 0 & 3 & 2 & -11 & 3 & 16 & -8 & -8 & 7 & 2 \\  0 & 0 & 0 & -1 & -3 & 9 & 3 & -25 & 8 & 28 & -20 & -14 & 15 & 1 & -5 \\  0 & 1 & -2 & -5 & 15 & 4 & -35 & 12 & 38 & -28 & -17 & 23 & -1 & -7 & 3 \\  2 & -3 & -9 & 18 & 11 & -41 & 7 & 43 & -27 & -18 & 22 & -1 & -6 & 2 & 0 \\  1 & -4 & 2 & 12 & -17 & -8 & 28 & -8 & -17 & 12 & 2 & -4 & 1 & 0 & 0 \\ \end{array} \right)\end{footnotesize}
\eeaa

\item{$\bar{P}_{([2],[2],[3])}({\bf 6_{1}^{3}}; a,q)=$}
\beaa
&&\tfrac{a^{1/2}(1-a) (1-a q) (1-a q^2) }{q^{1/2} (1-q)^3 (1-q^2)^3 (1-q^3)}\times\\
&&\begin{footnotesize}\left( \begin{array}{ccccccccccccccccccc}  0 & 0 & 0 & 0 & 0 & 0 & 0 & 0 & 0 & 0 & 0 & 0 & 0 & 0 & 0 & 0 & 1 & 0 & 0 \\  0 & 0 & 0 & 0 & 0 & 0 & 0 & 0 & 0 & 0 & 0 & 0 & -2 & -1 & 2 & 2 & -2 & -3 & 0 \\  0 & 0 & 0 & 0 & 0 & 0 & 0 & 0 & 1 & 4 & -5 & -6 & 5 & 9 & 2 & -10 & -3 & 7 & 2 \\  0 & 0 & 0 & 0 & 0 & -3 & 2 & 8 & -3 & -14 & -4 & 19 & 11 & -17 & -13 & 7 & 10 & -2 & -5 \\  0 & 0 & 2 & -5 & -3 & 13 & 6 & -16 & -18 & 15 & 29 & -11 & -25 & 2 & 15 & 4 & -8 & -2 & 3 \\  1 & 0 & -6 & 0 & 16 & 1 & -25 & -6 & 27 & 13 & -22 & -15 & 14 & 10 & -7 & -3 & 2 & 0 & 0 \\  1 & -3 & -1 & 10 & -3 & -13 & 3 & 12 & 3 & -13 & -3 & 10 & -1 & -3 & 1 & 0 & 0 & 0 & 0 \\ \end{array} \right)\end{footnotesize}
\eeaa

\item{$\bar{P}_{([2],[3],[3])}({\bf 6_{1}^{3}}; a,q)=$}
\beaa
&&\tfrac{a^{1/2}(1-a) (1-a q) (1-a q^2) }{q^{1/2} (1-q)^3  (1-q^2)^3 (1-q^3)^2}\times\\
&&\begin{tiny}\left( \begin{array}{cccccccccccccccccccccccccc}  0 & 0 & 0 & 0 & 0 & 0 & 0 & 0 & 0 & 0 & 0 & 0 & 0 & 0 & 0 & 0 & 0 & 0 & 0 & 0 & 0 & 0 & 0 & -1 & 0 & 0 \\  0 & 0 & 0 & 0 & 0 & 0 & 0 & 0 & 0 & 0 & 0 & 0 & 0 & 0 & 0 & 0 & 0 & 0 & 1 & 2 & 0 & -3 & 0 & 2 & 3 & 0 \\  0 & 0 & 0 & 0 & 0 & 0 & 0 & 0 & 0 & 0 & 0 & 0 & 0 & 0 & -3 & -1 & 2 & 7 & -1 & -12 & -5 & 6 & 8 & -2 & -7 & -2 \\  0 & 0 & 0 & 0 & 0 & 0 & 0 & 0 & 0 & 0 & 3 & 2 & -9 & -7 & 10 & 21 & -2 & -27 & -10 & 19 & 20 & -4 & -15 & -1 & 5 & 5 \\  0 & 0 & 0 & 0 & 0 & 0 & -1 & -3 & 9 & 7 & -16 & -22 & 12 & 44 & 3 & -49 & -25 & 30 & 37 & -10 & -28 & -3 & 11 & 6 & -4 & -3 \\  0 & 0 & 0 & 1 & -2 & -6 & 9 & 18 & -12 & -39 & -2 & 59 & 29 & -56 & -51 & 28 & 56 & -3 & -39 & -7 & 17 & 8 & -6 & -4 & 3 & 0 \\  0 & 2 & -2 & -8 & 3 & 21 & 5 & -37 & -27 & 43 & 52 & -27 & -64 & 0 & 56 & 15 & -33 & -16 & 13 & 11 & -5 & -4 & 2 & 0 & 0 & 0 \\  1 & -2 & -3 & 5 & 9 & -5 & -22 & 2 & 30 & 9 & -28 & -22 & 22 & 21 & -10 & -14 & 1 & 9 & -1 & -3 & 1 & 0 & 0 & 0 & 0 & 0 \\ \end{array} \right)\end{tiny}
\eeaa

\end{itemize}

\subsubsection{$ 6_3^3$ link}
\begin{figure}[h]
\centering{\includegraphics[scale=.8]{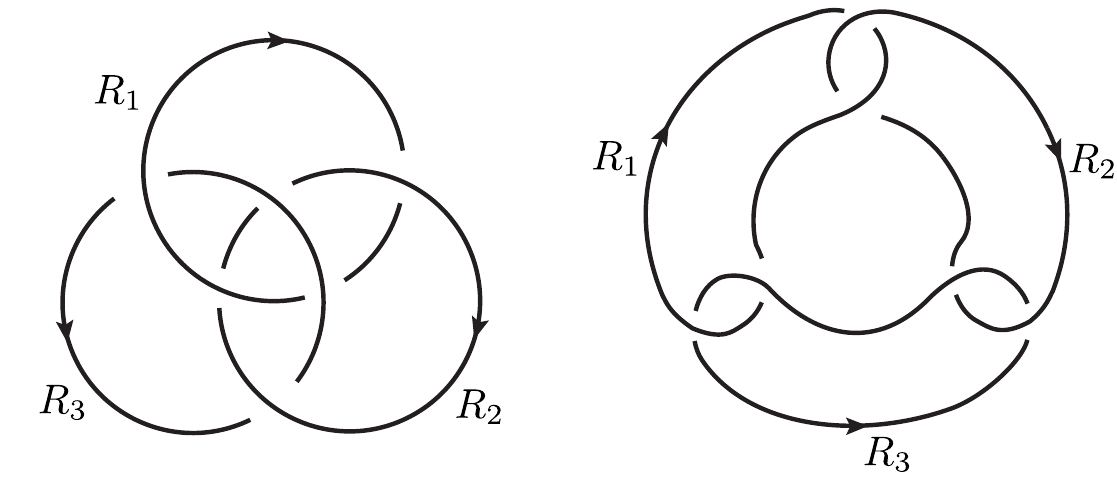}}
\caption{${\bf 6_3^3}$ link}
\end{figure}

\begin{eqnarray*}
\bar{P}_{(R_{1},R_{2},R_{3})}({\bf 6_3^3};\, a,q) & = &q^{\frac{{\ell}^{(1)}{\ell}^{(2)}}{N}+\frac{{\ell}^{(2)}{\ell}^{(3)}}{N}-\frac{{\ell}^{(1)}{\ell}^{(3)}}{N}} \sum_{l,x,y,z}\frac{1}{\epsilon_{l}^{R_{1},\bar{R}_{1}}\sqrt{\dim_{q}l}}\epsilon_{x}^{\bar{R}_{1},R_{2}}\sqrt{\dim_{q}x}\,\\
 &  & \times\epsilon_{y}^{R_{2},R_{3}}\sqrt{\dim_{q}y}\,\epsilon_{z}^{R_{1},R_{3}}\sqrt{\dim_{q}z}\, a_{lx}\!\left[\begin{footnotesize}\begin{array}{cc}
R_{1} & \bar{R}_{1}\\
R_{2} & \bar{R}_{2}
\end{array}\end{footnotesize}\right]\, a_{yl}\!\left[\begin{footnotesize}\begin{array}{cc}
R_{2} & R_{3}\\
\bar{R}_{3} & \bar{R}_{2}
\end{array}\end{footnotesize}\right]\,\\
 &  & \times  a_{zl}\!\left[\begin{footnotesize}\begin{array}{cc}
R_{1} & R_{3}\\
\bar{R}_{3} & \bar{R}_{1}
\end{array}\end{footnotesize}\right](\lambda_{x}^{(-)}(\bar{R}_{1},R_{2}))^{-2}\,(\lambda_{y}^{(+)}(R_{2},R_{3}))^{-2}\,(\lambda_{z}^{(+)}(R_{1},R_{3}))^{2}\end{eqnarray*}

The colored HOMFLY invariants of the link ${\bf 6_3^3}$ are symmetric under the interchange of the representations $R_1$ and $R_3$.
\begin{itemize}
\item{$\bar{P}_{([1],[1],[1])}({\bf 6_3^3}; a,q)=$}
\beaa
  \tfrac{(1-a)}{a^{5/2}q^{1/2} (1-q)^3}
\left( \begin{array}{ccccc}  0 & 2 & -3 & 2 & 0 \\  -1 & 1 & -2 & 1 & -1 \\  0 & 1 & -1 & 1 & 0 \\ \end{array} \right)
\eeaa

\item{$\bar{P}_{([1],[1],[2])}({\bf 6_3^3}; a,q)=\bar{P}_{([2],[1],[1])}({\bf 6_3^3}; a,q)=$}
\beaa
 \tfrac{(1-a) (1-a q) }{a^{3} q (1-q)^3 (1-q^2)}\begin{small}\left( \begin{array}{ccccccc}  0 & 0 & 2 & -2 & 0 & 1 & 0 \\  -1 & 1 & 0 & -2 & 0 & 1 & -1 \\  0 & 1 & -1 & 0 & 1 & 0 & 0 \\ \end{array} \right)\end{small}
\eeaa

\item{$\bar{P}_{([1],[1],[3])}({\bf 6_3^3}; a,q)= \bar{P}_{([3],[1],[1])}({\bf 6_3^3}; a,q)=$}
\beaa
\begin{small} \tfrac{(1-a) (1-a q) (1-a q^2)}{a^{7/2} q^{3/2} (1-q)^3 (1-q^2) (1-q^3)}\left( \begin{array}{ccccccccc}  0 & 0 & 0 & 2 & -2 & 1 & -1 & 1 & 0 \\  -1 & 1 & 0 & 0 & -2 & 0 & 0 & 1 & -1 \\  0 & 1 & -1 & 0 & 0 & 1 & 0 & 0 & 0 \\ \end{array} \right)
\end{small}
\eeaa

\item{$\bar{P}_{([1],[2],[1])}({\bf 6_3^3}; a,q)=$}
\beaa
\tfrac{(1-a) (1-a q) }{a^3 q^2(1-q)^3  (1-q^2)}
\begin{small}\left( \begin{array}{cccccc}  1 & -1 & 1 & 0 & -1 & 1 \\  -1 & 0 & 0 & -1 & 1 & -1 \\  0 & 1 & -1 & 1 & 0 & 0 \\ \end{array} \right)\end{small}
\eeaa

\item{$\bar{P}_{([1],[2],[2])}({\bf 6_3^3}; a,q) = \bar{P}_{([2],[2],[1])}({\bf 6_3^3}; a,q)=$}
\beaa
\tfrac{(1-a) (1-a q) }{a^{7/2}q^{7/2} (1-q)^3  (1-q^2)^2}\begin{small}\left( \begin{array}{cccccccccc}  0 & 0 & -2 & 2 & 2 & -5 & 1 & 3 & -2 & 0 \\  1 & -1 & 1 & 3 & -3 & -1 & 4 & -1 & -1 & 1 \\  -1 & 0 & 1 & -2 & -1 & 1 & 0 & -1 & 0 & 0 \\  0 & 1 & -1 & 0 & 1 & 0 & 0 & 0 & 0 & 0 \\ \end{array} \right)\end{small}
\eeaa

\item{$\bar{P}_{([1],[2],[3])}({\bf 6_3^3}; a,q) = \bar{P}_{([3],[2],[1])}({\bf 6_3^3}; a,q)=$}
\beaa
\tfrac{(1-a) (1-a q) (1-a q^2) }{a^4 q^5 (1-q)^3  (1-q^2)^2 (1-q^3)}\begin{small}\left( \begin{array}{ccccccccccccc}  0 & 0 & 0 & -2 & 1 & 2 & 0 & -3 & -1 & 3 & 0 & -1 & 0 \\  1 & -1 & 0 & 1 & 3 & -1 & -3 & 1 & 2 & 1 & -1 & -1 & 1 \\  -1 & 0 & 1 & 0 & -2 & -1 & 0 & 1 & 0 & -1 & 0 & 0 & 0 \\  0 & 1 & -1 & 0 & 0 & 1 & 0 & 0 & 0 & 0 & 0 & 0 & 0 \\ \end{array} \right)\end{small}\eeaa

\item{$\bar{P}_{([1],[3],[1])}({\bf 6_3^3}; a,q)$ }
\beaa
\tfrac{(1-a q) (1-a) (1-a q^2)}{a^{7/2} q^{7/2} (1-q)^3 (1-q^2) (1-q^3)}\begin{small}
\left( \begin{array}{cccccccc}  1 & 0 & -1 & 1 & -1 & 2 & -2 & 1 \\  -1 & 0 & -1 & 1 & -1 & 1 & -1 & 0 \\  0 & 1 & -1 & 1 & 0 & 0 & 0 & 0 \\ \end{array} \right)
\end{small}\eeaa

\item{$\bar{P}_{([1],[3],[2])}({\bf 6_3^3}; a,q) = \bar{P}_{([2],[3],[1])}({\bf 6_3^3}; a,q)=$}
\beaa
\tfrac{(1-a)(1-a q) (1-a q^2) }{a^4 q^4  (1-q)^3 (1-q^2)^2 (1-q^3)}\begin{small}
\left( \begin{array}{cccccccccccc}  0 & -1 & 1 & -1 & 0 & 1 & 0 & -1 & -1 & 1 & 1 & -1 \\  1 & 0 & 0 & 1 & 1 & -1 & -1 & 1 & 2 & -1 & -1 & 1 \\  -1 & 0 & 0 & 0 & -2 & 0 & 1 & 0 & -1 & 0 & 0 & 0 \\  0 & 1 & -1 & 0 & 1 & 0 & 0 & 0 & 0 & 0 & 0 & 0 \\ \end{array} \right)\end{small}\eeaa

\item{$\bar{P}_{([1],[3],[3])}({\bf 6_3^3}; a,q) = \bar{P}_{([3],[3],[1])}({\bf 6_3^3}; a,q)=$}
\bee\tfrac{(1-a) (1-a q) (1-a q^2)}{a^{9/2}q^{17/2}   (1-q)^3 (1-q^2)^2 (1-q^3)^2} \begin{footnotesize}
\left( \begin{array}{cccccccccccccccccc}  0 & 0 & 0 & 0 & 2 & -2 & -1 & 1 & 3 & 0 & -5 & 0 & 4 & 1 & -1 & -3 & 2 & 0 \\  0 & -1 & 1 & -1 & -2 & -1 & 3 & 2 & -4 & -4 & 2 & 4 & 0 & -4 & 0 & 1 & 1 & -1 \\  1 & 0 & 0 & 0 & 2 & 3 & -2 & -1 & -1 & 4 & 1 & -1 & -1 & 0 & 1 & 0 & 0 & 0 \\  -1 & 0 & 0 & 1 & -2 & -1 & -1 & 1 & 0 & 0 & -1 & 0 & 0 & 0 & 0 & 0 & 0 & 0 \\  0 & 1 & -1 & 0 & 0 & 1 & 0 & 0 & 0 & 0 & 0 & 0 & 0 & 0 & 0 & 0 & 0 & 0 \\ \end{array} \right)
\end{footnotesize}\eee
\item{$\bar{P}_{([2],[1],[2])}({\bf 6_3^3}; a,q)=$}
\beaa 
\begin{small}\tfrac{(1-a) (1-a q) }{a^{7/2} q^{3/2} (1-q)^3  (1-q^2)^2}
\left( \begin{array}{cccccccccc}  0 & 0 & 0 & 0 & 0 & -2 & 1 & 2 & -2 & 0 \\  0 & 0 & 1 & 1 & -3 & 1 & 5 & -3 & -1 & 2 \\  -1 & 1 & 1 & -3 & 0 & 2 & -2 & -1 & 1 & -1 \\  0 & 1 & -1 & -1 & 2 & 0 & -1 & 1 & 0 & 0 \\ \end{array} \right)\end{small}\eeaa

\item{$\bar{P}_{([2],[1],[3])}({\bf 6_3^3}; a,q) = \bar{P}_{([3],[1],[2])}({\bf 6_3^3}; a,q)=$}
\beaa
\tfrac{(1-a) (1-a q) (1-a q^2) }{a^4 q^2 (1-q)^3 (1-q^2)^2 (1-q^3)}\begin{small}
\left( \begin{array}{ccccccccccccc}  0 & 0 & 0 & 0 & 0 & 0 & 0 & -2 & 1 & 1 & 0 & -1 & 0 \\  0 & 0 & 0 & 1 & 1 & -3 & 0 & 3 & 2 & -1 & -2 & 1 & 1 \\  -1 & 1 & 1 & -1 & -2 & 0 & 2 & 0 & -2 & -1 & 0 & 1 & -1 \\  0 & 1 & -1 & -1 & 1 & 1 & 0 & -1 & 0 & 1 & 0 & 0 & 0 \\ \end{array} \right)\end{small}\eeaa

\item{$\bar{P}_{([2],[2],[2])}({\bf 6_3^3}; a,q)=$}
\beaa 
\tfrac{(1-a) (1-a q) }{a^5 q^4 (1-q)^3(1-q^2)^3}\begin{small}
\left( \begin{array}{cccccccccccccc}  0 & 0 & 0 & 0 & 0 & 3 & -3 & -4 & 7 & 1 & -5 & 1 & 1 & 0 \\  0 & 0 & -1 & -1 & 2 & 0 & -5 & 1 & 4 & -3 & -2 & 2 & 0 & -1 \\  1 & -1 & 0 & 3 & 0 & -2 & 2 & 2 & 0 & 0 & 0 & 0 & 1 & 0 \\  -1 & 0 & 2 & -2 & -2 & 2 & -1 & -2 & 1 & 0 & -1 & 0 & 0 & 0 \\  0 & 1 & -1 & -1 & 2 & 0 & -1 & 1 & 0 & 0 & 0 & 0 & 0 & 0 \\ \end{array} \right)\end{small}\eeaa

\item{$\bar{P}_{([2],[2],[3])}({\bf 6_3^3}; a,q) = \bar{P}_{([3],[2],[2])}({\bf 6_3^3}; a,q)=$}
\bee\tfrac{(1-a) (1-a q) (1-a q^2)}{a^{11/2}q^{11/2}  (1-q)^3 (1-q^2)^3 (1-q^3)}\begin{footnotesize}
\left( \begin{array}{ccccccccccccccccc}  0 & 0 & 0 & 0 & 0 & 0 & 0 & 3 & -2 & -4 & 3 & 2 & 1 & -2 & -2 & 2 & 0 \\  0 & 0 & 0 & -1 & -1 & 2 & 2 & -4 & -5 & 2 & 6 & 0 & -6 & -2 & 4 & 1 & -2 \\  1 & -1 & -1 & 2 & 2 & 0 & -3 & -1 & 6 & 3 & -3 & -3 & 2 & 3 & -1 & -1 & 1 \\  -1 & 0 & 2 & 0 & -3 & -1 & 2 & 1 & -2 & -2 & 0 & 1 & 0 & -1 & 0 & 0 & 0 \\  0 & 1 & -1 & -1 & 1 & 1 & 0 & -1 & 0 & 1 & 0 & 0 & 0 & 0 & 0 & 0 & 0 \\ \end{array} \right)
\end{footnotesize}\eee
\item{$\bar{P}_{([2],[3],[2])}({\bf 6_3^3}; a,q)=$}
\beaa
\tfrac{(1-a) (1-a q) (1-a q^2) }{a^{11/2}q^{15/2}  (1-q)^3 (1-q^2)^3 (1-q^3)}\begin{small}
\left( \begin{array}{ccccccccccccccccc}  0 & 0 & 0 & 0 & 2 & -1 & -4 & 4 & 3 & -3 & -3 & 1 & 5 & -2 & -3 & 2 & 0 \\  0 & -1 & 1 & 0 & -4 & 0 & 4 & 1 & -5 & -3 & 4 & 2 & -2 & -2 & 1 & 1 & -1 \\  1 & 0 & -1 & 1 & 2 & 2 & -1 & -2 & 2 & 1 & 1 & 0 & -1 & 0 & 1 & 0 & 0 \\  -1 & 0 & 1 & 0 & -2 & -1 & 1 & -1 & -1 & 1 & 0 & -1 & 0 & 0 & 0 & 0 & 0 \\  0 & 1 & -1 & -1 & 2 & 0 & -1 & 1 & 0 & 0 & 0 & 0 & 0 & 0 & 0 & 0 & 0 \\ \end{array} \right)
\end{small}\eeaa

\item{$\bar{P}_{([2],[3],[3])}({\bf 6_3^3}; a,q)=\bar{P}_{([3],[3],[2])}({\bf 6_3^3}; a,q)  =$}
\beaa
&&\tfrac{(1-a) (1-a q) (1-a q^2) }{a^6q^{10}(1-q)^3 (1-q^2)^3 (1-q^3)}\times \\
&&\begin{footnotesize}
\left( \begin{array}{ccccccccccccccccccccccc}  0 & 0 & 0 & 0 & 0 & 0 & 0 & 0 & -3 & 2 & 4 & 0 & -7 & -4 & 9 & 5 & -5 & -6 & 1 & 5 & -1 & -1 & 0 \\  0 & 0 & 0 & 0 & 1 & 1 & -2 & 0 & 3 & 4 & -3 & -7 & 3 & 8 & 2 & -7 & -4 & 5 & 3 & -1 & -2 & 0 & 1 \\  0 & -1 & 1 & 0 & -2 & -2 & 1 & 2 & -2 & -6 & 0 & 3 & 2 & -4 & -3 & 2 & 1 & -1 & -1 & 0 & 1 & -1 & 0 \\  1 & 0 & -1 & 0 & 2 & 2 & 0 & -2 & 2 & 3 & 2 & -1 & -2 & 2 & 2 & 0 & -1 & 0 & 1 & 0 & 0 & 0 & 0 \\  -1 & 0 & 1 & 1 & -2 & -2 & 1 & 1 & -1 & -2 & -1 & 1 & 0 & 0 & -1 & 0 & 0 & 0 & 0 & 0 & 0 & 0 & 0 \\  0 & 1 & -1 & -1 & 1 & 1 & 0 & -1 & 0 & 1 & 0 & 0 & 0 & 0 & 0 & 0 & 0 & 0 & 0 & 0 & 0 & 0 & 0 \\ \end{array} \right) \end{footnotesize}\eeaa

\item{$\bar{P}_{([3],[1],[3])}({\bf 6_3^3}; a,q)=$}
\beaa
&&  \tfrac{(1-a) (1-a q) (1-a q^2) }{a^{9/2} q^{5/2}(1-q)^3  (1-q^2)^2 (1-q^3)^2}\times\\
&&\begin{footnotesize}
\left( \begin{array}{cccccccccccccccccc}  0 & 0 & 0 & 0 & 0 & 0 & 0 & 0 & 0 & 0 & 0 & 0 & 2 & -1 & 0 & -2 & 2 & 0 \\  0 & 0 & 0 & 0 & 0 & 0 & 0 & -1 & -1 & 1 & 2 & -1 & -3 & -3 & 3 & 1 & 0 & -2 \\  0 & 0 & 0 & 1 & 0 & 0 & -3 & 1 & 4 & 2 & -2 & -4 & 3 & 4 & 0 & -1 & -1 & 2 \\  -1 & 1 & 1 & 0 & -3 & -1 & 3 & 2 & -2 & -3 & 0 & 2 & -1 & -1 & -1 & 1 & -1 & 0 \\  0 & 1 & -1 & -1 & 0 & 2 & 1 & -2 & -1 & 1 & 1 & 0 & -1 & 1 & 0 & 0 & 0 & 0 \\ \end{array} \right)
\end{footnotesize}\eeaa

\item{$\bar{P}_{([3],[2],[3])}({\bf 6_3^3}; a,q)=$}
\beaa
&& \tfrac{(1-a) (1-a q) (1-a q^2) }{a^6 q^7  (1-q)^3(1-q^2)^3 (1-q^3)^2}\times\\
&&\begin{footnotesize}
\left( \begin{array}{ccccccccccccccccccccccc}  0 & 0 & 0 & 0 & 0 & 0 & 0 & 0 & 0 & 0 & 0 & 0 & -3 & 1 & 4 & 1 & -4 & -4 & 3 & 3 & -1 & -1 & 0 \\  0 & 0 & 0 & 0 & 0 & 0 & 0 & 1 & 1 & 1 & -5 & -1 & 7 & 7 & -4 & -11 & 2 & 9 & 2 & -4 & -3 & 2 & 1 \\  0 & 0 & 0 & -1 & 0 & 0 & 2 & 0 & -4 & -4 & 1 & 6 & 0 & -8 & -4 & 3 & 4 & -2 & -3 & 0 & 1 & 0 & -1 \\  1 & -1 & -1 & 1 & 3 & 0 & -3 & -2 & 3 & 4 & 2 & -2 & -1 & 1 & 3 & 1 & 0 & 0 & 0 & 0 & 1 & 0 & 0 \\  -1 & 0 & 2 & 1 & -3 & -3 & 2 & 4 & -1 & -4 & -1 & 2 & 0 & -2 & -1 & 1 & 0 & -1 & 0 & 0 & 0 & 0 & 0 \\  0 & 1 & -1 & -1 & 0 & 2 & 1 & -2 & -1 & 1 & 1 & 0 & -1 & 1 & 0 & 0 & 0 & 0 & 0 & 0 & 0 & 0 & 0 \\ \end{array} \right)\end{footnotesize}\eeaa

\item{$\bar{P}_{([3],[3],[3])}({\bf 6_3^3}; a,q)$ }
\beaa
&&\tfrac{(1-a) (1-a q) (1-a q^2) }{a^{15/2} q^{23/2} (1-q)^3  (1-q^2)^3 (1-q^3)^3} \times\\
&&\begin{tiny}
\left( \begin{array}{ccccccccccccccccccccccccccccc}  0 & 0 & 0 & 0 & 0 & 0 & 0 & 0 & 0 & 0 & 0 & 0 & 0 & 4 & -3 & -4 & -2 & 7 & 8 & -7 & -7 & -2 & 7 & 5 & -4 & -1 & -2 & 2 & 0 \\  0 & 0 & 0 & 0 & 0 & 0 & 0 & 0 & -1 & -1 & -1 & 3 & 2 & -3 & -9 & -2 & 11 & 9 & -7 & -15 & -1 & 12 & 6 & -6 & -7 & 2 & 3 & 1 & -2 \\  0 & 0 & 0 & 0 & 1 & 0 & 0 & -1 & 1 & 3 & 3 & 1 & -4 & -3 & 6 & 9 & 4 & -11 & -6 & 5 & 10 & 1 & -6 & -2 & 3 & 1 & 0 & -1 & 1 \\  0 & -1 & 1 & 0 & -1 & -2 & 0 & 2 & 0 & -3 & -4 & -4 & 1 & 1 & 2 & -5 & -6 & -1 & 4 & 2 & -3 & -4 & 1 & 1 & 1 & -2 & 0 & 0 & 0 \\  1 & 0 & -1 & -1 & 2 & 3 & -1 & -2 & 0 & 3 & 2 & 1 & 3 & 1 & -1 & -1 & 2 & 4 & 0 & -1 & -1 & 1 & 1 & 0 & 0 & 0 & 0 & 0 & 0 \\  -1 & 0 & 1 & 2 & -2 & -3 & 0 & 3 & 1 & -3 & -2 & 1 & 0 & -1 & -2 & 1 & 0 & 0 & -1 & 0 & 0 & 0 & 0 & 0 & 0 & 0 & 0 & 0 & 0 \\  0 & 1 & -1 & -1 & 0 & 2 & 1 & -2 & -1 & 1 & 1 & 0 & -1 & 1 & 0 & 0 & 0 & 0 & 0 & 0 & 0 & 0 & 0 & 0 & 0 & 0 & 0 & 0 & 0 \\ \end{array} \right)\end{tiny}\eeaa

\end{itemize}


\section{Conclusions}\label{sec:conclusion}
In this paper, we evaluate colored HOMFLY invariants carrying symmetric representations of various non-tours knots and links by using the multiplicity-free $SU(N)$ quantum Racah coefficients \cite{Nawata:2013ppa} in the context of Chern-Simons theory. This method provides a powerful tool to demonstrate explicit computations of colored HOMFLY invariants.

From the observation in \S\ref{sec:links}, we predict the following properties of multi-colored HOMFLY invariants of links.
For an $s$-component link $\cal L$, an unreduced colored HOMFLY invariant ${\overline P}_{([n_1],\cdots,[n_s])}({\cal L};a,q)$ 
contains the unknot factor ${\overline P}_{[n_{\rm max}]}(\bigcirc;a,q)$ colored by the highest rank $n_{\rm max}=\max (n_1,\cdots,n_s)$.
Therefore, it is reasonable to define the reduced colored HOMFLY invariants  ${P}_{([n_1],\cdots,[n_s])}({\cal L};a,q)$ by
\bee
{P}_{([n_1],\cdots,[n_s])}({\cal L};a,q)={\overline P}_{([n_1],\cdots,[n_s])}({\cal L};a,q)/{\overline P}_{[n_{\rm max}]}(\bigcirc;a,q)
\eee
 for symmetric representations. Furthermore, if we normalize by 
\beaa
\frac{1}{(a;q)_{n_{\max}}}\left[\prod_{i=1}^s (q;q)_{n_i}\right]{\overline P}_{([n_1],\cdots,[n_s])}({\cal L};a,q)~,
\eeaa
then it becomes a Laurent polynomial with respect to the variables $(a,q)$. Moreover, the Laurent polynomials obey
the exponential growth property
\beaa
&&\lim_{q\to1}\frac{1}{(a;q)_{kn_{\max}}}\left[\prod_{i=1}^s (q;q)_{kn_i}\right]{\overline P}_{[kn_1],\cdots,[kn_s]}({\cal L};a,q)\\
&=&\left[\lim_{q\to1}\frac{1}{(a;q)_{n_{\max}}}\left[\prod_{i=1}^s (q;q)_{n_i}\right]{\overline P}_{([n_1],\cdots,[n_s])}({\cal L};a,q)\right]^k
\eeaa
where ${\rm gcd}(n_1,\cdots,n_s)=1$.

The direct line to study further is to categorify these invariants. Especially, colored HOMFLY homologies for thick knots are known only for the knots $\bf 8_{19}$ and $\bf 9_{42}$ \cite{Gukov:2011ry}. To distinguish between generic and particular properties which can be accidentally valid for simple knots, it is important to obtain explicit expressions of colored HOMFLY homologies of ten-crossing thick knots in \S\ref{sec:thick}. The colored HOMFLY homology for links will be studied in \cite{Gukov:2013}.

Although the closed form expression \cite{Nawata:2013ppa} of the multiplicity-free $SU(N)$ quantum Racah coefficients extends the scope for calculations of colored HOMFLY polynomials to some extent, we have not succeeded in obtaining the  invariants for the knot ${\bf 10_{161}}$ and the link ${\bf 7_6^2}$. In addition, the information about colored HOMFLY polynomials beyond symmetric representations are very limited \cite{Anokhina:2012rm}. To deal with more complicated links and non-symmetric representations,  further study has to be undertaken for the $SU(N)$ quantum Racah coefficients with multiplicity structure.
We hope to report this issue in future.


\section*{Acknowledgement}
The authors would like to thank Andrei Mironov, Alexei Morozov and Andrey Morozov for sharing Maple files. S.N. is indebted to Petr Dunin-Barkowski, Sergei Gukov, Kenichi Kawagoe, Alexei Sleptsov, Marko Sto$\check{\text{s}}$i$\acute{\text{c}}$, Piotr Su{\l}kowski and Miguel Tierz for valuable discussions and correspondences. In addition, S.N. is grateful to IIT Bombay for its warm hospitality.  S.N. and Z. would like to thank Indian String Meeting 2012 at Puri for providing a stimulating academic environment.  The work of S.N. is partially supported by the ERC Advanced
Grant no.~246974, {\it "Supersymmetry: a window to non-perturbative physics"}.

\appendix
\section{Fusion matrices}\label{sec:fusion}
In this appendix, we shall show the explicit expressions of the fusion matrices for some representations, which are determined by the quantum $6j$-symbols in the companion paper \cite{Nawata:2013ppa}.
\Yboxdim6pt
\begin{enumerate}
\item
$R_1=\yng(2),~~R_2=\yng(3)$
\bee
{\tiny
a_{ts}\left[\begin{footnotesize}
\begin{array}{cc} 
R_1& \bar{R}_1
\\ 
R_2& \bar{R}_2
\end{array}\end{footnotesize}\right]=
\tfrac1{\sqrt{K_{23}}}\left(
\renewcommand{\arraystretch}{1.5}
\begin{array}{c|c c c }
&\overset{{\kappa}_{0}}{s=\yng(1)}
&\overset{{\kappa}_{1}}{\young({\mdot}~~)}
&\overset{{\kappa}_{2}}{\young({\mdot}{\mdot}~~~)}
\\
\hline
t=0
& {\scriptsize\sqrt{\dim_q {\kappa}_{0}}}
&{\scriptsize-\sqrt{\dim_q {\kappa}_{1}}} 
&{\scriptsize\sqrt{\dim_q {\kappa}_{2}}}
\\ 
{\young({\mdot}~)}
&-x_1\sqrt{\tfrac{[N-1][N][N+1][N+2][N+3]}{[2][3]}}
&x_2\sqrt{\tfrac{[N][N+1][N+3]}{[3]}}
&x_3\sqrt{\tfrac{[N+2][N+3][N+4]}{[2]}}
\\ 
{\young({\mdot}{\mdot}~~)}
&y_1\sqrt{\tfrac{[N-1][N][N+2][N+3][N+4]}{[3]}}
&y_2\sqrt{\tfrac{[N][N+3][N+4]}{[2][3]}}
&y_3\sqrt{[N+1][N+2][N+3]}
\end{array}
\right)
}
\eee
where  $K_{23}=\dim_q R_1\dim_q R_2=\frac{[N]^2[N+1]^2[N+2]}{[3][2]^2}$ and the quantum dimensions are given by
\bee
\begin{array}{lll}
\dim_{q}{\kappa}_{0}=[N],~~&
\dim_{q}{\kappa}_{1}=\tfrac{[N-1][N][N+2]}{[2]},~~&
\dim_{q}{\kappa}_{2}=\tfrac{[N-1][N]^2[N+1][N+4]}{[3][2]^2}
\end{array}
\eee
and the variables $x_i's$ and $y_i's$ are given by
\bee
\renewcommand{\arraystretch}{1.5}
\begin{array}{ccc}
x_1=\tfrac{[2]}{[N+2]},~~~&x_2=\tfrac{([N-1][N+4])-[2]}{[2][N+3]},~~~&x_3=\tfrac{[N][N+1]}{[N+2][N+3]} \\
y_1=\tfrac{[N]}{[2][N+2]},~~~&y_2=\tfrac{[N][2]}{[N+3]},~~~&y_3=\tfrac{[N]}{[N+2][N+3]}.
\end{array}
\eee
{\small
\begin{equation*}
$${$a_{ts}\left[\begin{footnotesize}
\begin{array}{cc} 
R_1 
& R_2
\\ 
\bar{R}_1
& \bar{R}_2
\end{array}\end{footnotesize}\right]=$
$\left({\renewcommand{\arraystretch}{1.2} 
\begin{tabular}{c|c c c}  
{} 
t& $s=\overset{ {\kappa}_{0}}{\yng(1)}$
&$\overset{ {\kappa}_{1}}{\Yboxdim6pt\young(.~~)}$
&$\overset{ {\kappa}_{2}}{\Yboxdim6pt\young({.}{.}~~~)}$\\ 
\hline 
{$\rho_{[3,2]}$}
& {\tiny$z_1\sqrt{\frac{[N-1][N][N+1][N+2]}{[4][3][2]}}$}
&{\tiny$-\frac{[2]}{[N+1]}\sqrt{\frac{[N][N+1]}{[4][3]}}$} 
&{\tiny$\frac{[2]}{[N+2]}\sqrt{\frac{[N+2][N+4]}{[4][2]}}$}
\\ 
 {$\rho_{[4,1]}$} 
&{\tiny$-z_1[2]\sqrt{\frac{[N-1][N+1][N+2][N+3]}{[2][3][5]}}$}
&{\tiny$z_2\sqrt{\frac{[N+1][N+3]}{[3][5]}}$}
&{\tiny$z_3\sqrt{\frac{[N][N+2][N+3][N+4]}{[2][5]}}$}
\\ 
{$\rho_{[5]}$}
&{\tiny$z_1[3]\sqrt{\frac{[N+1][N+2][N+3][N+4]}{[5][4][3][2]}}$}
&{\tiny$z_4[3]\sqrt{\frac{[N-1][N+1][N+3][N+4]}{[5][4][3]}}$}
&{\tiny$z_3\sqrt{\frac{[N-1][N][N+2][N+3]}{[5][4][2]}}$}\\
  \end{tabular} }\right)$~,}$$
\end{equation*}
}where
$\begin{array}{ccc}
\rho_{[3,2]}=\Yboxdim6pt\yng(3,2),
&\rho_{[4,1]}=\Yboxdim6pt\yng(4,1),
&\rho_{[5]}=\Yboxdim6pt\yng(5)
\end{array}$ 
and the variables $z_i$'s are
\begin{eqnarray*}
z_1=\tfrac{[2]}{[N+1][N+2]},&~&z_2=z_1\left(\tfrac{[3][N+3][N+4]-[N][N-1]}
{[N]+[3][N+4]}\right),\\
z_3=\tfrac{[2]}{[N+2][N+3]},&~&z_4=\tfrac{[2]}{[N+1][N+3]}.
\end{eqnarray*}
The quantum dimension of each representation for $t$ are
\bee
\renewcommand{\arraystretch}{1.5}
\begin{array}{ll}
\dim_q \rho_{[3,2]}=\tfrac{[N-1][N]^2[N+1][N+2]}{[4][3][2]},&~~
\dim_q \rho_{[4,1]}=\tfrac{[N-1][N][N+1][N+2][N+3]}{[5][3][2]}\\
\dim_q \rho_{[5]}=\tfrac{[N][N+1][N+2][N+3][N+4]}{[5]!}.&
\end{array}
\eee

{\small
\begin{equation*}
$${$a_{ts}\left[\begin{footnotesize}
\begin{array}{cc} 
R_1 
& R_2
\\ 
\bar{R}_2
& \bar{R}_1
\end{array}\end{footnotesize}\right]=$
$\tfrac1{\sqrt{K_{23}}}\left({\renewcommand{\arraystretch}{1.2} 
\begin{tabular}{c|c c c}  
{} 
t& $s=0$
&$\overset{ {\tilde\rho}_{1}}{\Yboxdim6pt\young(.~)}$
&$\overset{ {\tilde\rho}_{2}}{\Yboxdim6pt\young({.}{.}~~)}$\\ 
\hline 
{$\rho_{[3,2]}$}
& {\tiny$\sqrt{\dim_q \rho_{[3,2]}}$}
&{\tiny$-\tfrac{[N][N+1]}{[3]}\sqrt{\frac{[N+3]}{[4]}}$} 
&{\tiny$\tfrac{[N]}{[3]}\sqrt{\frac{[N+1][N+3][N+4]}{[2][4]}}$}
\\ 
 {$\rho_{[4,1]}$} 
&{\tiny$-\sqrt{\dim_q \rho_{[4,1]}}$}
&{\tiny$-([N-2]+[N]-[N+4])\frac{[N+1]}{[2][3]}\sqrt{\frac{[N]}{[5]}}$}
&{\tiny$\tfrac{[2][N]}{[3]}\sqrt{\frac{[N][N+1][N+4]}{[2][5]}}$}
\\ 
{$\rho_{[5]}$}
&{\tiny$\sqrt{\dim_q \rho_{[5]}}$}
&{\tiny$[N + 1] \sqrt{\frac{[N-1] [N][N + 4]}{[4] [5]}}$}
&{\tiny$[N]\sqrt{\frac{[N-1] [N][N+1]}{[2][4] [5]}}$}\\
  \end{tabular} }\right)$.}$$
\end{equation*}
}

\item
$R=\yng(3)$

\bee
a_{ts}\left[
\begin{array}{cc} 
R& \bar{R}
\\ 
R& \bar{R}
\end{array}\right]=\frac{1}{K_{33}}
\left(
\renewcommand{\arraystretch}{1.5}
\scriptsize{
\begin{array}{c|c c c c}
&\overset{\tilde{\rho}_{0}}{s=0}
&\overset{\tilde{\rho}_{1}}{\young({\mdot}~)}
&\overset{\tilde{\rho}_{2}}{\young({\mdot}{\mdot}~~)}
&\overset{\tilde{\rho}_{3}}{\young({\mdot}{\mdot}{\mdot}~~~)}
\\
\hline
t=\tilde {\rho}_{0}
& {\scriptsize-\sqrt{\dim_q\tilde {\rho}_{0}}}
&{\scriptsize\sqrt{\dim_q\tilde {\rho}_{1}}} 
&{\scriptsize-\sqrt{\dim_q\tilde {\rho}_{2}}}
&{\scriptsize\sqrt{\dim_q\tilde {\rho}_{3}}}
\\ 
\tilde {\rho}_{1}
& {\scriptsize\sqrt{\dim_q\tilde {\rho}_{1}}} 
&a
&x
&y
\\ 
\tilde {\rho}_{2}
&{\scriptsize-\sqrt{\dim_q\tilde {\rho}_{2}}}
&x
&b
&z
\\
\tilde {\rho}_{3}
&{\scriptsize\sqrt{\dim_q\tilde {\rho}_{3}}}
&y
&z
&c
\end{array}
}
\right)
\eee
where $K_{33}=\dim_qR=\frac{[N][N+1][N+2]}{[2][3]}$ and the quantum dimensions of the representations are given by
\bee
\begin{array}{ll}
\dim_{q}\tilde{\rho}_{0}=1,~~~ 
&\dim_{q}\tilde{\rho}_{1}=[N-1][N+1],\\
\dim_{q}\tilde{\rho}_{2}=\tfrac{[N-1][N]^2[N+3]}{[2]^2},~~~
&\dim_{q}\tilde{\rho}_{3}=\tfrac{[N-1][N]^2[N+1]^2[N+5]}{[3]^2[2]^2}
\end{array}
\eee
and we also have
\bee
\begin{array}{lll}
\renewcommand{\arraystretch}{1.5}
a=\tfrac{[N+1](1-[2][N-1][N+4])}{[3][N+3]},&~~b=\tfrac{[N](-1+[2][N][N+5])}{[3][N+4]},&~~c=\tfrac{[N][N+1]}{[N+3][N+4]} \\
x=\tfrac{[N](-[2]^2+[N-1][N+5])}{[2][3][N+3]}\sqrt{[N+1][N+3]},&~~y=\tfrac{[N][N+1]}{[N+3][2]}\sqrt{[N+1][N+5]},& \\
z=\tfrac{[N][N+1]}{[N+3][N+4]}\sqrt{[N+3][N+5]}.&&
\end{array}
\eee

\begin{equation*}
a_{ts}\left[
\begin{array}{cc} 
R& R
\\ 
\bar{R}& \bar{R}
\end{array}\right]=\frac{1}{K_{33}}
\left(
\renewcommand{\arraystretch}{1.5}
\tiny{
\begin{array}{c|c c c c}
&\overset{\tilde{\rho}_{0}}{s=0}
&\overset{\tilde{\rho}_{1}}{\young({\mdot}~)}
&\overset{\tilde{\rho}_{2}}{\young({\mdot}{\mdot}~~)}
&\overset{\tilde{\rho}_{3}}{\young({\mdot}{\mdot}{\mdot}~~~)}
\\
\hline
t=\overset{{\rho}_{[6]}}{\yng(6)}
&-\sqrt{\dim_q{\rho}_{[6]}}
&\tfrac{[N][N+1]}{[2][3]}\sqrt{\tfrac{[N+1][N+2]}{[4]}}
&-\tfrac{[N][N+1]}{[2][3]}\sqrt{\tfrac{[N+2][N+3]}{[4]}}
&\tfrac{[N][N+1]}{[2][3]}\sqrt{\tfrac{[N+2][N+5]}{[4]}}
\\
\overset{{\rho}_{[5,1]}}{\yng(5,1)}
&\sqrt{\dim_q {\rho}_{[5,1]}}
&x_1
&x_2
&x_3
\\
\overset{{\rho}_{[4,2]}}{\yng(4,2)}
&-\sqrt{\dim_q {\rho}_{[4,2]}}
& y_1
&y_2
&y_3
\\
\overset{{\rho}_{[3,3]}}{\yng(3,3)}
&\sqrt{\dim_q {\rho}_{[3,3]}}
&z_1
&z_2
&z_3
\end{array}
}
\right)
\end{equation*}

The quantum dimensions of the representations are given by
\bee
\renewcommand{\arraystretch}{1.5}
\begin{array}{ll}
\dim_{q}{\rho}_{[3,3]}=\tfrac{[N-1][N]^2[N+1]^2[N+2]}{[2]^2[3]^2[4]},~~&
\dim_{q}{\rho}_{[4,2]}=\tfrac{[N-1][N]^2[N+1][N+2][N+3]}{[2]^2[4][5]},\\
\dim_{q}{\rho}_{[5,1]}=\tfrac{[N-1][N][N+1][N+2][N+3][N+4]}{[2][3][4][6]},~~&
\dim_{q}{\rho}_{[6]}=\tfrac{[N][N+1][N+2][N+3][N+4][N+5]}{[2][3][4][5][6]}.
\end{array}
\eee
We also have
{\small
\bee
\renewcommand{\arraystretch}{1.5}
\begin{array}{lll}
x_1=\tfrac{[N][N+1]([N-1]-[2][N+4])}{[2][3][N+3]}\sqrt{\tfrac{[N+2][N+3]}{[4][5]}},~~&x_2=\tfrac{[N](-[2][N]+[N+5])}{[2][3]}\sqrt{\tfrac{[N+1][N+2]}{[4][5]}}\\
x_3=\tfrac{[N][N+1]}{[2][3]}\sqrt{\tfrac{[N+1][N+2][N+3][N+5]}{[4][5]}},~~&y_1=\tfrac{[N+1]([N+3]-[2][N])}{[N+3]}\sqrt{\tfrac{[N][N+2][N+3][N+4]}{[2][3][4][6]}}\\
y_2= \tfrac{[N]([N+5][N+2]+[2N+4])}{[N+4][N+2]}\sqrt{\tfrac{[N][N+1][N+2][N+4]}{[2][3][4][6]}},~~&y_3=\tfrac{[N][N+1]}{[N+4][N+3]}\sqrt{[N][N+1][N+2][N+3][N+4]}\\
z_1=\frac{[3][N+1]}{[N+3]}\sqrt{\frac{[N-1][N][N+2][N+3][N+4][N+5]}{[2][3][4][5][6]}},~~&z_2=\frac{[3][N]}{[N+4]}\sqrt{\frac{[N-1][N][N+1][N+2][N+4][N+5]}{[2][3][4][5][6]}}\\
z_3=\frac{[N][N+1]}{[N+3][N+4]}\sqrt{\frac{[N-1][N][N+1][N+2][N+3][N+4]}{[4]}}.
\end{array} 
\eee
}
\end{enumerate}

\bibliography{CS}{}
\bibliographystyle{JHEP}

\end{document}